%% file: paper.tex
\documentclass[letterpaper,12pt,leqno]{article}
\usepackage{paper}
\usepackage{adjustbox}
\usepackage{bigfoot}
\usepackage{pdfpages}
\usepackage{changepage}
\newcommand{\citetitlemanual}[1]{\textit{Title of the Document}}
\hypersetup{pdftitle={Paper Title}}
\newcommand{\bib}{bibliography.bib}

\begin{document}
\title{\Large Why Do Contract Workers Earn Less?\\Evidence from India's Auto Industry}
\author{Davide Luparello \thanks{\href{mailto: dluparello@psu.edu}{dluparello@psu.edu} }}
\affil{{\normalsize \it The Pennsylvania State University} }
\date{October 26, 2025\\
\textcolor{blue}{\textbf{JOB MARKET PAPER}} \\
\textcolor{blue}{{\href{https://drive.google.com/file/d/114D7KnUx4t69ELQCk_Jjl0UdGMaU3d5N/view?usp=sharing}{\small[click here for the most updated version]}}}}                      
\begin{titlepage}\maketitle
\begin{abstract}
Contract workers constitute half of India's automotive employment but earn substantially less than permanent workers. Using ASI data (2002-2019), I develop an estimator of labor supply and demand schedules to explain this wage premium. The model features worker-type-specific discrete choice labor supply, nested CES production, Nash-Bertrand competition for contract workers, and plant-union bargaining for permanent workers. I find the premium stems entirely from higher productivity rather than differential monopsony power. While a lump-sum transfer offsetting wage markdowns would increase welfare by 14\% for permanent and 12\% for contract workers, it would simultaneously increase the premium by 14\%, exacerbating inequality.
\newline
JEL CODES: L11, L13, L62
\end{abstract}

\end{titlepage}

\doublespacing
\section{Introduction}\label{s:introduction}

\input{introduction.tex}

\section{Data and Stylized Facts}\label{s:data}
\input{data.tex}

\section{Empirical Framework for Workers' Supply and Demand}\label{s:theory}

\input{theory.tex}
\section{Empirical Strategy}\label{s:empirics}
\input{empirics.tex}

\section{Results and Discussion}\label{s:results}
\input{results.tex}
\section{Concluding Remarks}\label{s:conclusion}

\input{conclusion.tex}
\input{figures_tables.tex}
\clearpage
\bibliography{\bib}
\clearpage
\appendix
\include{paper_appendixa}

\end{document}

%% file: introduction.tex
Workers hired via contractors and staffing companies constitute half of total employment in the Indian automotive sector, by now the world's fourth-largest automobile producer. However, despite possessing similar qualifications and performing comparable tasks to permanent staff---particularly after 2001---these contract workers earn substantially less \citep{duvisac2019reconstituting, bertrand2021contract}\footnote{By law, contract workers should be engaged in tasks that are non-perennial and not regularly performed by permanent workers. However, a 2001 Supreme Court ruling (SAIL) eliminated mandatory reclassification requirements for contract workers in ``core'' production activities. \cite{bertrand2021contract} document a marked increase in both the use and employment share of contract workers by large plants following this decision. \cite{cox2012eyes}, \cite{sundar2012organizing}, and \cite{fudge2012challenging} describe the SAIL decision as ``de facto'' deregulation.}. This paper addresses the following question: What explains the wage premium of permanent over contract workers in the Indian automotive industry between 2002 and 2019?

Permanent workers may earn a wage premium through two distinct channels. First, selection effects in the hiring process as well as tenure may lead to permanent workers being more productive, thereby justifying higher compensation. Second, unlike permanent employees, contract workers typically have short-term contracts with third-party agencies, earn significantly lower wages with minimal allowances, and lack both job security and collective bargaining power \citep{duvisac2019reconstituting}. Institutional differences in wage-setting mechanisms, combined with varying labor supply elasticities across worker types, may enable producers to exercise differential monopsony power, resulting in higher wage markdowns that generate the wage premium.

Using the Annual Survey of Industries (2002--2019), this paper introduces an estimator of workers' supply and demand schedules that enables the decomposition of the permanent worker wage premium into productivity and monopsony channels. On the supply side, I model workers of each type (permanent or contract) as choosing among workplaces within a discrete choice framework \citep{azar2022estimating, rubens2024exploiting}. Worker type is fixed at the moment of workplace choice, and each type has a nested logit supply function with type-specific wage sensitivity and within-nest correlation of alternatives, where nests are defined as Indian states. This specification allows for heterogeneous labor supply elasticities across worker types. In terms of conduct, the contract worker market features Nash-Bertrand competition, while permanent workers are modeled as unionized at the plant level, with unions and plants bargaining over the surplus consisting of the plant's profits and permanent workers' welfare. These differences in labor supply elasticities and market structure enable firms to exercise differential monopsony power across worker types, contributing to the observed wage premium.

Worker-type specific demand derives from a production model featuring a nested CES framework where contract and permanent workers are modeled as distinct factors of production, thus allowing for differentiated marginal revenue products that embody productivity differences between worker types. Producers also incur non-wage, unobservable, and variable employment costs, including overhead, training costs, and other expenses not captured by payroll. The difference between each worker type's marginal revenue product and their respective non-wage marginal employment costs generates worker-type specific marginal values of production, which determine input demand for each worker type.

My empirical strategy proceeds through three stages. First, I estimate labor supply elasticities for both permanent and contract workers via GMM using the nested logit framework. Identification relies on the orthogonality of instrumental variables to unobservable workplace amenities. Variation in the number of plants active in each nest over time identifies the nesting parameter, while labor demand shifters serve as instruments for wage sensitivity parameters.

Second, I estimate production function parameters and unobservable components, including plant-level productivity and markups. Following \cite{doraszelski2018measuring}, I construct GMM estimating equations by exploiting ratios of the model's equilibrium first-order conditions and the production function, which I combine with productivity evolution equations. Identification relies on the orthogonality of instrumental variables to productivity innovations. To identify the CES substitution and share parameters, I employ plant cost shifters and lagged input allocation as instruments, based on the model's timing assumptions. The presence of multiple unobservables—markups, productivity, and markdowns—violates the scalar unobservability assumption \citep{olley1996dynamics}. Thus, I employ Kalman filtering techniques to disentangle productivity from measurement error \citep{hong2024search}. This approach assumes that productivity innovations and measurement error are serially and cross-sectionally uncorrelated, each maintaining constant variance.

Third, building on these production and labor supply estimates, I recover the bargaining power of permanent worker unions in plant-level negotiations. I exploit the first-order condition for optimal permanent worker allocation, which yields an equation where the bargaining parameter and the non-wage marginal cost of employing permanent workers are the only unknowns. For identification, I use the lagged interaction between state-level strike intensity and plant exposure to strikes as an instrument. Contemporaneous strike activity correlates positively with plant surplus—since larger surpluses expand bargaining scope—but may also correlate with non-wage marginal costs per worker (e.g., via legal expenses), raising endogeneity concerns. The lagged measure addresses this issue by exploiting the persistence of strike activity while avoiding contemporaneous confounding factors.

I find that the wage premium between permanent and contract workers is entirely attributable to the higher productivity of permanent workers. Permanent workers generate substantially greater marginal revenue product—74 USD per workday versus 55 USD for contract workers—despite comparable non-wage marginal employment costs (33 USD per workday for both types). However, plants simultaneously exercise greater monopsony power over permanent workers, stemming from their lower estimated labor supply elasticity, as unions have limited bargaining power relative to plants. Consequently, plants retain 20–25\% of the marginal value permanent workers generate compared to only 15\% for contract workers. The higher marginal value of production more than compensates for these higher markdowns, generating the observed wage premium for permanent workers. This productivity heterogeneity has important policy implications: a lump-sum transfer to workers offsetting wage markdowns would increase welfare by 14\% for permanent workers and 12\% for contract workers, but would simultaneously increase the compensation premium of permanent workers by 14\%, thereby exacerbating compensation inequalities.

This paper makes two principal contributions to the literature. First, it contributes to the examination of permanent and contract workers in India. Among recent research investigating this feature of the Indian labor market\footnote{Among others, \cite{chaurey2015labor, saha2013trade, kapoor2017informality, bertrand2021contract, chatterjee2024no}.}, \cite{bertrand2021contract} demonstrate how expanded contract labor utilization enables plants to circumvent regulatory constraints and minimize workforce adjustment costs, thereby accelerating job creation, enhancing product innovation, and reducing the average product of labor in larger establishments. \cite{chatterjee2024no} quantify heterogeneous adjustment costs between worker types, estimating lower hiring costs for contract workers and substantially higher firing costs for permanent workers. This paper proposes a distinct analytical framework that abstracts from these dynamic costs and instead leverages a tractable static model to investigate differences in productivity, labor supply elasticity, and market conduct across worker types as drivers of the permanent worker wage premium\footnote{My results complement research documenting lower wages following domestic outsourcing beyond India: \cite{dube2010does} for US janitors and guards, \cite{drenik2023paying} for Argentina, \cite{felix2024reallocation} for Brazilian security guards, and \cite{weil2014fissured} for the broader US labor market. Related work establishes links to increased firm rent capture \citep{appelbaum2017domestic} and heightened wage inequality \citep{bilal2021outsourcing} in France, as well as similar patterns across food, cleaning, security, and logistics services in Germany \citep{goldschmidt2017rise}.}.

Second, I introduce a markdown estimator that integrates heterogeneous labor supply elasticities and factor market conduct across worker types, while controlling for imperfect output market competition, factor-biased technological change, and unobservable input costs. This methodological advancement extends the production approach to markup estimation, which has typically operated under Hicks-neutrality assumptions \citep{loecker2012markups, morlacco2019market, brooks2021exploitation, mertens2022micro, rubens2023market}. The estimator also builds on research examining directed technical change under competitive factor markets \citep{doraszelski2018measuring, zhang2019non, demirer2020production, raval2023testing, hong2024search} and studies employing \cite{berry1994estimating} share inversion to measure labor monopsony power \citep{azar2022estimating, rubens2024exploiting}. Unlike \cite{rubens2024exploiting}, my framework permits heterogeneous conduct assumptions across production inputs, better reflecting institutional realities. Finally, I employ a multi-step estimation approach that resolves the circular dependency noted by \cite{rubens2025labor} in a similar class of worker-plant bargaining models: estimating production parameters requires observed bargaining weights, while estimating bargaining weights requires observed production parameters.

The paper proceeds as follows. Section \ref{s:data} introduces the primary data sources and presents stylized facts about the automotive industry. Section \ref{s:theory} develops the theoretical framework, covering labor supply functions, production technology, and market conduct assumptions, culminating in the surplus maximization problem that integrates these elements. Section \ref{s:empirics} outlines the empirical strategy for estimating the unknown parameters and latent variables governing labor supply, demand, and market conduct. Section \ref{s:results} presents the parameter estimates and analyzes the determinants of the permanent versus contract worker wage premium. Section \ref{s:conclusion} concludes and discusses directions for future research.

%% file: data.tex
\subsection{Production Data}

Production data are drawn from the 2002--2019 Annual Survey of Industries (ASI), a panel dataset compiled by the Indian Ministry of Statistics and Programme Implementation (MOSPI) that covers formal manufacturing establishments. For each establishment-year observation, the data include establishment location by state, as well as detailed information on production inputs and outputs at the item code level\footnote{Before 2011, item codes follow the Annual Survey of Industries Commodity Classification (ASICC), a hierarchical system organized by product categories and subcategories, typically with 5-digit codes. After 2011, the classification follows the National Product Classification for Manufacturing Sector (NPCMS), which replaced ASICC and uses a typical 7-digit code structure.}. On the output side, I observe revenue and quantities sold. On the input side, I obtain costs and quantities of both domestic and imported intermediates. The data also provide comprehensive labor information, including man-days worked and total payroll (comprising wages, bonuses, welfare expenditures, and provident fund premiums) disaggregated by worker type (supervisors, permanent workers, and contract workers), as well as net opening book values of production capital (plant and machinery). I calculate daily remuneration rates for each worker type by dividing total compensation by man-days worked.

Following \cite{orr2022within}, I construct yearly establishment-level price indexes for output and intermediates using Cobb-Douglas aggregation of item-code level prices\footnote{Specifically, let $P_{jt}^k$ denote the price of output item $k$. I construct the establishment-level output price index, $P_{jt}$, as
\begin{equation*}
    P_{jt} = \prod_k\left(\frac{P^k_{jt}}{\gamma^k_{jt}}\right)^{\gamma^k_{jt}}
\end{equation*}
where $\gamma^k_{jt}$ represents the revenue share of item $k$. I construct the establishment-level intermediate price index, $P_{jt,M}$, analogously from item-level intermediate prices $P_{jt,M}^k$, using weights $\gamma^k_{jt,M}$ that capture each material's expenditure share.}. Dividing establishment-level revenue and intermediate costs by their respective price indexes yields the corresponding quantity indexes. All nominal variables are deflated using India KLEMS deflators for the transport equipment industry, with 2005 as the base year. I apply output deflators to revenue and payroll variables, material input price deflators to intermediate expenditure, and capital-specific deflators to capital stock. INR/USD exchange rates are sourced from FRED.

The ASI dataset exhibits well-documented limitations that merit consideration. Beyond its restriction to formal manufacturing, the dataset employs differential establishment sampling: establishments exceeding 100 workers appear annually (Census Sector), while smaller establishments enter through random stratified sampling by industry and state (Sample Sector). To ensure consistent coverage, I restrict the sample to manufacturers of commercial vehicles, personal vehicles, two- and three-wheelers, and producers of engines, chassis, and other motor vehicles, capturing end-producer vehicle assemblers\footnote{These establishments operate under the following 5-digit NIC industry codes: 98 NIC---34101, 34102, 34103, 34104, 34105, 35911, 35912, 35913, 34106, 34107; 04 NIC---34101, 34102, 34103, 34104, 34105, 35911, 35912, 35913, 34106, 34107, 34109; `08 NIC---29102, 29101, 30911, 30912, 29103, 29104, 29109.}. Additionally, the ASI permits firms with multiple plants within a state to file joint returns \citep{anand2025multiplying}, causing some observations to reflect aggregated information across multiple plants rather than single-establishment data. However, this limitation do not impact my analysis since I focus on intensive production margins and their implications for the scale-invariant wage premium.

Following \cite{ruhl2017new}, I eliminate observations exhibiting extraordinary variation in input usage, input expenditure, output revenue, and quantity sold\footnote{For output, managers, permanent and contract workers, and intermediate inputs, the threshold is 180 percent change. For capital stock, the threshold is 100 percent change.}. After applying these filters, the final sample comprises 86 individual plants spanning 383 establishment-year observations\footnote{Appendix \ref{a:Vars} presents boxplot distributions of relevant production variables in the final sample.}.

\subsection{Permanent and Contract Workers}
Contract workers in India are not directly employed by companies but are hired through third-party agencies or contractors \citep{CEDA2024contractualisation}. This arrangement is governed by the Contract Labour (Regulation and Abolition) Act of 1970, which applies to establishments employing more than 20 workers on a contractual basis. Under this system, no direct employment relationship exists between the principal employer (the auto manufacturer) and the contract workers. Instead, the contractor is responsible for hiring workers and managing all conditions of employment \citep{Dubey_contract_labour}\footnote{The contract worker expenditures reported in the ASI represent payments to contractors or staffing agencies, which likely extract additional rent from workers. Consequently, my markdown estimates for contract workers abstract from this double marginalization and represent lower bounds from the workers' perspective.}. The Act requires both principal employers and contractors to obtain registrations and licenses, and mandates that wages and benefits be paid to contract workers in a timely and adequate manner \citep{Dubey_contract_labour}.

Contract workers cannot legally be engaged in core activities that are necessary to the functioning of the establishment—their engagement must be temporary and not of a continuing nature \citep{Dubey_contract_labour}. However, the Act permits contract labor for certain activities that would otherwise be considered core, including sanitation work (sweeping, cleaning, waste disposal) and security services \citep{Dubey_contract_labour}. A landmark 2001 Supreme Court judgment—Steel Authority of India Ltd. vs. National Union Waterfront Workers (SAIL)—significantly altered this landscape by eliminating mandatory reclassification requirements for contract workers in core production activities. This ruling created explicit incentives for firms to substitute contract labor for permanent employees, effectively establishing a "de facto" deregulation without any formal changes to labor laws \citep{bertrand2021contract}.

Contract labor dominates India's automotive industry, with the transport equipment sector exhibiting among the highest contract labor intensities across Indian manufacturing \citep{CEDA2024contractualisation}. Contract workers perform virtually all stages of vehicle production—from parts fabrication through final assembly to warehouse operations. Beyond a narrow set of highly skilled or sensitive tasks, automakers deploy contract labor for most shop-floor operations at 30\%--70\% lower wages \citep{ET2011_contract_labour}. Employment terms sharply distinguish contract workers: they hold temporary agreements with staffing agencies, earn substantially lower wages with minimal allowances, and lack both employment security and union representation \citep{ET2011_contract_labour, IndustriALL2018_indian_auto}.

Figure \ref{fig:con_vs_perm} examines the compensation and utilization patterns of permanent and contract workers over time. Panel \ref{fig:con_vs_permA} presents average relative wage trajectories for contract versus permanent workers across establishments. A persistent wage gap favors permanent workers, who earn approximately twice the contract worker wage. Panel \ref{fig:con_vs_permB} displays the proportion of total worker-days contributed by contract labor across establishments. Despite receiving systematically lower wages, contract workers constitute a substantial fraction of total labor input, accounting for 40--50\% of recorded worker-days between 2004 and 2019.

\begin{center}
[Figure \ref{fig:con_vs_perm} goes here]
\end{center}

\subsection{Segmented Labor Markets}\label{sec:geo_conc}
Automotive producers cluster into location-specific production hubs, with the most prominent located in the states of Haryana (headquarters of Maruti Suzuki), Tamil Nadu (home to Chennai, recognized as the ``Detroit of Asia''), and Maharashtra (home to Mahindra \& Mahindra). Geographic distance and mobility frictions—including linguistic barriers \citep{manjunath2024language} and location-specific social networks \citep{bhattacharyya1985role}—constrain workers' labor supply elasticities by limiting their ability to move between auto plants across production clusters. 

Figure \ref{fig:prod_loc} examines the spatial concentration of Indian automotive production. Panel \ref{fig:prod_loc_panA} displays each state's mean contribution to total industry revenue across the sample period. Panel \ref{fig:prod_loc_panB} documents the geographic distribution and count of active production facilities by state as recorded in 2019. I observe pronounced spatial concentration: Haryana, Tamil Nadu, and Maharashtra collectively generate 85\% of annual industry revenue, on average.

\begin{center}
[Figure \ref{fig:prod_loc} goes here]
\end{center}

Spatial agglomeration also enables automotive assemblers to exercise competitive pressure throughout their supply chains by exploiting the fragmented structure of the auto parts industry, which is populated by numerous small enterprises \citep{diebolt2016synoptic, uchikawa2011small} and characterized by dense supplier networks surrounding assembly hubs \citep{diebolt2016comparative}.

\subsection{Workers and Unions in India’s Auto Industry}\label{sec:workers}
Permanent workers in India's automotive industry have the legal right to form and join trade unions under the Trade Unions Act of 1926. In many auto companies, an in-house union already exists, and permanent workers can typically join by signing a membership form and paying a nominal fee. The union's constitution specifies eligibility—usually, any permanent employee of the company can become a member. 

While contract workers are not legally prohibited from joining or forming trade unions, they rarely participate as members in official factory-level unions \citep{duvisac2019reconstituting}. Several factors explain this absence. First, many existing trade unions historically restricted membership to permanent workers, excluding contract labor from their constitutions \citep{NewsClick2022_haryana_union}. Second, managements implicitly and sometimes explicitly discourage unionization of the contractual workforce, pressuring permanent unions not to admit contract workers \citep{NewsClick2022_haryana_union}. Third, the precarious employment status of contract workers themselves deters union participation. Unlike permanent employees—who enjoy stronger job security and can only be terminated for cause or retrenched with notice and compensation under the Industrial Disputes Act of 1947—contract workers can be easily terminated or replaced, making union activism particularly risky for them.

Indian autoworker unions tend to operate as plant- or enterprise-level organizations rather than industry-wide federations. Maruti Suzuki exemplifies this institutional structure: despite shared ownership, the company's Manesar and Gurgaon plants maintained distinct unions until 2012, when they consolidated under the Maruti Suzuki Workers' Union to coordinate actions. While national federations such as the All India Trade Union Congress (AITUC) exist, their influence remains limited to providing support and advice to plant-level unions \citep{duvisac2019reconstituting}.

This fragmented structure has left unions vulnerable to systematic obstruction from both state authorities and corporate management. State governments deploy multiple administrative mechanisms to impede union formation, including withholding essential registration documents, rescinding recognition from established unions, and strategically reclassifying firms as public utilities to eliminate workers' legal strike rights \citep{duvisac2019reconstituting}. Management opposition manifests through direct confrontation with organizing efforts. The Maruti Suzuki Manesar case illustrates this dynamic: following workers' union registration in 2011, management immediately imposed suspensions, triggering labour disputes that culminated in a July 2012 workplace incident resulting in mass arrests of 147 workers on murder charges \citep{nair2021regimes}. 

Despite these obstacles, unions have achieved limited success in wage improvements for permanent employees. Early strikes, such as the 2005 Honda action, secured significant wage increases \citep{nair2021regimes}. After the 2014 union election at Maruti Suzuki, the reconstituted union negotiated agreements covering wages, workload, leave policies, and bonus structures—yet these gains accrued exclusively to permanent workers \citep{duvisac2019reconstituting}. However, state repression and pro-business reforms have progressively weakened bargaining power, as state priorities in India have shifted from preserving industrial peace to enhancing firm competitiveness \citep{duvisac2019reconstituting}.

Variation in strike intensity thus provides useful additional information on bargaining activity between unions and plants. I source annual workdays lost due to industrial disputes (strikes and lockouts) from 2006 to 2020 by state from official government publications. These statistics encompass establishments in both the Central and State Spheres\footnote{The Central Sphere comprises industries and establishments under national government authority, including railways, mines, oil fields, major ports, and corporations established through central legislation. The State Sphere covers employments and industries within each state that fall outside the Central Sphere.}. I compile these data from the \textit{Indian Labour Statistics} (\citeyear{ILS0708,ILS20}), \textit{Statistics on Industrial Disputes, Closures, Retrenchments, and Lay-offs in India} (\citeyear{IDR08,IDR09,IDR10,IDR11,IDR12,IDR13,IDR14,IDR15,IDR16,IDR17}), and the \textit{Indian Labour Year Book} (\citeyear{ILYB22}).

Figure \ref{fig:lab_disp} examines cross-sectional and temporal variation in labor dispute intensity across Indian states where automotive plants are located. I construct this measure as the ratio of state-year man-days lost to industrial disputes relative to total state-year man-days worked in formal manufacturing by permanent workers. Panel \ref{fig:lab_disp_panA} presents state-level averages across the sample period, while Panel \ref{fig:lab_disp_panB} displays annual cross-state averages. Over the sample period, dispute intensity declines from 0.6\% to 0.2\%, yielding moderate persistence with an autocorrelation coefficient of 0.22. Across states, intensity ranges from 0.2\% in Gujarat to 1.5\% in Andhra Pradesh.

\begin{center}
[Figure \ref{fig:lab_disp} goes here]
\end{center}

%% file: theory.tex
In this section, I develop the empirical framework for labor supply and demand in India's automotive sector. On the supply side, permanent and contract workers make discrete choices among employment opportunities. On the demand side, I specify automotive plants' production technology as a nested CES production function in which permanent and contract workers are distinct inputs. These plants also face unobservable non-wage employment costs that influence their labor demand decisions. I model wage determination through two distinct mechanisms: Bertrand-Nash wage setting for contract workers and bilateral Nash bargaining between plants and unions for permanent workers.

\subsection{Auto Workers Labor Supply}\label{sec:workers_sup}

Labor market segmentation, mobility frictions\footnote{The sector's organization into localized production hubs generates segmented labor markets through inter-cluster mobility frictions, including local social capital \citep{bhattacharyya1985role}, linguistic barriers \citep{manjunath2024language}, and relocation costs. See Section \ref{sec:geo_conc} for details.}, and discrete workplace decisions generate upward-sloping worker supply curves. I treat workers' employment classification—contract or permanent—as exogenously fixed at the point of workplace selection\footnote{I assume that workers cannot choose their employment type when selecting workplaces. Endogenizing this choice would allow wage changes to affect labor supply through both workforce compositional shifts and cross-plant mobility. By excluding the compositional margin, I obtain a lower bound on labor supply elasticities. I leave this extension for future research. Anecdotally, official labor regulations contain no provision for regularizing contract workers into permanent positions. The Ministry of Labour explicitly stated in 2013 that ``there is no provision of regularisation under the Contract Labour Act, 1970, and therefore no proposal to regularise the contract workers''. Such conversions are likely negligible in practice.}. To derive labor supply curves for each worker type, I define the choice set as automotive sector establishments, with non-automotive manufacturing employment serving as the outside option. I nest these workplace choices at the state level.

Following \cite{azar2022estimating} and \cite{rubens2024exploiting}, I specify the indirect utility for worker $i$ of type $X$ employed at establishment $j$ in nest $r$ at time $t$ as:
\begin{equation}
\small
U_{ixjt}=\underbrace{\alpha_{x}t_{trend} +\gamma_{xt} W_{jt,x}+\xi_{xjt}}_{\delta_{xjt}}+\zeta_{ixrt}+(1-\eta_x)\epsilon_{ixjt}\quad  x\in\{D,C\}
\end{equation}
where nests $r$ correspond to states. The mean utility component $\delta_{xjt}$ comprises three establishment-specific elements: a time trend, daily wages $W_{jt,x}$, and unobserved establishment characteristics ("amenities"), $\xi_{xjt}$. The idiosyncratic terms $\zeta_{ixrt}$ and $\epsilon_{ixjt}$ represent worker preference heterogeneity across nests and establishments, respectively, under Type-I extreme value standard distributional assumptions \citep{cardell1997variance}. I allow the wage sensitivity parameter $\gamma_{xt}$ to vary across time and employment type. The nesting parameters $0<\eta_x<1$ govern the degree of within-nest substitutability among establishments for type-$x$ workers.

Utility maximization yields choice probabilities consistent with the nested logit model \citep{mcfadden1981econometric}. Adopting the market share notation of \cite{berry1994estimating} and normalizing against the outside option, I derive the labor supply equation:
\begin{equation}
\small
\log\left(\frac{s_{xjt}}{s_{x0t}}\right) = \alpha_{x}t_{trend} + \gamma_{xt} W_{jt,x}  + \eta_x\log(s_{xj|rt}) + \xi_{xjt}\quad  x\in\{D,C\}
\label{eq:lab_supply}
\end{equation}
where $s_{xjt}$ denotes the share of type-$x$ workers employed at establishment $j$ in period $t$, $s_{x0t}$ represents their share in non-automotive manufacturing, and $s_{xj|rt}$ captures their conditional share at establishment $j$ within state $r$. Establishment amenities $\xi_{xjt}$ correlate with both wages and conditional shares, generating endogeneity that I address through instrumental variables. I present the identification strategy in Section \ref{s:empirics}.

Following \cite{rubens2024exploiting}, I derive the inverse labor supply elasticity faced by plant $j$ in period $t$ for each worker type from equation \eqref{eq:lab_supply}:
{\small
\begin{equation}\label{eq:inv_elas_D}
\frac{\partial W_{jt,x}}{\partial X_{jt}}\frac{X_{jt}}{W_{jt,x}}=\frac{1-{\eta}_x}{{\gamma}_{xt}W_{jt,x} \left( 1 - {\eta}_x s_{xj|rt}-(1-{\eta}_x)s_{xjt}\right)}\quad  x\in\{D,C\}.
\end{equation}}
This elasticity depends on the wage rate $W_{jt,x}$, the unconditional and conditional employment shares $s_{xjt}$ and $s_{xj|rt}$, the wage sensitivity parameter $\gamma_{xt}$, and the nesting parameter $\eta_x$.

\subsection{Production Technology}
An automotive plant combines capital ($K_{jt}$), intermediate inputs ($M_{jt}$), and labor ($L_{jt}$) to generate output quantity ($Q_{jt}$) using a constant elasticity of substitution (CES) production technology exhibiting constant returns to scale:
\begin{equation}
\small
Q_{jt} = \left(\tilde{\alpha}_K K_{jt}^{\sigma^O} + \tilde{\alpha}_M M_{jt}^{\sigma^O} + \tilde{\alpha}_L (\exp(\omega_{jt}^{L}) L_{jt})^{\sigma^O} \right)^\frac{1}{\sigma^O}\exp(\omega_{jt}^{H}) \label{eq:prod}
\end{equation}
The parameter $\sigma^O$ captures the elasticity of substitution across all factors, while $\tilde{\alpha}_K$, $\tilde{\alpha}_M$, and $\tilde{\alpha}_L$ denote the respective factor share parameters. Following \cite{doraszelski2018measuring}, I model productivity heterogeneity through two channels: $\omega_{jt}^{H}$ represents Hicks-neutral productivity that augments all factors proportionally, while $\omega^L_{jt}$ captures labor-augmenting productivity that specifically enhances labor effectiveness\footnote{Hicks-neutral technological innovations include lean supply chain management practices \citep{kumar2023assessment}, Advanced High Strength Steel manufacturing techniques \citep{perka2022advanced}, Six Sigma quality control methodologies \citep{sharma2018six}, and integrated automotive clusters generating locational efficiency gains \citep{okada2007industrial}. Labor-augmenting innovations include advanced automation processes \citep{kulkarni2019recent} and worker skill enhancement programs \citep{kumar2023assessment}.}.

I model labor as a nested CES structure. At the top level, supervisors ($S_{jt}$) combine with workers ($E_{jt}$):
\begin{equation}
\small
L_{jt} = \left( \tilde{\alpha}_S S_{jt}^{\sigma^M} + \tilde{\alpha}_E E_{jt}^{\sigma^M} \right)^\frac{1}{\sigma^M}
\end{equation}
The parameters $\tilde{\alpha}_S$ and $\tilde{\alpha}_E$ denote the factor share parameters, while $\sigma^M$ governs the elasticity of substitution between supervisors and workers. In turn, I specify workers as a CES aggregate of permanent ($D_{jt}$) and contract ($C_{jt}$) employees:
\begin{equation}
\small
E_{jt} = \left(\tilde{\alpha}_C C_{jt}^{\sigma^I} + \tilde{\alpha}_D D_{jt}^{\sigma^I} \right)^\frac{1}{\sigma^I}\label{eq:blue_collar}
\end{equation}
Here, $\sigma^I$ measures substitutability between worker types, while $\tilde{\alpha}_C$ and $\tilde{\alpha}_D$ represent their respective factor shares\footnote{
The factor share parameters capture mean productivity differentials across worker categories, while I exclude worker-type-specific technical change. This specification choice reflects the empirical reality that contractors supply workers across the entire skill distribution—from college-educated professionals to casual laborers—while providing essential operational training \citep{barnes2015labour}.
}.

I classify inputs by their adjustment flexibility. Intermediate inputs, supervisors, and both permanent and contract workers constitute variable inputs, while capital remains predetermined. I depart from prior literature in two respects. First, I treat supervisors as variable rather than fixed inputs. Empirical patterns justify this specification: supervisor allocation and expenditure exhibit pooled cross-sectional elasticities to output of 0.60 (0.03) and 0.77 (0.04), respectively, while year-over-year changes show elasticities of 0.47 (0.08) and 0.66 (0.09) (plant-clustered standard errors in parentheses). Both levels and changes reveal substantial co-movement with output, inconsistent with fixed-cost behavior. Second, I abstract from dynamic adjustment costs in labor demand. While recent research \citep{bertrand2021contract, chatterjee2024no} incorporates hiring and firing frictions, I adopt a static framework to isolate productivity and monopsony differentials across worker types.

\subsection{Technical Change Dynamics and Measurement Error}
I model technical change as independent, exogenous AR(1) processes for both Hicks-neutral ($H$) and labor-augmenting ($L$) productivity components:
\begin{equation}
\small
\label{eq:productivity}
    \omega_{jt}^{i} = \iota_{i,t} + \rho_i \omega_{jt-1}^{i} + \xi_{jt}^{i}\quad~i\in\{H,L\}
\end{equation}
where $\iota_{i,t}$ captures aggregate time effects, and $\xi_{jt}^{i}$ represents (log) productivity innovations.
\begin{assumption}[Productivity Innovations]\label{asst}
Let $\mathcal{I}_{jt}$ denote the information set available to plant $j$ when allocating variable inputs at time $t$. Plant $j$ observes current productivity innovations but faces uncertainty regarding future innovations when making variable input decisions. Log productivity innovations have zero unconditional mean. Formally, $\forall j$ and $\forall t$:
\begin{equation}
\small
\mathbb{E}(\xi^i_{jt})=0;\quad \mathbb{E}(\xi^i_{jt}|\mathcal{I}_{jt})=\xi^i_{jt};\quad \mathbb{E}(\xi^i_{jt+1}|\mathcal{I}_{jt})=0\quad  i\in\{H,L\}.
\end{equation}
Hicks-neutral productivity innovations $\xi_{jt}^H$ are serially uncorrelated with finite, constant variance:
\begin{equation}
\small
    \mathbb{E}(\xi_{jt}^H\times\xi_{j\tau}^H)=\begin{cases}
        (\sigma^H)^2 & \text{if } \tau=t\\
        0 & \text{if } \tau\neq t
    \end{cases}
\end{equation}
\end{assumption}
This information structure follows standard practice in the production function literature \citep{olley1996dynamics, ackerberg2015identification, doraszelski2018measuring}, wherein plants observe current productivity realizations but face uncertainty about future innovations.

The econometrician does not perfectly observe plant output. After output realizes at the end of each period, plants record this information with measurement error. I denote observed plant output as $\tilde{Q}_{jt}$:
\begin{equation}
\tilde{Q}_{jt}=Q_{jt}\exp(\varepsilon_{jt})
\label{eq:obs_prod}
\end{equation}
where $\varepsilon_{jt}$ represents (log) measurement error.
\begin{assumption}[Measurement Error]\label{assm}
The (log) measurement error $\varepsilon_{jt}$ has zero mean and remains unobserved by the plant when allocating variable inputs. The measurement error is orthogonal to Hicks-neutral productivity at all leads and lags, serially uncorrelated, and has finite, constant variance. Formally, $\forall j$ and $\forall t$:
\begin{equation}
\small
    \mathbb{E}(\varepsilon_{jt})=0; \quad \mathbb{E}(\varepsilon_{jt}|\mathcal{I}_{jt})=0; \quad \mathbb{E}(\varepsilon_{jt}\times\xi_{j\tau}^H)=0\quad \forall \tau; \quad \mathbb{E}(\varepsilon_{jt}\times\varepsilon_{j\tau})=\begin{cases}
        (\sigma^\varepsilon)^2 & \text{if } \tau=t\\
        0 & \text{if } \tau\neq t
    \end{cases}
\end{equation}
\end{assumption}

The constant variance assumptions on Hicks-neutral productivity innovations and measurement errors is key to my identification strategy. By restricting both $\xi_{jt}^H$ and $\varepsilon_{jt}$ to time-invariant dispersion, I can separately estimate productivity from measurement error through deconvolution via Kalman filtering. However, this approach necessarily precludes environments with time-varying productivity uncertainty\footnote{See \cite{luparello2023productivity} for an application examining how time-varying productivity dispersion affects input misallocation.}.

\subsection{Input and Output Markets}
The upward-sloping labor supply curves generate variable wage markdowns through distinct mechanisms for each worker type. I model contract worker wages $W_{jt,C}$ through Bertrand-Nash competition and permanent worker wages $W_{jt,D}$ through bilateral Nash bargaining between plants and unions. Supervisors exhibit lower substitutability and greater spatial mobility than production workers: their skills raise replacement costs, while higher wages offset relocation expenses more effectively. These characteristics place supervisors in competitive labor markets with competitively determined wages $W_{jt,S}$. Global supplier competition and dense supply networks surrounding assembly hubs justify treating material input markets as competitive, with prices $P_{jt,M}$. Plants operate in imperfectly competitive output markets, setting prices $P_{jt}$ at variable markups over marginal cost. These positive profits generate the surplus over which plants and permanent worker unions bargain.

\subsection{Equilibrium Allocations of Contract Workers, Supervisors, and Materials}
I model plants as profit-maximizing entities that select contract workers, supervisors, and materials conditional on the equilibrium allocation of permanent workers. In the contract labor market, plants face an upward-sloping labor supply curve, which they internalize when making hiring decisions. This market structure allows plants to exercise monopsony power, setting wages through Bertrand-Nash competition. The total cost of employment extends beyond wages to include unobservable non-wage costs $\Phi_{jt}$ associated with screening, training, search, and supervision activities. These costs apply to both contract and permanent workers. Given this framework, I specify the plant's conditional optimization problem as follows:
\begin{equation}
\small
\begin{aligned}
\label{eq:max_problem}
\max_{\{C_{jt},S_{jt},M_{jt}\}}&\quad\Pi_{jt} \\
\text{s.t.}\quad\Pi_{jt}&=P_{jt}(Q_{jt})Q_{jt}-W_{jt,C}(C_{jt})C_{jt} - W^*_{jt,D}D^*_{jt}-W_{jt,S}S_{jt}-P_{jt,M}M_{jt}-\Phi_{jt}\\
Q_{jt} &= \left(\tilde{\alpha}_K {K}_{jt}^{\sigma^O} + \tilde{\alpha}_M {M}_{jt}^{\sigma^O} + \tilde{\alpha}_L (\exp({\omega}_{jt}^{L}) {L}_{jt})^{\sigma^O} \right)^\frac{1}{\sigma^O}\exp({\omega}_{jt}^{H})\\
{L}_{jt} &= \left( \tilde{\alpha}_S {S}_{jt}^{\sigma^M}+ \tilde{\alpha}_E {E}_{jt}^{\sigma^M} \right)^\frac{1}{\sigma^M}\\
{E}_{jt} &= \left( \tilde{\alpha}_C {C}_{jt}^{\sigma^I} + \tilde{\alpha}_D (D^*_{jt})^{\sigma^I} \right)^\frac{1}{\sigma^I}\\
\omega_{jt}^{i} &= \iota_{i,t} + \rho_i \omega_{jt-1}^{i} + \xi_{jt}^{i}\quad~i\in\{H,L\}\\
U_{iCjt}&=\alpha_{C}t_{\text{trend}} +\gamma_{Ct} W_{jt,C}+\xi_{Cjt}+\zeta_{iCrt}+(1-\eta_C)\epsilon_{iCjt}
\end{aligned}
\end{equation}
where asterisks denote equilibrium allocations and wages for permanent workers. I derive the first-order conditions for contract workers, supervisors, and materials in Appendix \ref{a:FOCs_orig}.

\subsection{Modeling Permanent Worker Wage Bargaining}

I model permanent workers as organized into unions that bargain with individual plants over surplus division. In each period, unions and plants engage in bilateral wage negotiations, with each union bargaining separately and simultaneously with each plant. Wages are determined as Nash equilibria of these bilateral bargaining problems\footnote{This approach was first introduced by \cite{horn1988bilateral} to model firm-supplier bargaining. See \cite{grennan2013price} for an application to hospital bargaining in a medical device market.}. Formally, the permanent worker wage rate maximizes the Nash product of plant and union surplus, conditional on the optimal choices of contract workers, supervisors, materials, and contract worker wages, as well as the permanent worker wage rates and amenities offered by other plants. The bargaining problem is:
{\small
\begin{equation}
\begin{aligned}
\label{eq:nash_bargaining}
\max_{W_{jt,D}}&\quad\left(U^U_{jt}-U^U_{Ojt}\right)^{\beta}\left(\Pi_{jt}-\Pi_{Ojt}\right)^{1-\beta}
\end{aligned}
\end{equation}}
where $\beta \in [0,1]$ represents the union's bargaining power, $U^U_{jt}$ denotes union utility, and $\Pi_{jt}$ denotes plant profits. The terms $U^U_{Ojt}$ and $\Pi_{Ojt}$ represent the union and plant disagreement payoffs, respectively.

I specify the disagreement point by assuming that bargaining failure leads to permanent worker exit in the current period and immediate plant shutdown. This assumption reflects the institutional context in which legal plant operation requires permanent worker employment. Under this specification, disagreement profits equal zero: $\Pi_{Ojt}=0$.

I define union agreement utility as the aggregate expected utility, in monetary terms, that permanent workers derive from the automotive sector labor market at time $t$:
{\small
\begin{equation}
    U^U_{jt}=\frac{1}{\gamma_{Dt}}\int_{i\in D_t}\mathbb{E}\left[\max_g\max_{j\in B_g}U_{iDjt}\right]di=\frac{1}{\gamma_{Dt}}D_{t}\underbrace{\log\left(\sum_{g}\left(\sum_{j \in B_g} \exp\left(\frac{\delta_{Djt}}{1-\eta}\right)\right)^{(1-\eta)}\right)}_{IV_{Dt}}
\end{equation}}
where $IV_{Dt}$ denotes the inclusive value from permanent workers' discrete choice problem at time $t$\footnote{See Appendix \ref{a:Inc_Val} for a discussion on the inclusive value in the nested logit framework.}, and $D_{t}$ denotes permanent worker employment in the automotive sector at time $t$.

Under bargaining failure, permanent workers choose to work for alternative plants, but their choice set contracts as plant $j$ exits the market. I define union disagreement utility as the aggregate expected utility permanent workers obtain when plant $j$ is excluded from the choice set, holding wages and amenities at alternative plants fixed:
{\small
\begin{equation}
    U^U_{Ojt}=\frac{1}{\gamma_{Dt}}\int_{i \in D_t}\mathbb{E}\left[\max_{g}\max_{j'\in B_g, j'\neq j}U_{iDj't}\right]di=\frac{1}{\gamma_{Dt}}D_{t}\underbrace{\log\left(\sum_{g}\left(\sum_{j'\in B_g, j'\neq j} \exp\left(\frac{\delta_{Dj't}}{1-\eta}\right)\right)^{(1-\eta)}\right)}_{IV_{Dt\backslash j}}
\end{equation}}
where $IV_{Dt\backslash j}$ denotes permanent workers' inclusive value at time $t$ excluding plant $j$.

To characterize the bargaining equilibrium, I derive the first-order condition from the maximization problem \eqref{eq:nash_bargaining}:
{\small
\begin{equation}
\label{eq:Nash_bargaining_FOC}
    \begin{gathered}
    \underbrace{\left(\frac{\partial P_{jt}}{\partial Q_{jt}}\frac{Q_{jt}}{P_{jt}}+1\right)P_{jt}\frac{\partial Q_{jt}}{\partial D_{jt}}}_{
         \text{Marginal Revenue Product}  
   }=
   \underbrace{\left(\frac{\partial W_{jt,D}}{\partial D_{jt}}\frac{D_{jt}}{W_{jt,D}}+1\right)W_{jt,D}+\frac{\partial \Phi_{jt}}{\partial D_{jt}}}_{\text{Marginal Cost}}+\\
   \underbrace{\beta\left[\left(\frac{\partial P_{jt}}{\partial Q_{jt}}\frac{Q_{jt}}{P_{jt}}+1\right)P_{jt}\frac{\partial Q_{jt}}{\partial D_{jt}}-\left(\frac{\partial W_{jt,D}}{\partial D_{jt}}\frac{D_{jt}}{W_{jt,D}}+1\right)W_{jt,D}-\frac{\partial \Phi_{jt}}{\partial D_{jt}}-\frac{\Pi_{jt}}{D_{jt}}\frac{\partial (U^U_{jt}-U^U_{Ojt})}{\partial D_{jt}}\frac{D_{jt}}{{U^U_{jt}-U^U_{Ojt}}}\right]}_{\text{Nash-Bargaining Margin}}
   \end{gathered}
\end{equation}}
This condition equates the marginal revenue product of permanent workers to their marginal cost plus a Nash-bargaining margin. Union bargaining power generates this margin, inducing plants to employ more permanent workers than they would under monopsony. The plant internalizes union power $\beta$ through two offsetting channels. First, it recognizes the incremental value from additional permanent employment—the marginal revenue product net of marginal costs. Second, it faces an opportunity cost proportional to per-worker profits, scaled by the elasticity of union surplus with respect to permanent employment at plant $j$\footnote{I derive the closed-form expression of this term in Appendix \ref{a:proof_deriv}.}.

The parameter $\beta$ encompasses two limiting cases. When $\beta = 0$, unions possess no bargaining power and plants set wages through Bertrand-Nash competition as with contract workers. When $\beta = 1$, unions extract the entire bargaining surplus, driving permanent worker compensation above their marginal value of production.

%% file: empirics.tex
\subsection{Estimation of Workers Supply Functions}

To estimate the nested logit supply function for each type of worker, equation \eqref{eq:lab_supply}, I use mandays worked to construct the share variables. Moreover, I employ an instrumental variable approach since conditional shares and workers' wages may be endogenous to unobservable workplace amenities, $\xi_{xjt}$ for $x \in \{C, D\}$. I use two sets of instruments to address this endogeneity. First, I use the number of car assembly plants in each state-year pair as an instrument, which provides the variation needed to identify the nesting parameter \citep{rubens2024exploiting}. Second, I use the lagged log wage rate for supervisors and its interaction with a time trend. 

The validity of the second instrument rests on two key assumptions. First, both contract and permanent worker supplies are uncorrelated with managerial remuneration, which is set at supervisors' marginal revenue products in the absence of oligopsony power exercised by the plant. Second, managerial remuneration levels affect worker demand by shifting the plant's marginal production costs, thereby satisfying the relevance condition. To mitigate potential residual contemporaneous correlations between the managerial labor market and the markets for permanent and contract workers, I employ lagged values of managerial wages as a precautionary measure.

After estimating the parameters of the supply function for both worker types, I differentiate establishment $j$'s employment share $s_{xjt}$ with respect to its own wage rate to derive the inverse labor supply elasticity at the establishment level for each worker type, as in equation \eqref{eq:inv_elas_D}. Given Nash-Bertrand competition in the contract worker market, these inverse supply elasticity estimates allow me to directly infer wage markdowns for contract workers.

\subsection{Production Function Estimation}

The production function estimation strategy proceeds through four sequential steps. First, I normalize the CES production function using geometric means. In subsequent steps, I follow an approach similar to \citet{doraszelski2018measuring} to recover the production function parameters. In steps two and three, I construct GMM-estimable equations by exploiting, respectively, the labor-augmenting productivity process alongside the ratio of first-order conditions, and the Hicks-neutral productivity process combined with the production function. Step three generates an estimation residual that conflates Hicks-neutral productivity with measurement error. In the final step, I disentangle these components by employing Kalman filtering techniques to construct a quasi-maximum likelihood estimator for the variances of productivity innovation and measurement error, then applying the Kalman smoother with these variance estimates to derive point estimates of Hicks-neutral productivity and measurement error. Knowledge of planned output then allows me to infer output market markups and profits.

\subsubsection{Step 1 -- Production Function Normalization}

I normalize the production function using geometric means\footnote{Following established practice \citep{de1989quest, klump2000ces, klump2000economic, de2006conjecture, leon2010identifying, grieco2016production, harrigan2018techies, hong2024search}, the CES production function requires normalization for parameter interpretability. For a detailed derivation of the normalization procedure, see Appendix \ref{a:geo_means}.}, which enables factor share parameters to represent marginal returns at the baseline point where inputs, productivity, and prices equal their geometric means. Normalized variables take the form $\ddot{X}_{jt} = \frac{X_{jt}}{\bar{X}}$, where $\bar{X} = \left( \prod_{n=1}^N X_n \right)^{\frac{1}{N}}$ represents the geometric mean across observations. The normalized CES share parameters are denoted $\alpha_X$ for $X\in\{C,D,S,E,M,K,L\}$.

Under this normalization, I express the share parameters for contract and permanent workers, supervisors and the workers' aggregate, and materials, capital, and aggregate labor ($\alpha_C$, $\alpha_D$, $\alpha_S$, $\alpha_E$, $\alpha_M$, $\alpha_K$, and $\alpha_L$) as functions of observed mean expenditures and structural parameters:
\begin{equation}
    \small
\begin{aligned}
   \alpha_C &= \frac{\phi_C\bar{W}_C\bar{C}}{\phi_C\bar{W}_C\bar{C} + \phi_D\bar{W}_D\bar{D}},\quad \alpha_D = 1 - \alpha_C, \quad \alpha_S = \frac{\bar{W}_S\bar{S}}{\bar{W}_S\bar{S}+\phi_C\bar{W}_C\bar{C} + \phi_D\bar{W}_D\bar{D}},\quad \alpha_E = 1 - \alpha_S,\\
   \alpha_M &=  \frac{\bar{P}_M \bar{M}}{\bar{P}_M \bar{M} + \bar{W}_S \bar{S} + \phi_C\bar{W}_C \bar{C} + \phi_D\bar{W}_D \bar{D} + \tau\bar{P}_M \bar{M}},\\
   \alpha_K &= \frac{\tau\bar{P}_M \bar{M}}{\bar{P}_M \bar{M} + \bar{W}_S \bar{S} + \phi_C\bar{W}_C \bar{C} + \phi_D\bar{W}_D \bar{D} + \tau\bar{P}_M \bar{M}},\quad \alpha_L = 1-\alpha_M-\alpha_K.
   \label{eq:alpha_tau}
\end{aligned}
\end{equation}
The terms $\bar{W}_C\bar{C}$, $\bar{W}_D\bar{D}$, $\bar{W}_S\bar{S}$, and $\bar{P}_M\bar{M}$ denote observed geometric mean expenditures on contract workers, permanent workers, supervisors, and materials, respectively. The parameters $\phi_C$ and $\phi_D$ capture mean wedges between marginal revenue products and wages arising from non-wage costs and imperfect competition in the markets for contract and permanent workers, while $\tau$ captures capital's deviation from static optimality due to its quasi-fixed nature within the period.

The normalized input allocation problem\footnote{See the maximization problem in \eqref{eq:Input_problem_norm}.} generates a system of first-order conditions for contract workers, supervisors, and materials\footnote{Listed in Appendix \ref{a:FOCs}.}. These conditions, combined with the normalized productivity processes and production function, provide the foundation for estimating the production function parameters.

\subsubsection{Step 2 -- Estimating $\sigma^O$, $\sigma^M$, $\sigma^I$, $\phi_C$, $\phi_D$ and $\ddot{\omega}_{jt}^L$}

Taking the ratio of the first-order conditions for supervisors and material inputs, applying logarithmic transformation, substituting for $\alpha_L$ and $\alpha_M$ using \eqref{eq:alpha_tau}, and rearranging terms yields an explicit characterization for labor-augmenting productivity as a function of data and unknown parameters:
{\small
\begin{equation}
\begin{aligned}
&\ddot{\omega}^L_{jt}(\sigma^O,\sigma^M,\sigma^I, \phi_C, \phi_D) = -\frac{1}{\sigma^O}\log\alpha_S(\phi_C,\phi_D) + \frac{1}{\sigma^O}\log\left(\frac{{W}_{jt,S}{S}_{jt}}{{P}_{jt,M}{M}_{jt}}\right) \\
&+\frac{1}{\sigma^O}\log\left(\frac{\bar{P}_M\bar{M}}{\bar W_S \bar S + \phi_C\bar W_C \bar C + \phi_D\bar W_D \bar D}\right)+ \log\ddot{M}_{jt} - \left(\frac{\sigma^M}{\sigma^O}\right)\log\ddot{S}_{jt}\\
&+ \left(\frac{\sigma^M}{\sigma^O}-1\right)\log\left( \alpha_S(\phi_C,\phi_D) \ddot{S}_{jt}^{\sigma^M} + \alpha_E(\phi_C,\phi_D) \left[ \alpha_C(\phi_C,\phi_D) \ddot{C}_{jt}^{\sigma^I} + \alpha_D(\phi_C,\phi_D) \ddot{D}_{jt}^{\sigma^I} \right]^\frac{\sigma^M}{\sigma^I} \right)^\frac{1}{\sigma^M}.
\label{eq:charac_L}
\end{aligned}
\end{equation}}
Then, I substitute the empirical characterization \eqref{eq:charac_L} into the normalized Markov process for labor-augmenting productivity:
{\small
\begin{equation}
   \ddot{\omega}^L_{jt}(\sigma^O,\sigma^M,\sigma^I, \phi_C, \phi_D) = \ddot{\iota}_{L,t} + \rho_L \ddot{\omega}^L_{jt-1}(\sigma^O,\sigma^M,\sigma^I, \phi_C, \phi_D)+ \xi_{jt}^{L}.
   \label{eq:Markov_L}
\end{equation}}
This specification delivers a GMM estimator for the substitution elasticities $\sigma^O$, $\sigma^M$, $\sigma^I$, the mean wedges $\phi_C$, $\phi_D$ and the linear productivity process parameters $\ddot{\iota}_{L,t}$, and $\rho_L$. Identification, formally established in Appendix \ref{a:step1_proof}, rests on orthogonality between the labor-augmenting productivity innovation and a vector of instruments $\mathbf{Z}_L$:
{\small
\begin{equation}
   \mathbb{E}\left[\mathbf{Z}_L\otimes\xi^L_{jt}(\ddot{\iota}_{L,t},\sigma^O,\sigma^M,\sigma^I,\rho_L,{\phi}_C, {\phi}_D)\right]=\mathbf{0}.
\end{equation}}
This estimation step recovers substitution elasticities across the three production nests, the wedge parameters $\phi_C$ and $\phi_D$, and derives estimates of (normalized) labor-augmenting productivity $\ddot{\omega}_{jt}^L$.

In the empirical application, I estimate the coefficients of interest by concentrating out the parameters of the AR(1) process for labor-augmenting productivity\footnote{Specifically, I use a matrix of year dummies and $\ddot{\omega}^L_{jt-1}(\sigma^O,\sigma^M,\sigma^I, \phi_C, \phi_D)$ as instruments to identify the AR(1) coefficients.} and using as instruments functions of lagged supervisor wages and mandays, lagged permanent worker mandays, lagged contract worker mandays, and lagged materials.

The rationale for these instruments follows standard economic principles. Supervisor wage rates represent exogenous cost shifters that plants take as given, ensuring orthogonality to productivity innovations while correlating with productivity through input demand. I use lagged supervisor wages because temporal distance reduces endogeneity concerns relative to contemporaneous values \citep{olley1996dynamics, levinsohn2003estimating, ackerberg2006structural, doraszelski2018measuring}. Under Assumption \ref{asst}, plants allocate variable inputs while facing uncertainty about future productivity innovations. This timing structure ensures that lagged materials and mandays for supervisors, permanent workers, and contract workers remain orthogonal to current productivity innovations while correlating with labor-augmenting productivity through its temporal persistence.

\subsubsection{Step 3 -- Estimating $\tau$ and $\tilde{\omega}_{jt}$}
The third estimation step recovers the parameter $\tau$, which quantifies the mean deviation of capital allocation from its optimal static level and fully characterizes the remaining unobservable CES share parameters. This step also produces normalized estimates of Hicks-neutral productivity that incorporate measurement error: $\tilde{\omega}_{jt}=\widehat{\ddot{\omega}^H_{jt}+\varepsilon_{jt}}$.

Given the estimates $\hat{\sigma}^O$, $\hat{\sigma}^M$, $\hat{\sigma}^I$, $\hat{\phi}_C$, $\hat{\phi}_D$ and $\hat{\ddot{\omega}}^L_{jt}$, I construct the corresponding estimates for the workers nest, $\hat{\ddot{E}}_{jt}$, and the total labor nest, $\hat{\ddot{L}}_{jt}$. Let $\chi_{jt}$ a composite residual that combines the current productivity innovation $\xi^H_{jt}$ with current and lagged measurement errors:
\begin{equation}
\small
    \chi_{jt}=\xi^H_{jt} + \varepsilon_{jt} -\rho_H\varepsilon_{jt-1}
\end{equation}
Furthermore, let $f_{jt}(\tau)$ be the CES production component in normalized planned output:
\begin{equation}
\small
    f_{jt}(\tau)=\left(\alpha_K(\tau) \ddot{K}_{jt}^{\hat{\sigma}^O} + \alpha_M(\tau) \ddot{M}_{jt}^{\hat{\sigma}^O} + \alpha_L(\tau) (\exp(\hat{\ddot{\omega}}_{jt}^{L}) \hat{\ddot{L}}_{jt})^{\hat{\sigma}^O} \right)^\frac{1}{\hat{\sigma}^O}
\end{equation}
Combining the normalized observed output equation with the AR(1) process for normalized Hicks-neutral productivity yields the following specification: 
{\small
\begin{equation}
    \log \ddot{\tilde Q}_{jt} = \ddot{\iota}_{H,t}  + \log f_{jt}(\tau) + \rho_H \left(\log \ddot{\tilde Q}_{jt-1} - \log f_{jt-1}(\tau)\right) + \chi_{jt}.
\label{eq:NLLS}
\end{equation}}
where $\ddot{\tilde Q}_{jt}$ denotes the normalized plant-level observed output index.

This specification yields a GMM estimator for the parameter $\tau$ as well as the linear productivity process parameters $\ddot{\iota}_{H,t}$ and $\rho_H$. Formal identification, established in Appendix \ref{a:step3_proof}, relies on the orthogonality between the composite residual $\chi_{jt}$ and a vector of instruments $\mathbf{Z}_H$:
\begin{equation}\small
    \mathbb{E}\left[\mathbf{Z}_H\otimes\chi_{jt}(\ddot{\iota}_{H,t},\tau,\rho_H)\right]=\mathbf{0}.
\end{equation}

In the empirical application, I concentrate out the time effects in the Hicks–neutral productivity AR(1) process\footnote{Specifically, I use year dummies as instrument to identify these AR(1) coefficients.}, and identify the remaining parameters using the estimated labor-augmenting productivity at $t\!-\!1$ and the lagged capital stock as excluded instruments.

Labor-augmenting and Hicks-neutral productivities evolve independently, so \({\ddot{\omega}}^L_{jt-1}\) is uncorrelated with innovations to Hicks-neutral productivity. Moreover, under the timing in Assumption \ref{assm}, \(\ddot{\omega}^L_{jt}\) is predetermined with respect to output measurement error, implying orthogonality to \(\varepsilon_{jt}\) and \(\varepsilon_{jt+1}\), while remaining relevant through its effect on within-period input choices. 

Because capital is fixed within the period and additionally assuming that output measurement error is orthogonal to capital allocation regardless of timing, the lagged capital stock is orthogonal to Hicks–neutral productivity innovations and to \(\varepsilon_{jt}\) and \(\varepsilon_{jt-1}\). It remains relevant because it enters \(f_{jt-1}(\tau)\) and thereby shifts production decisions at \(t\).
 To mitigate short-run potential endogeneity in book values arising from strategic investment responses to productivity shocks, I use a perpetual inventory measure of capital—constructed with depreciation rates from \citet{orr2022within}—as the capital instrument. This measure exhibits strong correlation with book value of capital, yielding a slope coefficient of 0.88 and $R^2$ of 0.93 in linear regression.

\subsubsection{Step 4 -- Point Estimating $\ddot{\omega}^H_{jt}$, $\varepsilon_{jt}$, and Markups}

The third estimation step delivers a noisy estimate for Hicks-neutral productivity, $\tilde{\omega}_{jt}$. The main objective in this section is separating the observed noisy productivity $\tilde{\omega}_{jt}$ into point estimates for Hicks-neutral productivity $\omega_{jt}$ and measurement error $\varepsilon_{jt}$. I employ Kalman filtering techniques \citep{hamilton1994filter} to achieve this decomposition\footnote{
Standard control function methods \citep{olley1996dynamics, levinsohn2003estimating, ackerberg2015identification} prove ineffective in this setting. The failure of these methods stems from the violation of \textit{scalar unobservability} due to unobservable markups and markdowns, which under broad conduct assumptions may render input demand functions---such as those for investment---non-invertible with respect to Hicks-neutral productivity alone.
}.

I reformulate equation \eqref{eq:NLLS} as a linear state-space representation following \cite{hong2024search}, incorporating the estimated parameter values from Step 2. Specifically, I specify the state equation as 
\begin{equation}
\small
\label{eq:state_space_transition}
  \begin{bmatrix} {\xi^H_{jt+1}} \\ {\varepsilon_{jt+1}} \\ {\varepsilon_{jt}}\end{bmatrix}=\begin{bmatrix} 0 & 0 & 0 \\ 0 & 0 & 0 \\ 0 & 1 & 0 \end{bmatrix}\times\begin{bmatrix} {\xi^H_{jt}} \\ {\varepsilon_{jt}} \\ {\varepsilon_{jt-1}}\end{bmatrix}+\begin{bmatrix} 1 & 0 \\ 0 & 1 \\ 0 & 0 \end{bmatrix}\times\begin{bmatrix} {\xi^H_{jt+1}} \\ {\varepsilon_{jt+1}}\end{bmatrix}
\end{equation}
which governs the evolution of the unobserved state variables. The corresponding observation equation links these states to the observed productivity measure:
{\small
\begin{equation}
\label{eq:state_space_measurement}
       \tilde{\omega}_{jt} =  \left[\begin{array}{cccc} \hat{\rho}_{H}  \end{array}\right]\times\left[\begin{array}{cccc} \tilde{\omega}_{jt-1} \end{array}\right] + \begin{bmatrix} 1 & 1 & -\hat{\rho}_H \end{bmatrix} \times\left[\begin{array}{ccc} \xi^H_{jt} & \varepsilon_{jt} & \varepsilon_{jt-1} \end{array}\right]'+\hat{\ddot{\iota}}_{H,t}.
\end{equation}}

The filtering approach relies on equations \eqref{eq:state_space_transition} and \eqref{eq:state_space_measurement} and requires stochastic properties that follow directly from Assumptions \ref{asst} and \ref{assm}. These assumptions ensure that $[{\xi^H_{jt}} ~ {\varepsilon_{jt}}]'$ constitutes a vector white noise process with autocovariance function $\mathbb{E}\left([{\xi^H_{jt}} ~ {\varepsilon_{jt}}]'[{\xi^H_{j\tau}} ~ {\varepsilon_{j\tau}}]\right)$ equal to zero for all $\tau \neq t$, and equal to the diagonal variance matrix with elements $(\sigma^H)^2$ and $(\sigma^\varepsilon)^2$ when $\tau = t$. This white noise specification guarantees the crucial orthogonality condition $\mathbb{E}\left([ {\xi^H_{jt}} ~ {\varepsilon_{jt}}]'[ {\xi^H_{j\tau}} ~ {\varepsilon_{j\tau}}  ~ \varepsilon_{j\tau-1}]\right)=0$ for all $\tau=t-1,t-2,...,1$, establishing that current state equation disturbances remain uncorrelated with all past state realizations. Finally, the state equation exhibits stationarity through a transition matrix possessing one null eigenvalue with multiplicity 3, while the vector $\left[ \widetilde{\omega}_{jt-1} ~ ~\mathbf{I}\{year=t\}\right]'$ in the observation equation is exogenous in \cite{hamilton1994filter}'s sense.

Let the state vector $\zeta_{jt}\equiv[{\xi^H_{jt}} ~ {\varepsilon_{jt}} ~{\varepsilon_{jt-1}}]'$. The variances of productivity innovations and measurement errors, $(\sigma^H)^2$ and $(\sigma^\varepsilon)^2$, constitute the remaining unknown parameters requiring estimation. I construct a quasi-maximum likelihood estimator using the Kalman filter \citep{hamilton1994filter}, where the log likelihood function takes the form
\begin{equation}
\small
\label{eq:MLE}
            \sum^J_{j=1}\sum^T_{t=1}\log f_{\tilde{\omega}_{jt}|\Omega_{t-1}}( \tilde{\omega}_{jt}|\Omega_{t-1})
\end{equation}
where $\Omega_{t-1}$ represents the econometrician's information set available through time $t-1$. I use the Normal distribution to approximate the density function\footnote{\footnotesize
\begin{equation*}
    \begin{aligned}
           \tilde{\omega}_{jt}|\Omega_{t-1}\sim N\Big(&\left(\left[\begin{array}{cccc} \hat{\rho}_{H} & \hat{\beta}_{H,1} & \hat{\beta}_{H,2}  \end{array}\right]\times\left[\begin{array}{cccc} \tilde{\omega}_{jt-1} & IDA_{rt} & Imp_{jt-1} \end{array}\right]'+\hat{\ddot{\iota}}_{H,t}+\begin{bmatrix} 1 & 1 & -\hat{\rho} \end{bmatrix}\times\hat{\zeta}_{j,t|t-1}\left(\sigma^H,\sigma^\varepsilon\right)\right),\\
           &\left(\begin{bmatrix} 1 & 1 & -\hat{\rho} \end{bmatrix}\times\mathbf{P}_{j,t|t-1}\left(\sigma^H,\sigma^\varepsilon\right)\times\begin{bmatrix} 1 & 1 & -\hat{\rho} \end{bmatrix}'\right)\Big)
\end{aligned}
\end{equation*}
Here $\hat{\zeta}_{j,t|t-1}\left(\sigma^H,\sigma^\varepsilon\right)$ denotes the conditional forecast of the state vector at time $t$ given information through $t-1$, while $\mathbf{P}_{j,t|t-1}\left(\sigma^H,\sigma^\varepsilon\right)$ represents the corresponding mean squared error matrix. Normality of the innovation to productivity and measurement vector, $[{\xi^H_{jt}} ~ {\varepsilon_{jt}}]'$, is not required. Maximizing the pseudo log-likelihood \eqref{eq:MLE} yields consistent and asymptotically Normal estimates of $\sigma^H$ and $\sigma^\varepsilon$ \citep{hamilton1994filter}.
}. Upon obtaining estimates $\hat{\sigma}^H$ and $\hat{\sigma}^\varepsilon$, I apply smoothing techniques \citep{hamilton1994filter} to construct estimates of the state vector. This procedure delivers point estimates $\hat{\xi}^H_{jt}$ and $\hat{\varepsilon}_{jt}$ for all $t$ and $j$, enabling me to recover true Hicks-neutral productivity $\hat{\omega}^H_{jt}$. Appendix \ref{a:kalman} provides comprehensive details of the filtering and smoothing algorithms.

Upon inferring measurement error, $\hat{\varepsilon}_{jt}$, I recover planned output, $\hat{Q}_{jt}$, from equation \eqref{eq:obs_prod} by dividing observed output by $\exp(\hat{\varepsilon}_{jt})$. Using these recovered values and the estimated production function parameters, I then infer output market markups from the first-order condition for material inputs\footnote{The first-order condition for supervisors yields identical markup estimates by construction, consistent with \cite{raval2023testing}:
\begin{equation*}
   \hat{\mu}_{jt}=\frac{P_{jt}\hat{Q}_{jt}}{\left(\alpha_K(\hat{\tau}) \ddot{K}_{jt}^{\hat{\sigma}^O} + \alpha_M(\hat{\tau}) \ddot{M}_{jt}^{\hat{\sigma}^O} + \alpha_L(\hat{\tau}) (\exp(\hat{\ddot{\omega}}_{jt}^{L}) \hat{\ddot{L}}_{jt})^{\hat{\sigma}^O} \right)}\frac{\alpha_L(\hat{\tau})(\exp(\hat{\ddot{\omega}}_{jt}^{L}) \hat{\ddot{L}}_{jt})^{\hat{\sigma}^O}}{\left( \alpha_S \ddot{S}_{jt}^{\hat{\sigma}^M} + \alpha_E \hat{\ddot{W}}_{jt}^{\hat{\sigma}^M} \right)}\frac{\hat{\alpha}_S\ddot{S}_{jt}^{\hat{\sigma}^M}}{W_{S,jt}S_{jt}}
\end{equation*}}:
{\footnotesize
\begin{equation}
   \hat{\mu}_{jt}\equiv\widehat{\left(\frac{\partial P_{jt}}{\partial Q_{jt}}\frac{Q_{jt}}{P_{jt}}+1\right)^{-1}}=\frac{P_{jt}}{\hat{\mathcal{MC}}_{jt}}=\frac{P_{jt}\hat{Q}_{jt}}{\left(\alpha_K(\hat{\tau}) \ddot{K}_{jt}^{\hat{\sigma}^O} + \alpha_M(\hat{\tau}) \ddot{M}_{jt}^{\hat{\sigma}^O} + \alpha_L(\hat{\tau}) (\exp(\hat{\ddot{\omega}}_{jt}^{L}) \hat{\ddot{L}}_{jt})^{\hat{\sigma}^O} \right)}\frac{\alpha_M(\hat{\tau}) \ddot{M}_{jt}^{\hat{\sigma}^O}}{P_{M,jt}M_{jt}}\label{eq:markups}
\end{equation}}
where ${\mathcal{MC}}_{jt}$ denotes the marginal cost of production for plant $j$ at time $t$. Finally, I infer plant profits from markups and planned output:
{\small
\begin{equation}
 \hat{\Pi}_{jt} = \hat{Q}_{jt}\left(P_{jt}-\frac{P_{jt}}{\hat{\mu}_{jt}}\right). \label{eq:profits}
\end{equation}}

\subsection{Bargaining Parameter and Wage Markdown Estimation for Permanent Workers}

In the market for permanent workers, unions and plants engage in Nash bargaining. Therefore, knowledge of the inverse elasticity of labor supply is insufficient to infer markdowns, as markdowns depend on the unknown bargaining parameter $\beta$.

Rearranging equation \eqref{eq:Nash_bargaining_FOC} in terms of wage rates yields:
{\small
\begin{equation}
\label{eq:Nash_bargaining_FOC}
    \begin{aligned}
    &\left(\frac{\partial W_{jt,D}}{\partial D_{jt}}\frac{D_{jt}}{W_{jt,D}}+1\right)W_{jt,D}
   =\\
   &\left(\frac{\partial P_{jt}}{\partial Q_{jt}}\frac{Q_{jt}}{P_{jt}}+1\right)P_{jt}\frac{\partial Q_{jt}}{\partial D_{jt}}+\frac{\beta}{1-\beta}\left[\frac{\Pi_{jt}}{D_{jt}}\frac{\partial (U^U_{jt}-U^U_{Ojt})}{\partial D_{jt}}\frac{D_{jt}}{{U^U_{jt}-U^U_{Ojt}}}\right]-\frac{\partial \Phi_{jt}}{\partial D_{jt}}
   \end{aligned}
\end{equation}}
The left-hand side of this equation is observed by the econometrician since the wage rate $W_{jt,D}$ is data and one plus the inverse elasticity of labor supply for permanent workers, $\left(\frac{\partial W_{jt,D}}{\partial D_{jt}}\frac{D_{jt}}{W_{jt,D}}+1\right)$, is estimated from the labor supply model. On the right-hand side, the marginal revenue product of permanent workers, $\left(\frac{\partial P_{jt}}{\partial Q_{jt}}\frac{Q_{jt}}{P_{jt}}+1\right)P_{jt}\frac{\partial Q_{jt}}{\partial D_{jt}}$, is estimated from the production function model, as are profits $\Pi_{jt}$. Furthermore, permanent worker employment $D_{jt}$ is data, and the elasticity of union surplus to employment, $\frac{\partial (U^U_{jt}-U^U_{Ojt})}{\partial D_{jt}}\frac{D_{jt}}{{U^U_{jt}-U^U_{Ojt}}}$, has a closed form and is observed because it is a combination of data and estimated permanent worker supply parameters. This leaves only two unknowns: the bargaining parameter $\beta$ and the non-wage marginal employment cost for permanent workers, $\frac{\partial \Phi_{jt}}{\partial D_{jt}}$.

I assume the following functional form for (log) non-wage marginal costs:
{\small
\begin{equation}\label{eq:perm_marg_cost}
    \log\left(\frac{\partial \Phi_{jt}}{\partial D_{jt}}\right)=\theta_0+\theta_1\log D_{jt} + \varphi_{jt} 
\end{equation}}
where $\theta_0$ represents the constant component of log non-wage marginal costs, and $\theta_1$ captures the elasticity of non-wage marginal costs with respect to permanent worker employment. The term $\varphi_{jt}$ is a mean-zero, unobservable component measuring the extent to which log non-wage marginal costs heterogeneously deviate from the level predicted by employment.

This specification yields a GMM estimator for the bargaining parameter $\beta$ and the cost parameters $\theta_0$ and $\theta_1$. Identification, formally established in Appendix \ref{a:bargaining_proof}, relies on the orthogonality condition between the unobservable component $\varphi_{jt}$ and a vector of instruments $\mathbf{Z}_B$:
{\small
\begin{equation}
    \mathbb{E}\left[\mathbf{Z}_B\otimes\varphi_{jt}(\beta,\theta_0,\theta_1)\right]=\mathbf{0}.
\end{equation}}

In the empirical application, I estimate the coefficients of interest using as instruments the unit vector to identify the constant $\theta_0$, lagged supervisor wages and lagged prices of materials to identify $\theta_1$, and a measure of the lagged intensity of strikes to which a plant is subject to identify $\beta$. This strike intensity measure is constructed as:
{\small
\begin{equation}\label{eq:instrument}
    \text{Strike Intensity} = \log\left[\left(\frac{\text{\# of mandays lost for strikes}}{\text{\# of permanent worker mandays in manufacturing}}\right)_{rt-1}\times D_{jt-1} \right]
\end{equation}}
The first term varies at the state $r$ level and denotes an aggregate measure of strike intensity. The denominator includes the number of mandays worked by permanent workers in the manufacturing sector of state $r$. I interact this term with a measure of plant-level exposure, which I take to be the lagged permanent worker employment.

The rationale for these instruments is as follows. Supervisor wage rates and material prices represent exogenous cost shifters that plants take as given. These variables are orthogonal to the non-wage marginal cost random component, $\varphi_{jt}$, yet correlated with permanent worker demand. I use lagged supervisor wages and material prices because temporal distance mitigates endogeneity concerns \citep{olley1996dynamics, levinsohn2003estimating, ackerberg2006structural, doraszelski2018measuring}. 

Plants with higher profitability per permanent workday, $\frac{\Pi_{jt}}{D_{jt}}$, should be associated with higher contemporaneous strike intensity because workers have greater incentives to demand a larger share of surplus. However, current strike intensity may correlate with increased expenditure per worker (for example, for legal procedures), which in turn may imply correlation with $\varphi_{jt}$. To avoid this endogeneity concern, I use lagged strike intensity, which correlates with contemporaneous values through two channels: the autocorrelation of permanent employment (driven by productivity persistence) and the persistence of aggregate strike intensity at the state level.

%% file: results.tex
    \subsection{Workers Supply Function Estimates}
    
Table \ref{tab:bc_IV} reports instrumental variable estimates for equation \eqref{eq:lab_supply}, contrasting OLS and IV coefficients across worker types. The results reveal upward-sloping labor supply curves for both categories, with contract workers exhibiting substantially higher wage sensitivity (0.046) than permanent workers (0.009). The time-varying sensitivity parameter shows a modest decline in responsiveness over time, while the nesting parameter $\eta$ is statistically significant for permanent workers and satisfies the theoretical unit interval constraint for both worker types.

\begin{center}
[Table \ref{tab:bc_IV} goes here]
\end{center}

Table \ref{tab:bc_FS} presents the corresponding first-stage estimates, where Sanderson-Windmeijer F-statistics confirm instrument relevance. The number of active plants in each state, $N_{r,t}$, demonstrates the expected negative correlation with conditional employment shares, as increasing plant density within states compresses individual plant market shares.

\begin{center}
[Table \ref{tab:bc_FS} goes here]
\end{center}

Figure \ref{fig:elas_lab} presents revenue-weighted mean labor supply elasticity estimates, constructed using equation \eqref{eq:inv_elas_D}, across worker categories. Permanent workers consistently exhibit lower supply elasticities than contract workers. Panel \ref{fig:elas_labA} documents substantial cross-state heterogeneity, with permanent worker elasticities ranging from 2 to 8 and contract worker elasticities spanning 4 to 8. Haryana, Tamil Nadu, and Maharashtra exhibit the highest elasticities—states where automotive production is concentrated, providing workers with expanded employment opportunity sets. Panel \ref{fig:elas_labB} reveals divergent temporal dynamics: contract worker elasticities increased while permanent worker elasticities declined over time. Overall, spatial heterogeneity dominates time variation.

\begin{center}
[Figure \ref{fig:elas_lab} goes here]
\end{center}

\subsection{Production Function Estimates}

Table \ref{tab:CES_parameters} reports the production function parameters estimated across the four sequential estimation stages. Column "Normalization" presents the normalization parameters: the baseline capital deviation from its optimal static level $\tau$, and the baseline wage wedges $\phi_C$ and $\phi_D$. Column "Substitution" shows the CES substitution parameters: $\sigma^O$, $\sigma^M$, and $\sigma^I$. Column "Share" reports the share parameters $\alpha_L$, $\alpha_M$, $\alpha_K$, $\alpha_S$, $\alpha_E$, $\alpha_D$, and $\alpha_C$. Finally, column "Variance" presents the estimated standard deviations of productivity innovations and measurement error derived through Kalman filtering.

\begin{center}
[Table \ref{tab:CES_parameters} goes here]
\end{center}

The parameter estimates reveal distinct patterns in production technology. The baseline wage wedges $\phi_C$ and $\phi_D$ indicate labor market frictions for both contract and permanent workers, with frictions substantially higher in the permanent worker market. I estimate elasticities of substitution indicating that capital, materials, and labor function as substitutes in production, $\sigma^O$, as do managers and workers, $\sigma^M$. In contrast, I find that permanent and contract workers operate as complements, $\sigma^I$. However, these relationships exhibit substantial heterogeneity across manufacturing plants. Intermediate inputs account for the largest expenditure shares, $\alpha_M$, while labor represents a relatively modest fraction of total costs, $\alpha_L$. Within the labor aggregate, supervisors command most expenditure, $\alpha_S$, with permanent employees dominating worker expenditure, $\alpha_D$. Furthermore, I estimate highly persistent productivity processes (Table \ref{tab:prod}), with AR(1) coefficients, $\rho$, of 0.88 for Hicks-neutral productivity, $\omega^H$, and 0.91 for labor-augmenting productivity, $\omega^L$.

\begin{center}
[Table \ref{tab:prod} goes here]
\end{center}

Figure \ref{fig:outp_elas} presents density distributions of implied output elasticities across three production nests. I examine contract versus permanent workers in the first panel, finding that contract workers' elasticities concentrate near lower values while permanent workers command substantially higher output elasticity. The second panel contrasts workers and managers, where I observe substantial distributional overlap, indicating comparable marginal productivity across organizational hierarchies. The third panel displays outer nest production factors—capital, labor, and materials. I find labor elasticities clustering near zero, consistent with labor's modest contribution to marginal output. Capital and materials exhibit greater dispersion with higher central tendencies, confirming these inputs' dominant productive role and aligning with automotive assembly's characteristic materials and capital intensity.

\begin{center}
[Figure \ref{fig:outp_elas} goes here]
\end{center}

I document the evolution of production unobservables---productivity and markups---through revenue-weighted averages in Figure \ref{fig:prod_unob}. Panel \ref{fig:prod_unobA} presents mean deviations of Hicks-neutral and labor-augmenting productivity over time\footnote{By normalization, productivity point estimates represent plant-time deviations from the (log) geometric average: $\hat{\ddot{\omega}}^X_{jt}=\widehat{\omega^X_{jt}-\bar{\omega}^X_{jt}}$.}. I observe labor-augmenting productivity increasing steadily to approximately three times the geometric mean, while Hicks-neutral productivity declines to roughly one-third of its benchmark. 

Panel \ref{fig:prod_unobB} examines markup dynamics using the Lerner Index, $(P_{jt}-\mathcal{MC}_{jt})/P_{jt}$, which measures the proportion of price that exceeds marginal cost. This index declines from 40\% to 20\%, indicating that markups fall by half over the sample period. This pattern is consistent with intensified competition driven by sustained market entry: between 2002 and 2019, the total number of production units in the industry increased at an average annual rate of 10\%. As a robustness check, I construct an alternative Lerner Index using average variable costs, $(P_{jt}-\mathcal{AC}_{jt})/P_{jt}$\footnote{Average cost is defined as \begin{equation}
\mathcal{AC}_{jt}=\frac{(W_{jt,S}S_{jt}+W_{jt,D}D_{jt}+W_{jt,C}C_{jt}+P_{jt,M}M_{jt})}{\hat{Q}_{jt}}
\end{equation}}. This alternative measure exhibits a similar declining trend.

\begin{center}
[Figure \ref{fig:prod_unob} goes here]
\end{center}

\subsection{Bargaining Power Estimation}
Table \ref{tab:param_estimates} presents the bargaining and marginal cost parameters that I estimate using GMM on equations \eqref{eq:Nash_bargaining_FOC} and \eqref{eq:perm_marg_cost}. The parameter $\theta_1$ indicates that a 1\% increase in permanent worker employment raises their non-wage marginal cost by 0.02\%. The parameter $\theta_0$, representing the constant component of this marginal cost, equals approximately 890 INR per day in 2005 values (32 USD in 2024 values). The union bargaining parameter, $\beta$, equals 0.003, demonstrating that plants command virtually all bargaining power in negotiations with the permanent workers' union.

\begin{center}
[Table \ref{tab:param_estimates} goes here]
\end{center}

To analyze the instrument's predictive power, I present in Figure \ref{fig:Instr_anal}, Panel \ref{fig:FS}, a scatterplot of the log of the variable associated with the bargaining parameter (the "Bargaining Variable") from equation \eqref{eq:Nash_bargaining_FOC}---profit per permanent worker man-day scaled by the elasticity of union surplus to permanent worker employment, $\frac{\Pi_{jt}}{D_{jt}}\left(\frac{\partial (U^U_{jt}-U^U_{Ojt})}{\partial D_{jt}}\frac{D_{jt}}{(U^U_{jt}-U^U_{Ojt})}\right)$---against the strike intensity instrument ("Strike Intensity") from equation \eqref{eq:instrument}. The fitted regression line reveals the expected positive and significant correlation: plants experiencing higher strike intensity in period $t-1$ exhibit higher permanent worker profitability in period $t$.

The magnitude of the estimated $\beta$ parameter merits further examination. As a benchmark, Panel \ref{fig:OLS} presents a scatterplot of the left-hand side variable (the "Wage Variable") of equation \eqref{eq:Nash_bargaining_FOC}---the markdown-adjusted wages of permanent workers, $\left(\frac{\partial W_{jt,D}}{\partial D_{jt}}\frac{D_{jt}}{W_{jt,D}}+1\right)W_{jt,D}$---against the bargaining variable, along with the fitted regression line. Treating the regression slope as a naive estimator of the bargaining parameter ratio, $\beta/(1-\beta)$, yields $\beta = 0.001$, consistent with the GMM estimates. Figure \ref{fig:OLS_0} in Appendix \ref{sec:sec53_addfig} shows the same scatterplot after dropping observations with zero estimated operating profits; the slope coefficient is virtually unchanged. 

\begin{center}
[Figure \ref{fig:Instr_anal} goes here]
\end{center}

Thus, the magnitude of the estimated $\beta$ parameter stems from the substantial difference in scale between the wage and bargaining variables. In Figure \ref{fig:comp_analysis} in Appendix \ref{sec:wage_barg_comp}, I examine the empirical distributions of these variables' components to identify the source of this difference. The scale difference emerges predominantly from the gap between wages and profitability for permanent workers.

\subsection{Decomposing the Wage Premium Between Permanent and Contract Workers}

The model's first-order conditions yield a decomposition of the log wage premium between permanent and contract workers into two components: the difference in log marginal value of production (the gap between marginal revenue products, $MRPL^X_{jt}$, and non-wage marginal costs, $\frac{\partial\Phi_{jt}}{\partial X_{jt}}$) and the difference in negative log markdowns, $MDown^X_{jt}$. Formally,
{\small
\begin{equation}
    \Delta_{D,C}\log W_{jt,X}=\Delta_{D,C}\log\left(MRPL^X_{jt}-\frac{\partial\Phi_{jt}}{\partial X_{jt}}\right)+\Delta_{D,C}\left(-\log MDown^X_{jt}\right)
\end{equation}}
where $\Delta_{D,C}$ denotes the difference between permanent and contract workers.

I present in Figure \ref{fig:markdowns} revenue-weighted mean wage markdowns for permanent and contract workers. I calculate markdowns as $(MRPL^D_{jt}-W_{jt})/MRPL^D_{jt}$, the proportion of marginal product that plants retain rather than compensate to workers. Despite unionization, permanent workers face substantially higher markdowns than contract workers. This differential stems from permanent workers' lower supply elasticities. Contract worker markdowns remain stable at 15\%, while permanent worker markdowns span 20--25\% throughout the sample period. Reflecting unions' negligible estimated bargaining power, mean markdowns under Nash bargaining do not differ statistically from those under Nash-Bertrand competition. For this reason, and because the Nash bargaining sample differs due to instrument availability and dependence on previously estimated variables and parameters, I employ Nash-Bertrand markdowns for permanent workers in the remaining analysis.

\begin{center}
[Figure \ref{fig:markdowns} goes here]
\end{center}

Because permanent workers face higher markdowns, their wage premium must originate from superior marginal value of production. Figure \ref{fig:decomp}, Panel \ref{fig:decompA}, confirms this mechanism, displaying the empirical distributions of log marginal revenue products across worker types. Permanent workers exhibit substantially higher productivity, generating an average marginal revenue product of 74 USD per workday (in 2024 terms) compared to 55 USD for contract workers---a 30\% premium. Panel \ref{fig:decompB} reveals that non-wage marginal costs are similar across worker types, averaging 33 USD for both. 

\begin{center}
[Figure \ref{fig:decomp} goes here]
\end{center}

Therefore, the wage premium for permanent workers derives entirely from their productivity advantage. Their higher marginal revenue product more than offsets the elevated markdowns they face, ultimately delivering superior wages relative to contract workers.

\subsubsection{Welfare Effects of Offsetting Markdowns and Policy Implications}
Finally, I consider a counterfactual scenario in which lump-sum government transfers to workers fully offset wage markdowns, such that total compensation (wage plus transfer) equals the competitive wage. Crucially, since these transfers do not affect workers' labor supply elasticity, I can isolate the welfare effects of eliminating monopsony distortions without changing plants' equilibrium employment allocations. 

I measure welfare changes via the inclusive value, defined as the expected maximum utility each worker type derives from automotive labor market participation. To simulate counterfactual welfare under markdown-offsetting transfers, I recompute inclusive values using the counterfactual workers' utility function:
\begin{equation}\label{eq:counter}
\small
U^{Counter}_{ixjt}=\alpha_{x}t_{trend} +\gamma_{xt} W_{jt,x}+\gamma_{xt}T_{jt,x}+\xi_{xjt}+\zeta_{ixrt}+(1-\eta_x)\epsilon_{ixjt}\quad  x\in\{D,C\}
\end{equation}
where $T_{jt,x}$ represents the markdown-offsetting lump-sum government transfer. This transfer shifts labor supply outward but does not alter the slope of the labor supply curve faced by each plant.

Figure \ref{fig:Counter}, Panel \ref{fig:counterA}, shows that welfare increases substantially: by 14\% on average for permanent workers and 12\% for contract workers. However, Panel \ref{fig:counterB} reveals an important distributional consequence. This policy substantially widens the permanent worker wage premium, defined as the ratio of permanent to contract compensation rates, $(W_{jt,D}+T_{jt,D})/(W_{jt,C}+T_{jt,C})$. The premium rises by an average of 14\% above observed levels.

\begin{center}
[Figure \ref{fig:Counter} goes here]
\end{center}

These results are particularly relevant given renewed interest among global antitrust authorities in labor market power and its wage-depressing effects. While markdown-offsetting policies substantially increase workers' welfare---valuable in concentrated labor markets such as the Indian auto industry, with an average plant-level HHI of 1,788 based on worker employment---they may exacerbate inequality in compensation when workers differ in productivity. Despite generating positive welfare gains for all workers, such policies could intensify perceptions of distributional inequity between high- and low-productivity workers, potentially generating political resistance to otherwise welfare-improving interventions.

%% file: conclusion.tex
I decompose the permanent worker wage premium and find that it stems entirely from productivity differences. Despite facing higher wage markdowns—with plants retaining 20–25\% of permanent workers' marginal value versus 15\% for contract workers—permanent employees earn more because they generate substantially higher marginal revenue products (74 versus 55 USD per workday). This productivity advantage more than offsets the higher markdowns. The productivity heterogeneity has important policy implications: while a lump-sum transfer offsetting wage markdowns would increase welfare by 14\% for permanent workers and 12\% for contract workers, it would simultaneously widen the compensation premium by 14\%, exacerbating inequality between worker types.

This paper suggests important directions for future research. First, extending the model to incorporate adjustment costs in a tractable estimation framework would advance our understanding and measurement of market power in dynamic settings where producers are forward-looking. Second, the framework can be extended to incorporate intermediaries between contract workers and plants. Since contract worker expenditures represent payments to contractors or staffing agencies that likely extract additional rent, my markdown estimates for contract workers represent a lower bound from the workers' perspective. Both extensions would deepen our understanding of labor market power and compensation inequality.

%% file: figures_tables.tex
\begin{figure}[h]
    \centering
    \caption{Contract vs. Permanent Workers}
    \label{fig:con_vs_perm}
    \begin{subfigure}[h]{0.47\textwidth}
        \centering
        \includegraphics[width=\linewidth]{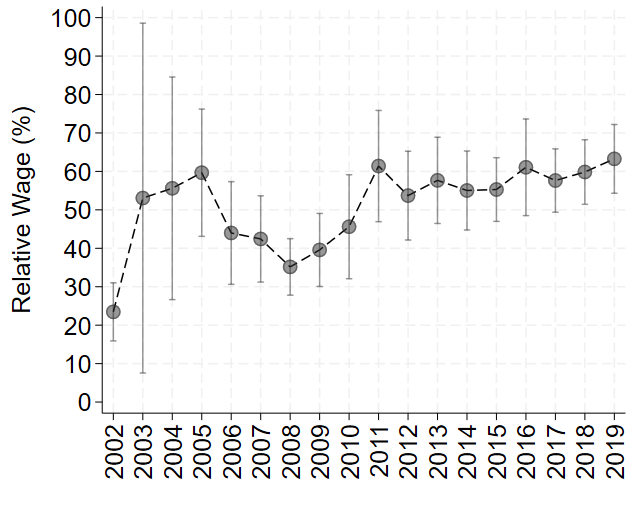}
        \caption{Contract/Permanent Relative Wage}
        \label{fig:con_vs_permA}
    \end{subfigure}
    \hfill
    \begin{subfigure}[h]{0.47\textwidth}
        \centering
        \includegraphics[width=\linewidth]{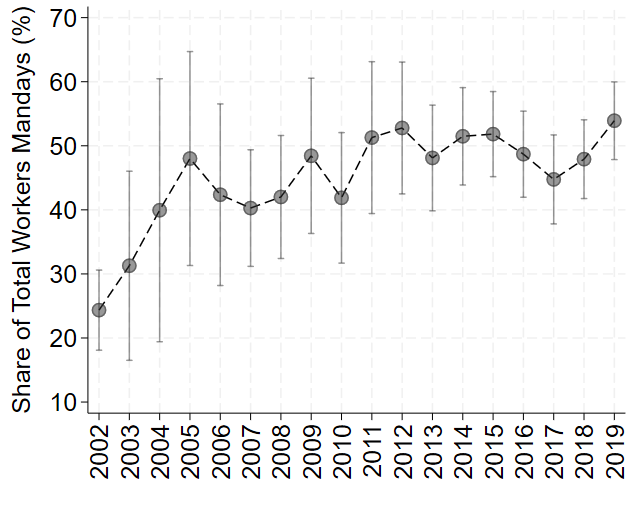}
        \caption{Contract Workers Production Intensity}
        \label{fig:con_vs_permB}
    \end{subfigure}
\caption*{\scriptsize \textit{Note:} Panel A displays average relative wage rates for contract versus permanent workers across establishments over time. Panel B presents the average share of total worker-days performed by contract workers across establishments over time. Both panels show 90\% confidence intervals. All averages are weighted by plant revenue.}
    \end{figure}

\begin{figure}[h]
    \centering
    \caption{Production Location}
    \label{fig:prod_loc}
    \begin{subfigure}[h]{0.45\textwidth}
        \centering
        \includegraphics[width=0.9\linewidth]{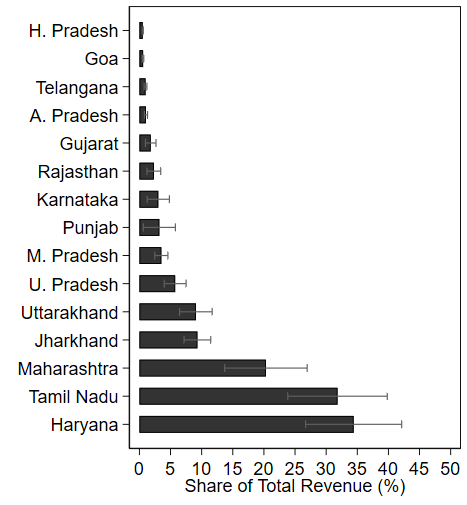}
        \caption{Average Revenue Share by State}
        \label{fig:prod_loc_panA}
    \end{subfigure}
    \hfill
    \begin{subfigure}[h]{0.45\textwidth}
        \centering
        \includegraphics[width=0.9\linewidth]{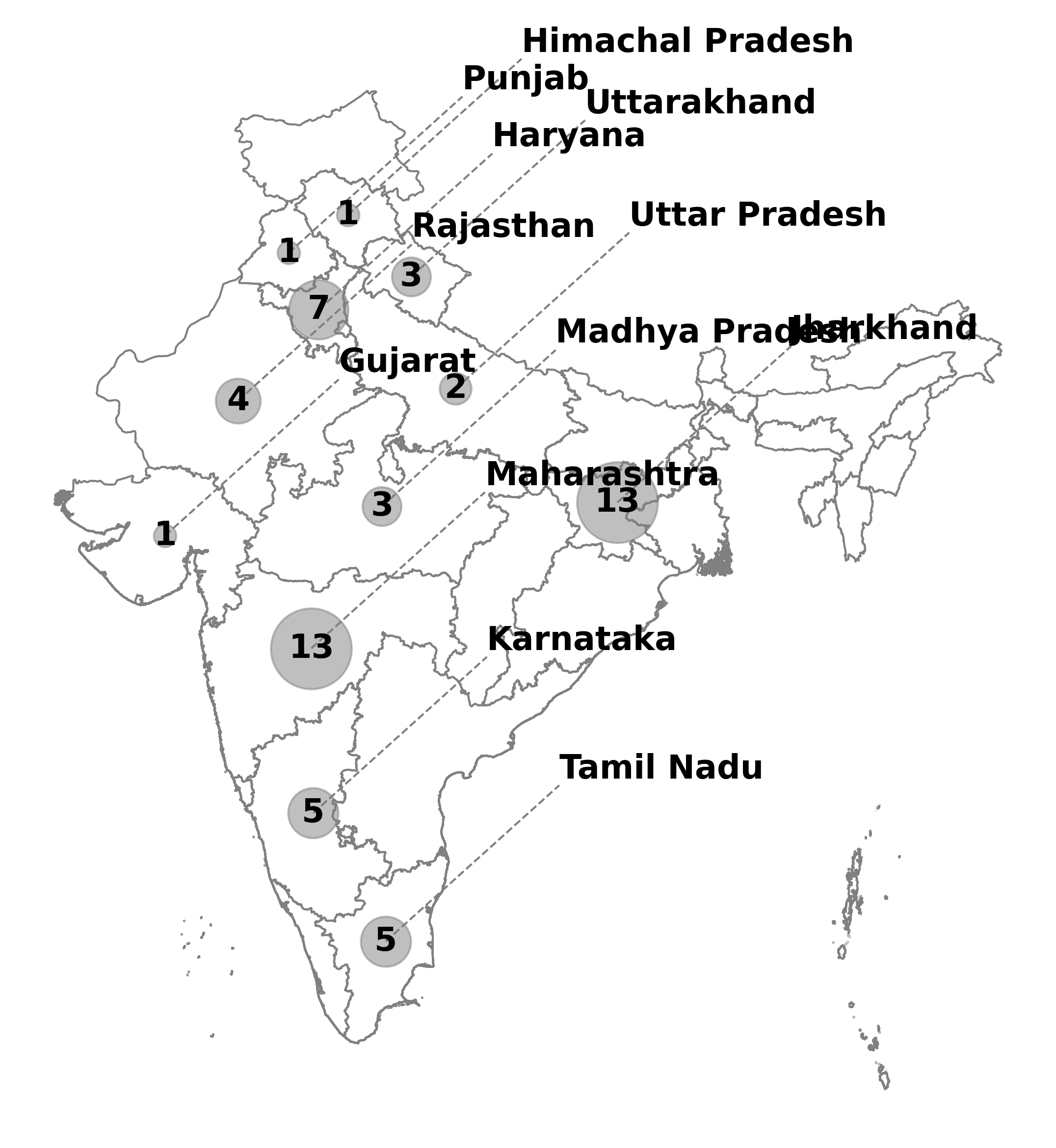}
        \caption{Production Units Location - 2019}
        \label{fig:prod_loc_panB}
    \end{subfigure}
\caption*{\scriptsize \textit{Note:} Panel A presents the mean share of total industry revenue generated by each state over the study period, along with 90\% confidence intervals. Panel B shows the number and location of active production units by state in 2019. No active production units were recorded in Goa, Telangana, or Andhra Pradesh in 2019.}
    \end{figure}

\begin{figure}[h]
    \centering
    \caption{Intensity of Labor Disputes}
    \label{fig:lab_disp}
    \begin{subfigure}[h]{0.45\textwidth}
        \centering
        \includegraphics[width=0.9\linewidth]{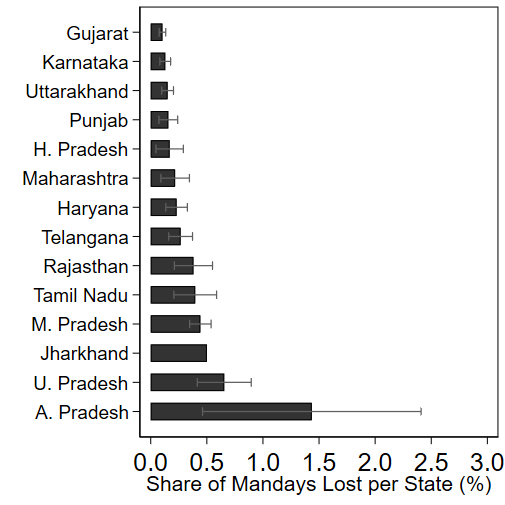}
        \caption{Variation by State}
            \label{fig:lab_disp_panA}
    \end{subfigure}
    \hfill
    \begin{subfigure}[h]{0.45\textwidth}
        \centering
        \includegraphics[width=0.9\linewidth]{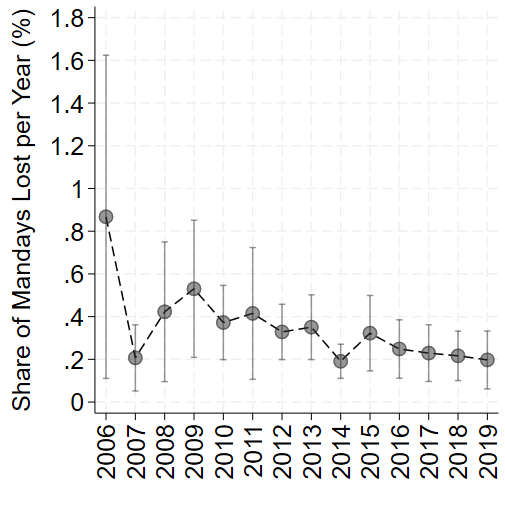}
        \caption{Variation by Year}
        \label{fig:lab_disp_panB}
    \end{subfigure}
    \caption*{\scriptsize \textit{Note:} This figure examines the intensity of labor disputes, measured as the share of total manufacturing employment time lost to strikes and lockouts. I construct this measure by computing the ratio of state-year mandays lost due to industrial disputes to total state-year mandays worked in the formal manufacturing sector by permanent workers. Panel A displays state-level averages across the sample period, with error bars denoting 90\% confidence intervals. Panel B presents the annual averages across all states with corresponding 90\% confidence intervals.}
    \end{figure}

\begin{table}[ht]
\caption{Workers Labor Supply Parameters Estimates}
\label{tab:bc_IV}
    \begin{adjustbox}{width=0.55\textwidth}
{\small
\begin{tabular}{l*{5}{c}}
\hline\hline
                           &     &\multicolumn{2}{c}{Contract}&\multicolumn{2}{c}{Permanent}\\
                                \cmidrule(lr){3-4}\cmidrule(lr){5-6}
           &     &\multicolumn{1}{c}{OLS}&\multicolumn{1}{c}{IV}&\multicolumn{1}{c}{OLS}&\multicolumn{1}{c}{IV}\\
\hline
Constant & $c$        &   -3.140&   -9.470&   -4.606&   -7.300\\
            &    &  (0.561)&  (2.330)&  (0.528)&  (1.103)\\
 Wage coefficient &$\gamma$             &    0.004&    0.046&   0.004&    0.009\\
              &  &  (0.003)&  (0.016)&    (0.001)&  (0.002)\\
Time-varying factor & $\gamma_t$          &   <-0.001&   -0.002&   <-0.001&   <-0.001\\
            &    &  (<0.001)&  (0.001)&   (<0.001)& (<0.001)\\
Nesting parameter & $\eta$  &    0.609&    0.149&   0.617&    0.519\\
            &    &  (0.043)&  (0.182)&   (0.032)&  (0.075)\\
Time trend & $\alpha$     &   -0.050&   -0.052&    0.041&    0.090\\
            &    &  (0.041)&  (0.193)&  (0.035)&  (0.078)\\
\hline
Observations  &  &      383&      272&      383&      272\\
\hline\hline
\end{tabular}
}
\end{adjustbox}
\caption*{\scriptsize \textit{Note:} This table presents OLS and IV second-stage estimates of equation \eqref{eq:lab_supply} for contract and permanent workers. Robust standard errors appear in parentheses.}
\end{table}

\begin{table}[ht]
    \centering
    \caption{First Stage Results}
    \label{tab:bc_FS}
    \begin{adjustbox}{width=0.74\textwidth}
    {\small
    \begin{tabular}{l*{6}{c}}
\hline\hline
                &\multicolumn{3}{c}{Contract}&\multicolumn{3}{c}{Permanent}\\
\cmidrule(lr){2-4}\cmidrule(lr){5-7}
                &\multicolumn{1}{c}{(1)}&\multicolumn{1}{c}{(2)}&\multicolumn{1}{c}{(3)}&\multicolumn{1}{c}{(4)}&\multicolumn{1}{c}{(5)}&\multicolumn{1}{c}{(6)}\\
                &\multicolumn{1}{c}{$W^C_{jt}$}&\multicolumn{1}{c}{$W^C_{jt}\times t_{trend}$}&\multicolumn{1}{c}{$\log(s_{j|rt,C})$}&\multicolumn{1}{c}{$W^D_{jt}$}&\multicolumn{1}{c}{$W^D_{jt}\times  t_{trend}$}&\multicolumn{1}{c}{$\log(s_{j|rt,D})$}\\
\hline
Constant        &   80.253& 2618.751&   -0.694& -488.482& 5850.181&   -4.804\\
                &(109.739)&(1427.400)&  (1.955)&(493.701)&(5974.267)&  (3.158)\\
$\log W^S_{jt-1}$       &   -0.721& -599.608&    0.084&  122.200&-1061.692&    0.632\\
                & (15.933)&(210.528)&  (0.264)& (68.769)&(838.630)&  (0.419)\\
$\log W^S_{jt-1}\times t_{trend}$       &    0.979&   55.481&    0.020&   -1.712&  179.114&   -0.000\\
                &  (1.184)& (17.371)&  (0.021)&  (5.105)& (72.869)&  (0.029)\\
$N_{r,t}$        &    3.090&   59.086&   -0.329&   28.298&  312.499&   -0.447\\
                &  (2.615)& (42.749)&  (0.030)&  (8.552)&(123.337)&  (0.042)\\
$t_{trend}$      &    6.726&   -0.572&   -0.174&   20.531& -625.752&    0.003\\
                &  (8.451)&(123.585)&  (0.156)& (37.053)&(530.417)&  (0.218)\\
\hline
Obs.    &      272&      272&      272&      272&      272&      272\\
\hline\hline
\end{tabular}
}
    \end{adjustbox}
    \caption*{\scriptsize \textit{Note:} This table reports first-stage estimates from the IV specification for equation \eqref{eq:lab_supply} for contract and permanent workers. Robust standard errors appear in parentheses. First-stage Sanderson-Windmeijer F-statistics confirm instrument relevance: for contract workers, F-statistics equal 10.11 ($W^C$), 25.40 ($W^C\times t$), and 36.68 ($\log(s_{j|rt,C})$); for permanent workers, F-statistics reach 55.59 ($W^D$), 75.58 ($W^D\times t$), and 66.17 ($\log(s_{j|rt,D})$).}
\end{table}

\begin{figure}[ht]
    \centering
    \caption{Elasticity of Labor Supply}
    \label{fig:elas_lab}
    \begin{subfigure}[h]{0.485\textwidth}
        \centering
        \includegraphics[width=\linewidth]{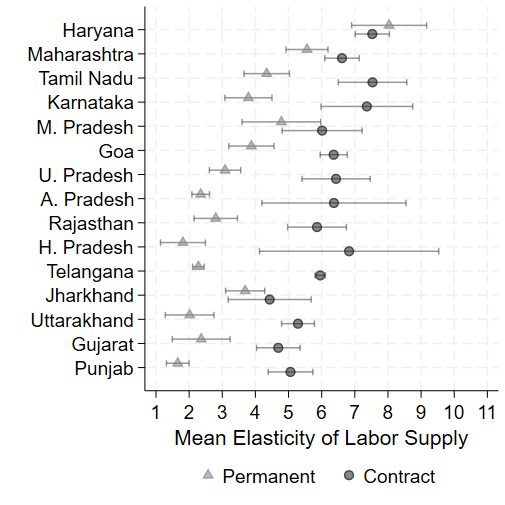}
        \caption{Variation by State}
            \label{fig:elas_labA}
    \end{subfigure}
    \hfill
    \begin{subfigure}[h]{0.485\textwidth}
        \centering
        \includegraphics[width=\linewidth]{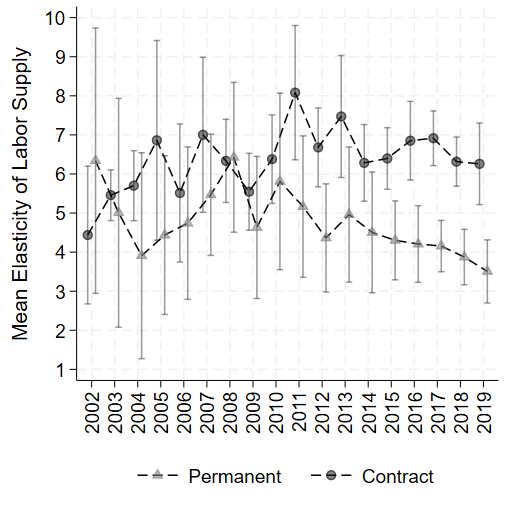}
        \caption{Variation by Year}
        \label{fig:elas_labB}
    \end{subfigure}
\caption*{\scriptsize \textit{Note:} This figure presents revenue-weighted mean elasticity of labor supply estimates for contract and permanent workers. Panel A reports average elasticities across states, while Panel B displays their temporal evolution over the sample period. Error bars indicate 95\% confidence intervals around each point estimate.}
    \end{figure}

\begin{table}[ht]
\centering
\caption{CES Production Function Parameter Estimates}
\label{tab:CES_parameters}
\begin{adjustbox}{width=\textwidth,center}
\begin{tabular}{l@{\hspace{0.15em}}c@{\hspace{0.15em}}c@{\hspace{0.15em}}c@{\hspace{0.15em}}c@{\hspace{0.15em}}c@{\hspace{0.15em}}c@{\hspace{0.15em}}c@{\hspace{0.15em}}c@{\hspace{0.15em}}c@{\hspace{0.15em}}c@{\hspace{0.15em}}c@{\hspace{0.15em}}c@{\hspace{0.15em}}c@{\hspace{0.15em}}c@{\hspace{0.15em}}c}
  \toprule
 & \multicolumn{3}{c}{\textbf{Normalization}} & \multicolumn{3}{c}{\textbf{Substitution}} & \multicolumn{7}{c}{\textbf{Share}} & \multicolumn{2}{c}{\textbf{Variance}} \\ 
  \cmidrule(lr){2-4} \cmidrule(lr){5-7} \cmidrule(lr){8-14} \cmidrule(lr){15-16}
 & $\log(\tau)$ & $\phi_C$ & $\phi_D$ & $\sigma^O$ & $\sigma^M$ & $\sigma^I$ & $\alpha_L$ & $\alpha_M$ & $\alpha_K$ & $\alpha_S$ & $\alpha_E$ & $\alpha_D$ & $\alpha_C$ & $\sigma^H$ & $\sigma^\varepsilon$ \\ 
  \midrule
Est. & -0.590 & 1.493 & 1.935 & 0.914 & 0.202 & -1.366 & 0.063 & 0.603 & 0.334 & 0.640 & 0.360 & 0.757 & 0.243 & 0.786 & 0.468 \\ 
  SE & (1.900) & (0.952) & (0.864) & (0.690) & (1.056) & (1.529) & (0.035) & (0.332) & (0.330) & (0.021) & (0.021) & (0.030) & (0.030) & (0.129) & (0.114) \\ 
   \bottomrule
   \vspace{0.005pt}
\end{tabular}
\end{adjustbox}
\caption*{\scriptsize \textit{Note:} This table reports CES production function parameter estimates obtained through the four-step sequential estimation procedure. Mean absolute correlations between instruments and residuals remain appropriately low (0.02 for Step 2, 0.05 for Step 3), supporting instrument exogeneity. Standard errors are generated using non-parametric bootstrap procedures (detailed in Appendix \ref{a:bootstrap_pf}).}
\end{table}

\begin{table}[ht]
\centering
\caption{Productivity Processes Estimates}
\label{tab:prod}
    \begin{adjustbox}{width=0.3\textwidth}
{\small
\begin{tabular}{lccc}
  \toprule
& & $\omega^H$ & $\omega^L$ \\ 
 \midrule
  Constant & $c$ & 0.383 & -0.132 \\ 
  & & (0.354) & (0.319) \\ 
 Persistency & $\rho$ & 0.881 & 0.911 \\ 
  & & (0.088) & (0.040) \\ 
   \midrule
   Year FE& & YES & YES \\
   \bottomrule
      \vspace{0.005pt}
\end{tabular}
}
\end{adjustbox}
\caption*{\scriptsize \textit{Note:} This table presents AR(1) process parameter estimates for plant-level productivity dynamics obtained through GMM estimation of equations \eqref{eq:Markov_L} and \eqref{eq:NLLS}. I employ non-parametric Bootstrap procedures to generate standard errors, reported in parentheses (see Appendix \ref{a:bootstrap_pf} for implementation details).}
\end{table}

\begin{figure}[ht]
    \centering
    \caption{Implied Output Elasticities}
    \label{fig:outp_elas}
    \begin{subfigure}[h]{0.320\textwidth}
        \centering
        \includegraphics[width=\linewidth]{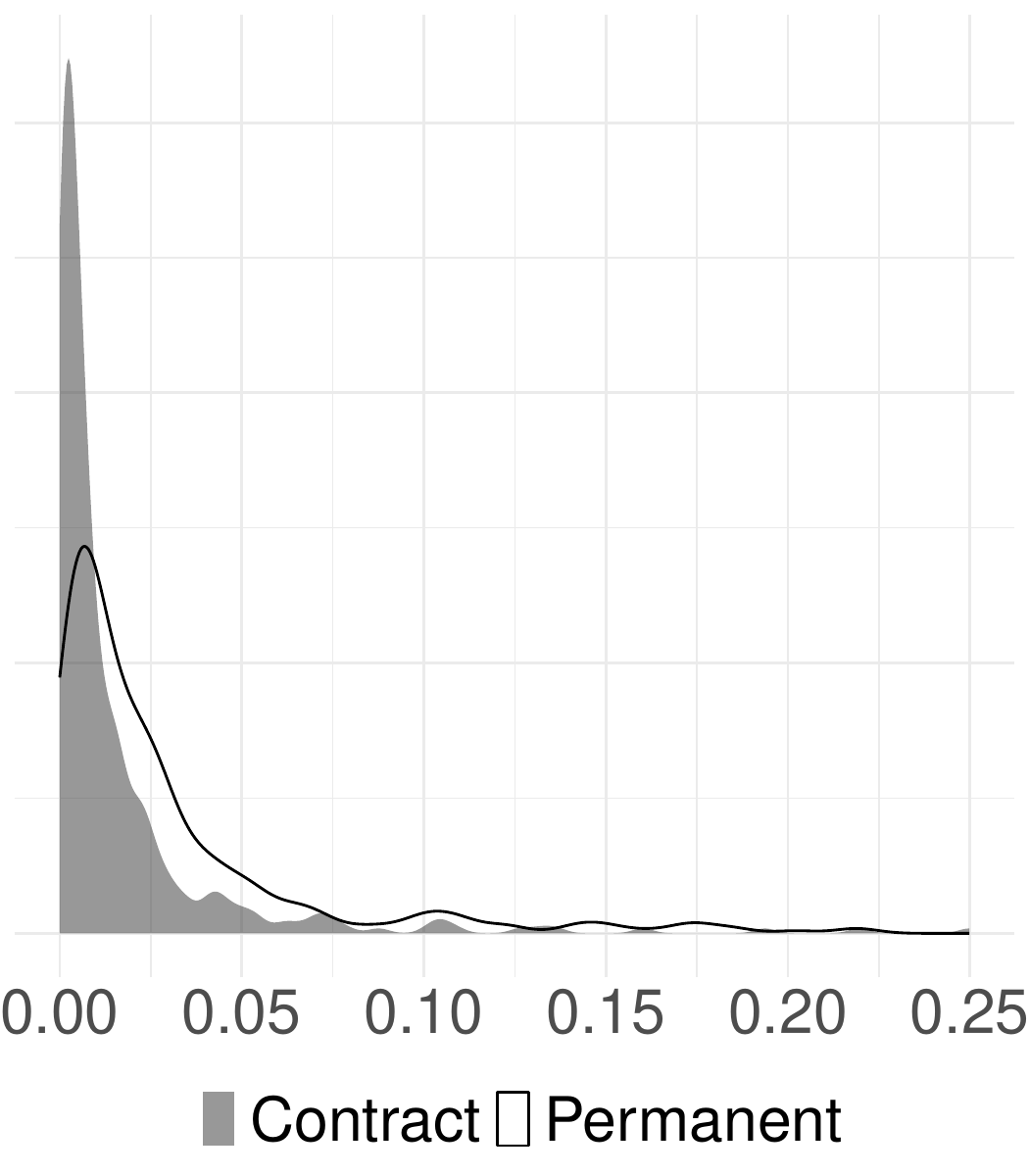}
        \caption{Worker Nest}
    \end{subfigure}
    \begin{subfigure}[h]{0.320\textwidth}
        \centering
        \includegraphics[width=\linewidth]{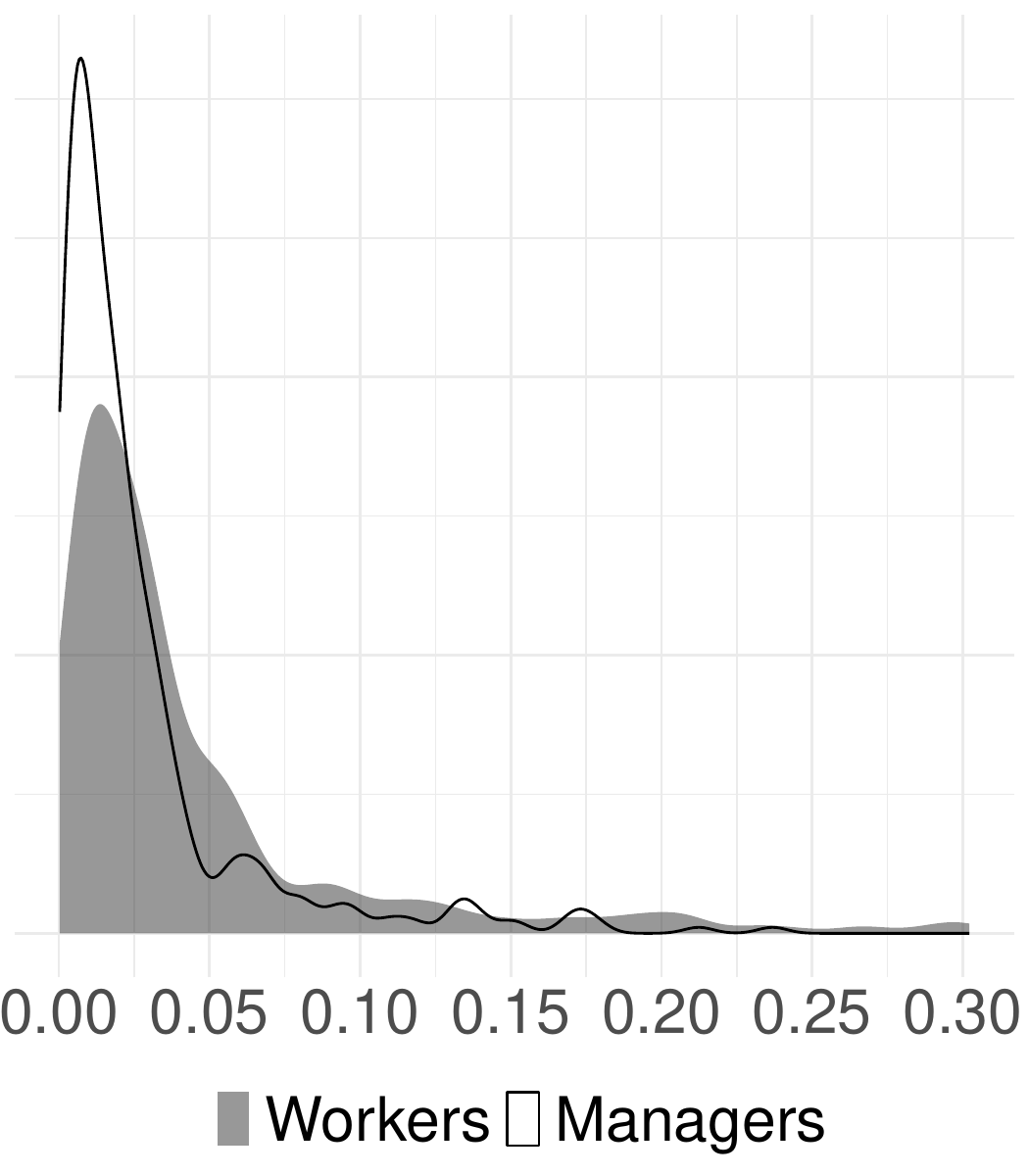}
        \caption{Labor Nest}
    \end{subfigure}
    \begin{subfigure}[h]{0.320\textwidth}
        \centering
        \includegraphics[width=\linewidth]{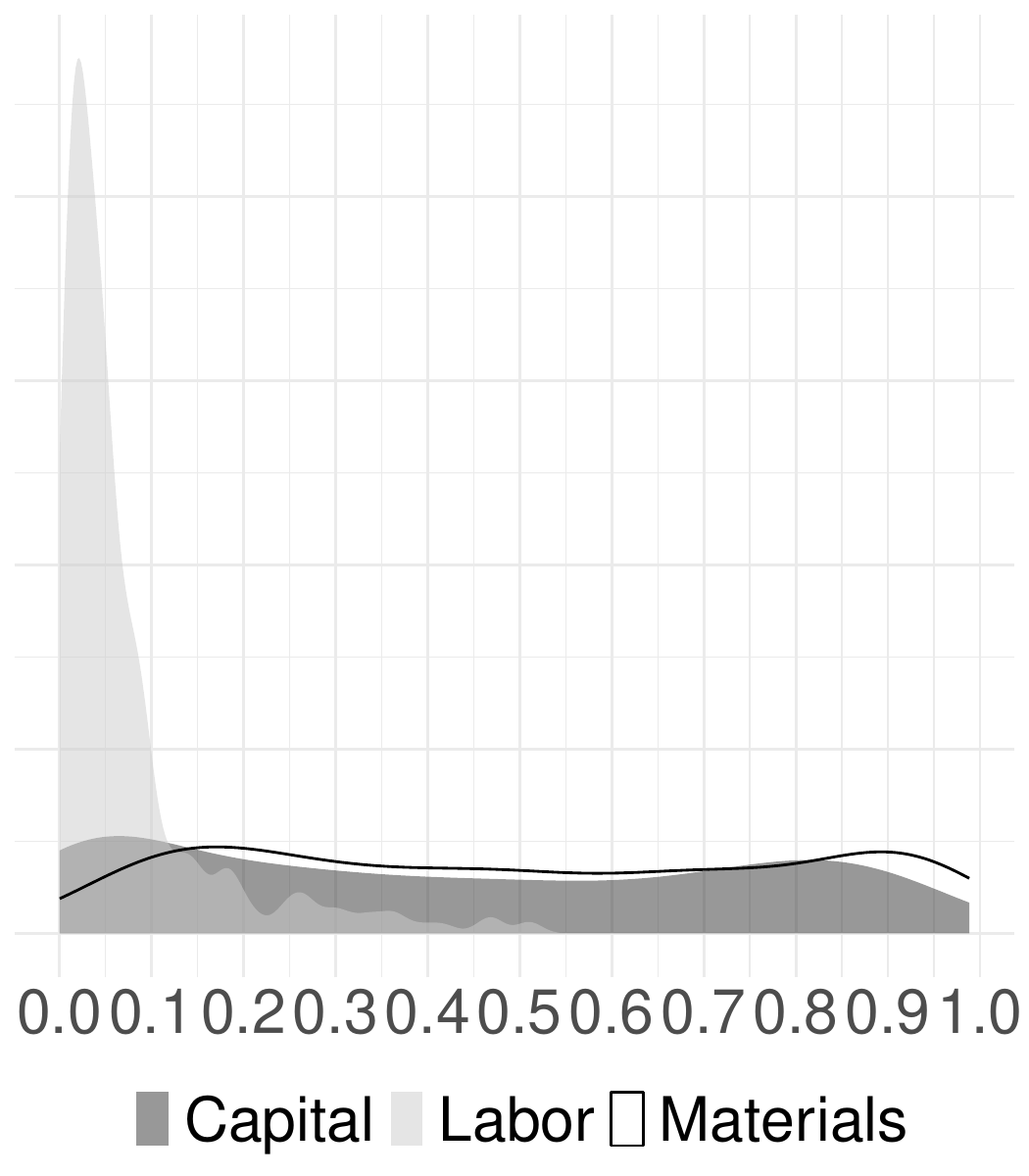}
        \caption{Outer Nest}
    \end{subfigure}
    \caption*{\scriptsize \textit{Note:} These figures present the density distributions of the estimated implied output elasticities across inputs, organized by nests within the CES production function.}
    \end{figure}

    \begin{figure}[ht]
    \centering
    \caption{Production Unobservables}
    \label{fig:prod_unob}
    \begin{subfigure}[h]{0.48\textwidth}
        \centering
        \includegraphics[width=\linewidth]{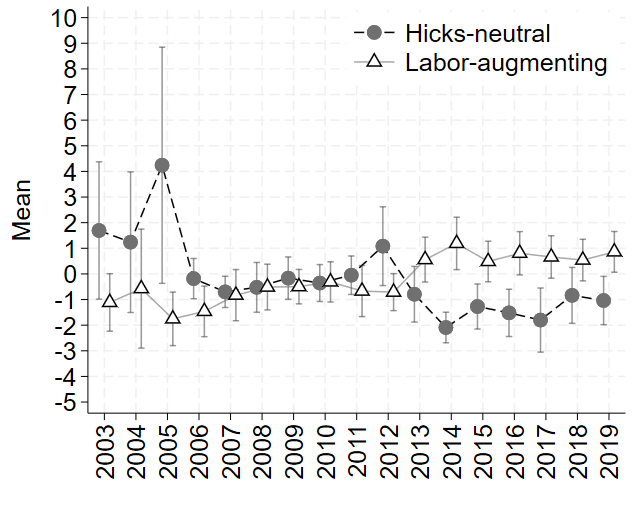}
        \caption{Productivity - Mean Deviations}
            \label{fig:prod_unobA}
    \end{subfigure}
    \hfill
    \begin{subfigure}[h]{0.48\textwidth}
        \centering
        \includegraphics[width=\linewidth]{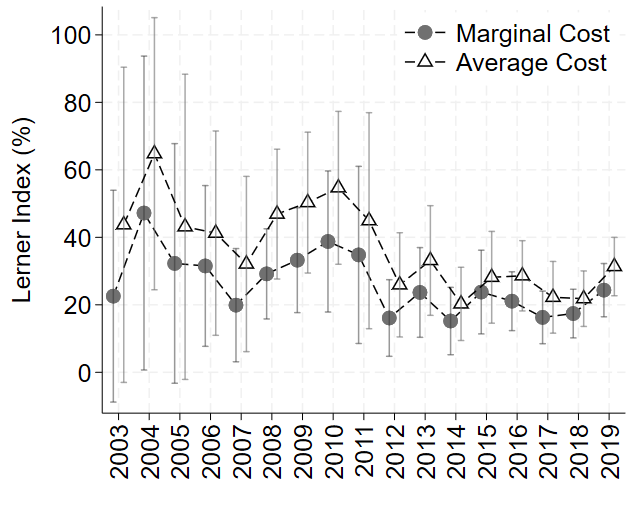}
        \caption{Output Markups}
        \label{fig:prod_unobB}
    \end{subfigure}
\caption*{\scriptsize \textit{Note:} Panel A presents revenue-weighted averages of estimated Hicks-neutral and labor-augmenting productivity deviations from their geometric means, with 95\% confidence intervals. Panel B displays the evolution of revenue-weighted average markups using the Lerner Index, computed as $(P-MC)/P$ for "Marginal Cost" and $(P-AC)/P$ for "Average Cost", where $P$ denotes price, $MC$ marginal cost, and $AC$ average cost. The index measures the proportion of price exceeding factor remuneration. For observations with negative estimated markups, I set both Lerner indices to zero. Both panels report 95\% confidence intervals.}
    \end{figure}

\begin{table}[ht]
\centering
\caption{Parameter Estimates} 
\label{tab:param_estimates}
    \begin{adjustbox}{width=0.3\textwidth}
\begin{tabular}{lccc}
  \toprule
 & $\beta$ & $\theta_1$ & $\theta_0$ \\ 
  \midrule
Est. & 0.003 & 0.020 & 6.788 \\ 
  SE & (0.001) & (0.161) & (1.880) \\ 
   \bottomrule
\end{tabular}
\end{adjustbox}
\caption*{\scriptsize \textit{Note:} This Table presents the bargaining and marginal cost parameters estimated via GMM on equations \eqref{eq:Nash_bargaining_FOC} and \eqref{eq:perm_marg_cost}. Standard errors are generated using non-parametric bootstrap procedures (detailed in Appendix \ref{a:bootstrap_barg}).}
\end{table}

\begin{figure}[ht]
\centering
\caption{Strike Intensity Instrument Analysis}\label{fig:Instr_anal}
\begin{subfigure}[h]{0.49\textwidth}
    \centering
    \includegraphics[width=\linewidth]{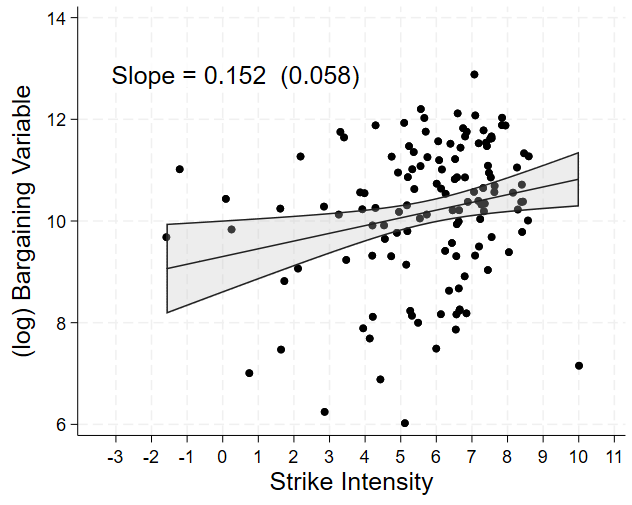}
    \caption{First Stage}\label{fig:FS}
\end{subfigure}
\hfill
\begin{subfigure}[h]{0.49\textwidth}
    \centering
    \includegraphics[width=\linewidth]{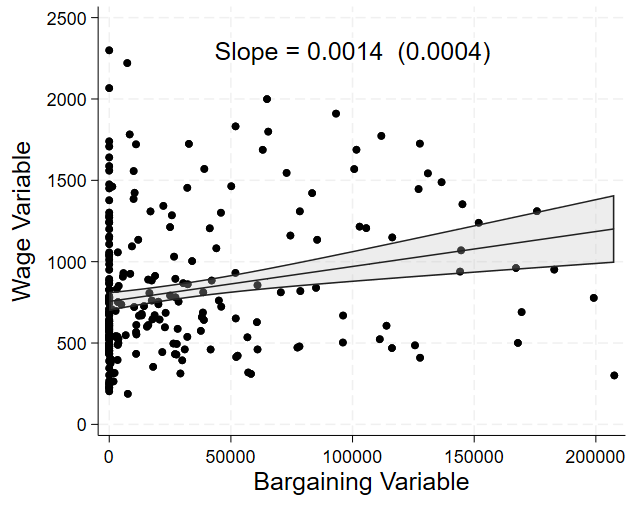}
    \caption{OLS - Second Stage}\label{fig:OLS}
\end{subfigure}
\caption*{\scriptsize \textit{Note:} Panel (A) plots the log of the variable associated with the bargaining parameter ("Bargaining Variable") from equation \eqref{eq:Nash_bargaining_FOC}, $\frac{\Pi_{jt}}{D_{jt}}\left(\frac{\partial (U^U_{jt}-U^U_{Ojt})}{\partial D_{jt}}\frac{D_{jt}}{(U^U_{jt}-U^U_{Ojt})}\right)$, against the strike intensity instrument ("Strike Intensity") from equation \eqref{eq:instrument}. Panel (B) plots the left-hand side variable ("Wage Variable") of equation \eqref{eq:Nash_bargaining_FOC}, $\left(\frac{\partial W_{jt,D}}{\partial D_{jt}}\frac{D_{jt}}{W_{jt,D}}+1\right)W_{jt,D}$, against the bargaining variable. Fitted regression lines with 95\% confidence intervals are displayed. The first-stage F-statistic for panel (A) regression equals 6.78.}
\end{figure}

    \begin{figure}[ht]
    \centering
        \caption{Workers' Wage Markdowns}
    \includegraphics[width=1.0\linewidth]{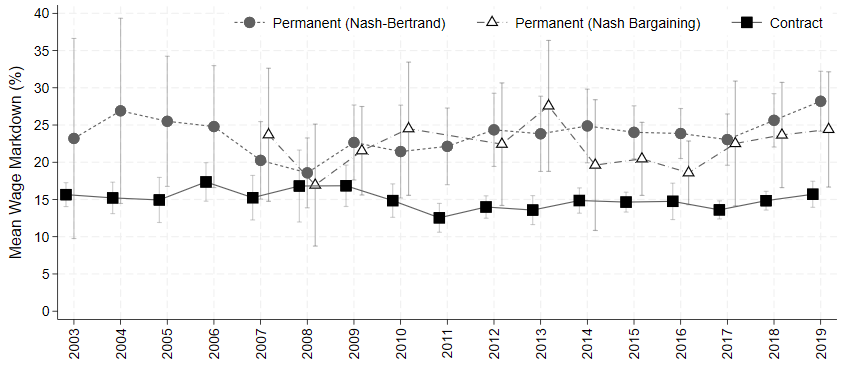}
    \label{fig:markdowns}
    \caption*{\scriptsize \textit{Note:} I present revenue-weighted mean wage markdowns for permanent workers under Nash-Bertrand and Nash bargaining conduct, and for contract workers. Markdowns equal $(MRPL-W)/MRPL$, the proportion of marginal product that plants retain. Error bars denote 95\% confidence intervals.}
\end{figure}

\begin{figure}[ht]
    \centering
\caption{Empirical Distributions of MRPL and Non-wage Marginal Costs}
    \label{fig:decomp}
    \begin{subfigure}[h]{0.47\textwidth}
        \centering
        \includegraphics[width=\linewidth]{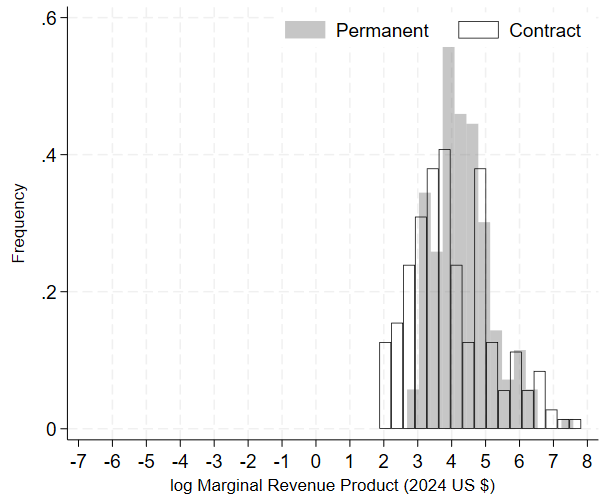}
        \caption{}
            \label{fig:decompA}
    \end{subfigure}
    \hfill
    \begin{subfigure}[h]{0.47\textwidth}
        \centering
        \includegraphics[width=\linewidth]{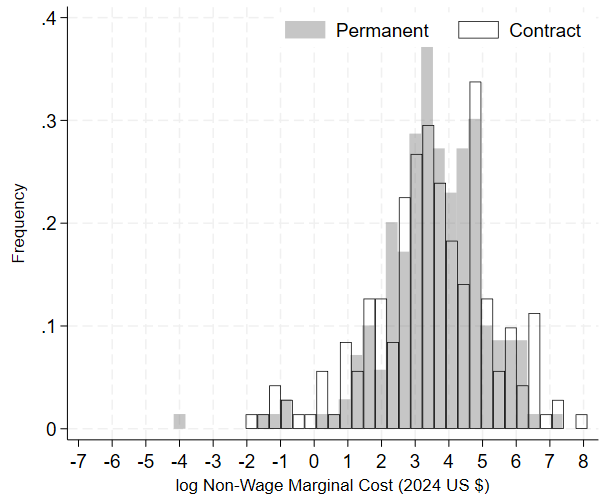}
        \caption{}
        \label{fig:decompB}
    \end{subfigure}
\caption*{\scriptsize \textit{Note:} This figure presents empirical distributions of log marginal revenue product (Panel A) and log non-wage marginal cost (Panel B) for permanent and contract workers.}
    \end{figure}

\begin{figure}[ht]
    \centering
\caption{Workers Welfare and Wage Premium Increase after Offsetting Markdowns}
    \label{fig:Counter}
    \begin{subfigure}[h]{0.47\textwidth}
        \centering
        \includegraphics[width=\linewidth]{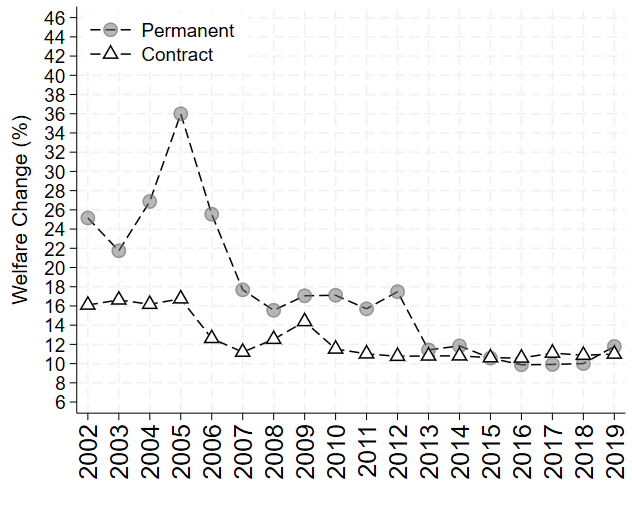}
        \caption{Workers Welfare}
            \label{fig:counterA}
    \end{subfigure}
    \hfill
    \begin{subfigure}[h]{0.47\textwidth}
        \centering
        \includegraphics[width=\linewidth]{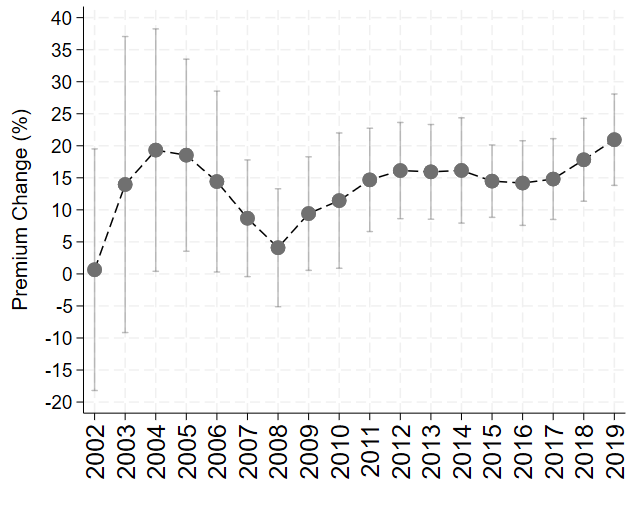}
        \caption{Permanent Worker Compensation Premium}
        \label{fig:counterB}
    \end{subfigure}
\caption*{\scriptsize \textit{Note:} Panel A presents the percentage increase in welfare for permanent and contract workers after receiving a markdown-offsetting lump-sum transfer. Welfare changes are computed as the difference between counterfactual workers' inclusive values, based on equation \eqref{eq:counter}, and observed inclusive values. Panel B presents the revenue-weighted average percentage increase in the permanent worker compensation premium after receiving the lump-sum transfer relative to observed levels, along with 95\% confidence intervals.}
    \end{figure}

%% file: paper_appendixa.tex
\section{Boxplot Distributions of Plant-year Production Variables} \label{a:Vars}
\begin{figure}[h!]
    \centering
    \begin{subfigure}[b]{0.49\textwidth}
        \centering
        \includegraphics[width=\linewidth]{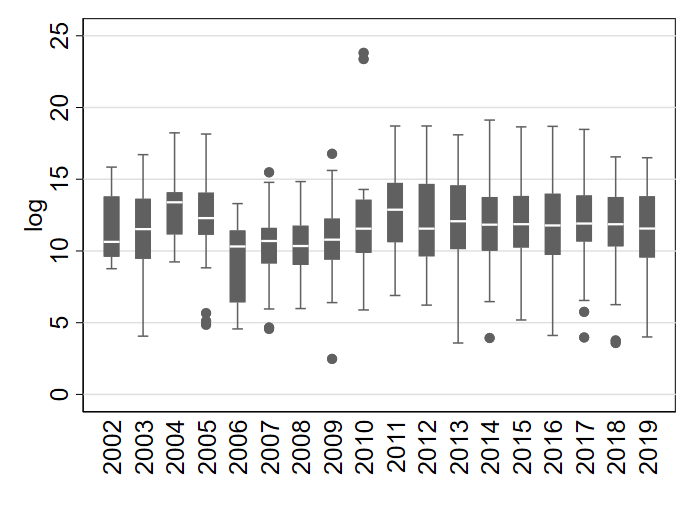}
        \caption*{Output Quantity}
    \end{subfigure}
    \hfill
    \begin{subfigure}[b]{0.49\textwidth}
        \centering
        \includegraphics[width=\linewidth]{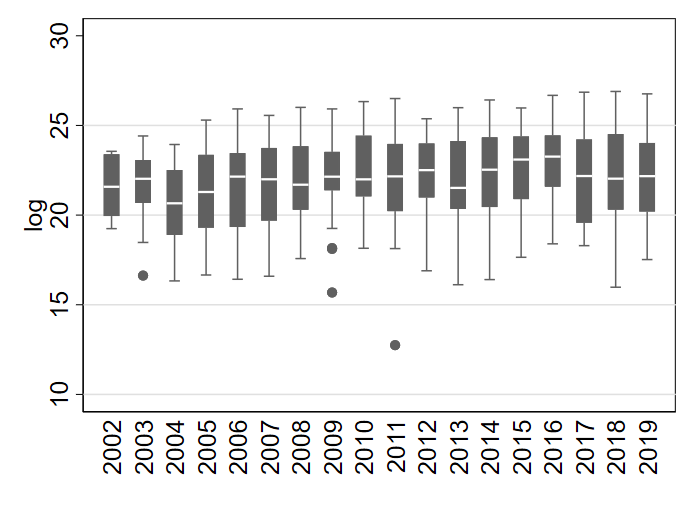}
        \caption*{Revenue}
    \end{subfigure}

    \vspace{0.1cm}
    \begin{subfigure}[b]{0.49\textwidth}
        \centering
        \includegraphics[width=\linewidth]{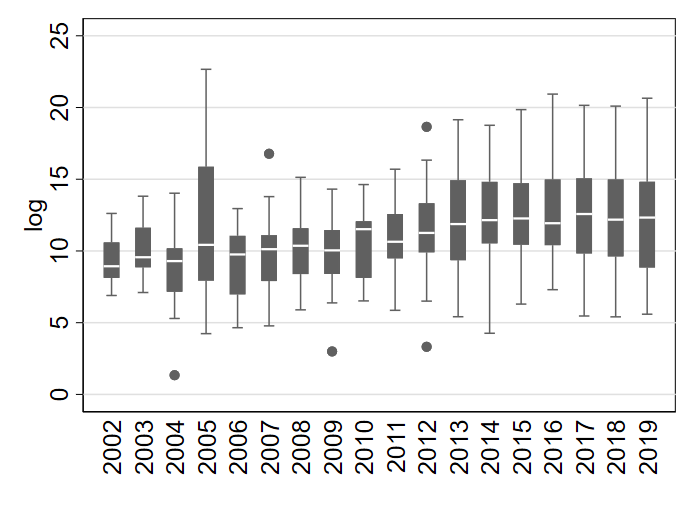}
        \caption*{Intermediates Quantity}
    \end{subfigure}
    \hfill
    \begin{subfigure}[b]{0.49\textwidth}
        \centering
        \includegraphics[width=\linewidth]{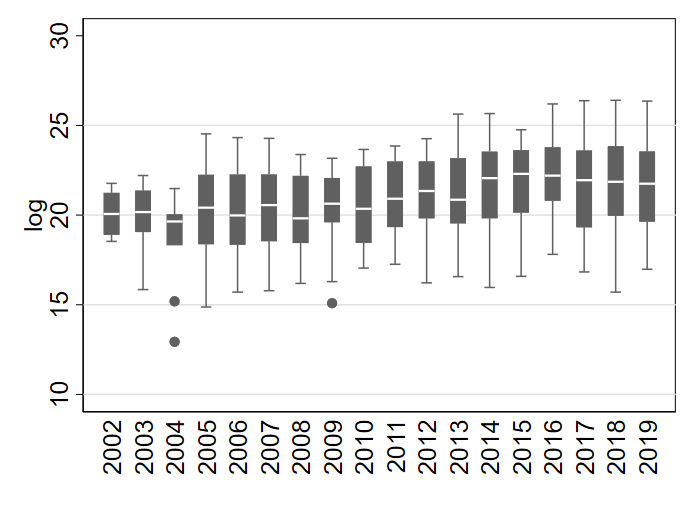}
        \caption*{Intermediates Expenditure}
    \end{subfigure}
    
    \vspace{0.1cm}
    \begin{subfigure}[b]{0.49\textwidth}
        \centering
        \includegraphics[width=\linewidth]{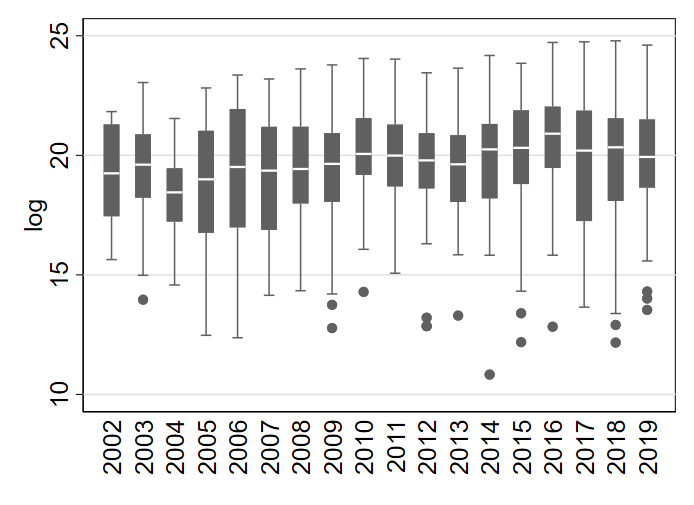}
        \caption*{Capital}
    \end{subfigure}
\end{figure}

\begin{figure}[h]
    \centering
    \begin{subfigure}[b]{0.49\textwidth}
        \centering
        \includegraphics[width=\linewidth]{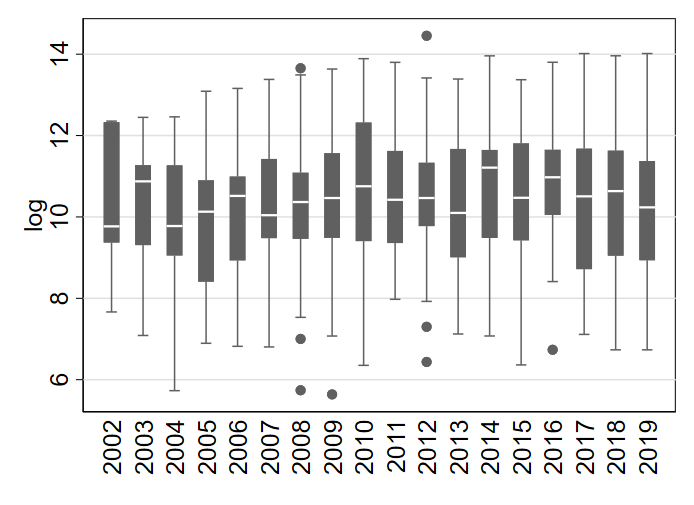}
        \caption*{Managers Mandays Worked}
    \end{subfigure}
    \hfill
    \begin{subfigure}[b]{0.49\textwidth}
        \centering
        \includegraphics[width=\linewidth]{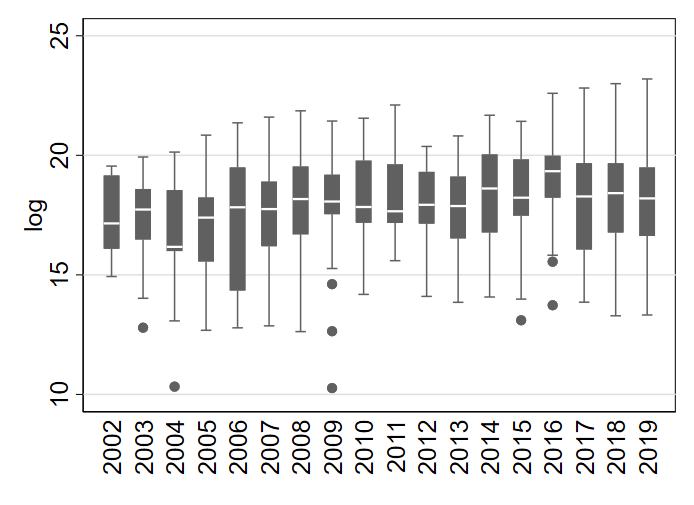}
        \caption*{Managers Payroll}
    \end{subfigure}

    \vspace{0.1cm}
    \begin{subfigure}[b]{0.49\textwidth}
        \centering
        \includegraphics[width=\linewidth]{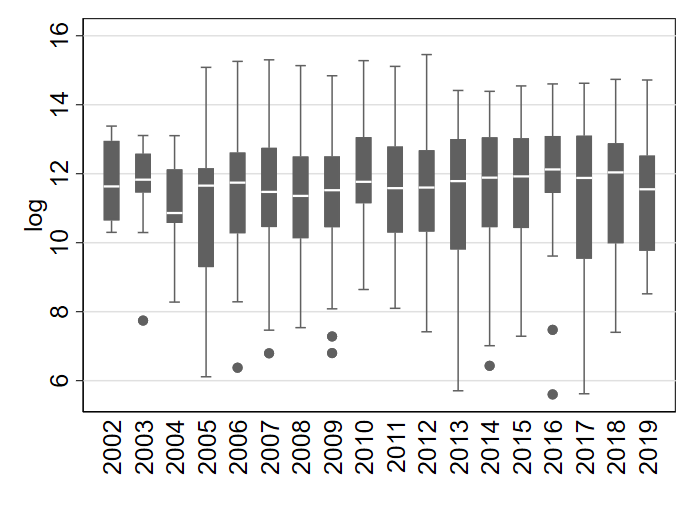}
        \caption*{Permanent Workers Mandays Worked}
    \end{subfigure}
    \hfill
    \begin{subfigure}[b]{0.49\textwidth}
        \centering
        \includegraphics[width=\linewidth]{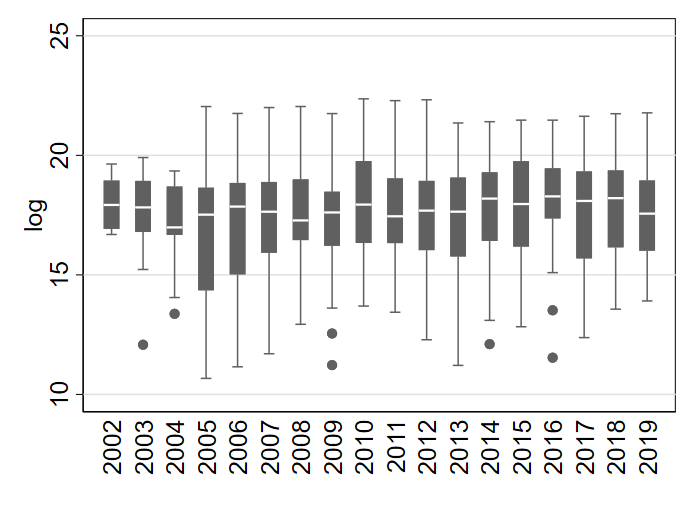}
        \caption*{Permanent Workers Payroll}
    \end{subfigure}
    
    \vspace{0.1cm}
    \begin{subfigure}[b]{0.49\textwidth}
        \centering
        \includegraphics[width=\linewidth]{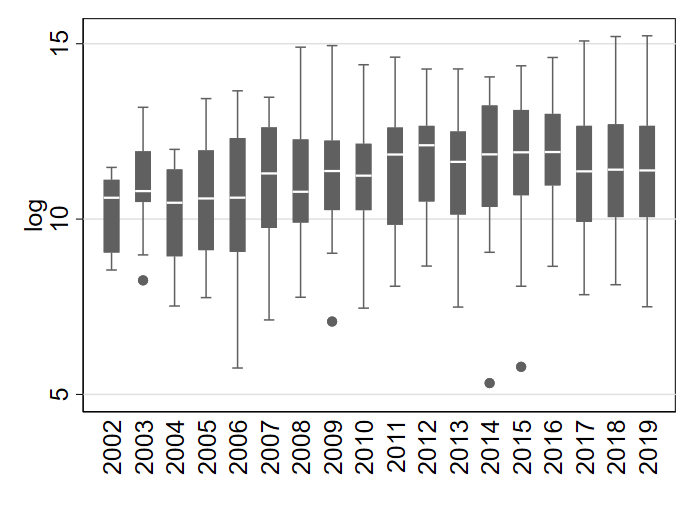}
        \caption*{Temporary Workers Mandays Worked}
    \end{subfigure}
    \hfill
    \begin{subfigure}[b]{0.49\textwidth}
        \centering
        \includegraphics[width=\linewidth]{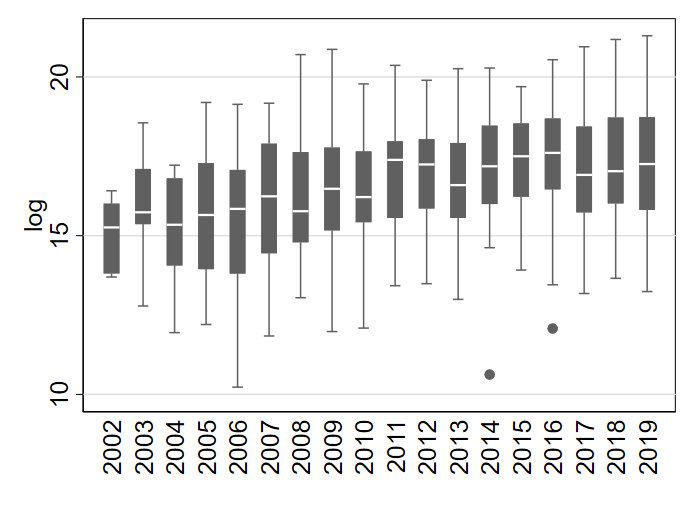}
        \caption*{Temporary Workers Payroll}
    \end{subfigure}
    \caption*{\footnotesize \textit{Note:} The figure presents boxplot distributions of production variables employed in the analysis across years. Dots denote outliers.}
\end{figure}


\clearpage
\section{Original First Order Conditions}\label{a:FOCs_orig}
For contract workers
\begin{equation}
\small
\begin{aligned}
C:\quad\underbrace{\left(\frac{\partial P_{jt}}{\partial Q_{jt}}\frac{Q_{jt}}{P_{jt}}+1\right)P_{jt}\frac{\partial Q_{jt}}{\partial C_{jt}}}_{\text{MRP(C)}}=\underbrace{\left[\left(\frac{\partial W_{jt,C}}{\partial C_{jt}}\frac{\partial C_{jt}}{W_{jt,C}}+1\right)+\frac{\partial\Phi_{jt}}{\partial C_{jt}}\frac{1}{W_{jt,C}}\right]W_{jt,C}}_{\text{MC(C)}}
\end{aligned}
\end{equation}
where
\begin{equation}
\small
\begin{aligned}
\frac{\partial Q_{jt}}{\partial C_{jt}}=\frac{Q_{jt}}{\left(\tilde{\alpha}_K {K}_{jt}^{\sigma^O} + \tilde{\alpha}_M {M}_{jt}^{\sigma^O} + \tilde{\alpha}_L (\exp({\omega}_{jt}^{L}) {L}_{jt})^{\sigma^O} \right)}\frac{\tilde{\alpha}_L (\exp({\omega}_{jt}^{L}) {L}_{jt})^{\sigma^O}}{\left( \tilde{\alpha}_S {S}_{jt}^{\sigma^M}+ \tilde{\alpha}_E {E}_{jt}^{\sigma^M} \right)}\frac{\tilde{\alpha}_E {E}_{jt}^{\sigma^M}}{\left( \tilde{\alpha}_C {C}_{jt}^{\sigma^I} + \tilde{\alpha}_D {D}_{jt}^{\sigma^I} \right)}\tilde{\alpha}_C {C}_{jt}^{\sigma^I-1}
\end{aligned}
\end{equation}
\noindent\rule{\textwidth}{2pt} 
For supervisors
\begin{equation}
\small
\begin{aligned}
S:\quad\underbrace{\left(\frac{\partial P_{jt}}{\partial Q_{jt}}\frac{Q_{jt}}{P_{jt}}+1\right)P_{jt}\frac{\partial Q_{jt}}{\partial S_{jt}}}_{\text{MRP(S)}}=\underbrace{W_{jt,S}}_{\text{MC(S)}}
\end{aligned}
\end{equation}
where
\begin{equation}
\small
\begin{aligned}
\frac{\partial Q_{jt}}{\partial S_{jt}}=\frac{Q_{jt}}{\left(\tilde{\alpha}_K {K}_{jt}^{\sigma^O} + \tilde{\alpha}_M {M}_{jt}^{\sigma^O} + \tilde{\alpha}_L (\exp({\omega}_{jt}^{L}) {L}_{jt})^{\sigma^O} \right)}\frac{\tilde{\alpha}_L (\exp({\omega}_{jt}^{L}) {L}_{jt})^{\sigma^O}}{\left( \tilde{\alpha}_S {S}_{jt}^{\sigma^M}+ \tilde{\alpha}_E {E}_{jt}^{\sigma^M} \right)}\tilde{\alpha}_S {S}_{jt}^{\sigma^M-1}
\end{aligned}
\end{equation}
\noindent\rule{\textwidth}{2pt}
For materials
\begin{equation}
\small
\begin{aligned}
M:\quad\underbrace{\left(\frac{\partial P_{jt}}{\partial Q_{jt}}\frac{Q_{jt}}{P_{jt}}+1\right)P_{jt}\frac{\partial Q_{jt}}{\partial M_{jt}}}_{\text{MRP(M)}}=\underbrace{P_{jt,M}}_{\text{MC(M)}}
\end{aligned}
\end{equation}
where
\begin{equation}
\small
\begin{aligned}
\frac{\partial Q_{jt}}{\partial M_{jt}}=\frac{Q_{jt}}{\left(\tilde{\alpha}_K {K}_{jt}^{\sigma^O} + \tilde{\alpha}_M {M}_{jt}^{\sigma^O} + \tilde{\alpha}_L (\exp({\omega}_{jt}^{L}) {L}_{jt})^{\sigma^O} \right)}\tilde{\alpha}_M {M}_{jt}^{\sigma^O-1}
\end{aligned}
\end{equation}

\clearpage
\section{Derivation of the Inclusive Value in Nested Logit Models}\label{a:Inc_Val}

\subsection{Utility Specification}
In a nested logit framework, the indirect utility of agent $i$ selecting choice $j$ within group $g$ is modeled as:
\begin{equation}
    U_{ijg} = \delta_j + \zeta_{ig} + (1 - \eta) \epsilon_{ij}.
    \label{eq:utility}
\end{equation}
Here, $\delta_j$ denotes the mean utility derived from choice $j$, which is common across all agents. The parameter $\eta$, $0<\eta<1$, is a nesting parameter that captures the degree of correlation of choices within groups, with higher values indicating stronger within-group substitutability. The model incorporates two stochastic components: $\zeta_{ig}$ represents a group-specific preference shock for agent $i$, capturing unobserved factors common to all choices within group $g$, and $\epsilon_{ij}$ represents an idiosyncratic shock specific to agent $i$'s evaluation of choice $j$. 

The random components $\zeta_{ig}$ and $\epsilon_{ij}$ are assumed to be mutually independent. Specifically, $\epsilon_{ij}$ follows a standard Type-I extreme value distribution, while $\zeta_{ig}$ has a unique distribution chosen such that the composite error term
\begin{align}
\zeta_{ig} + (1 - \eta)\epsilon_{ij}, \label{eq:assg} 
\end{align}
is distributed as Type-I extreme value \citep{cardell1997variance}. This distributional structure is crucial for the tractability of the nested logit model, as it ensures that choice probabilities at both the within-group and across-group levels follow closed-form logit expressions.

\subsection{Inclusive Value: Definition and Derivation}

The inclusive value, also known as the log-sum or expected maximum utility, is a central quantity in discrete choice models. It measures the expected utility an agent obtains from an optimally chosen alternative within a choice set, before the realization of idiosyncratic shocks. In welfare analysis, changes in the inclusive value directly quantify changes in agent surplus. For nested logit models, inclusive values arise at two levels: the group level, capturing the attractiveness of alternatives within each group, and the overall level, representing the value of the entire choice environment. I derive these inclusive values in two steps, beginning with the group-level value and then aggregating to obtain the overall measure.

\subsubsection{Step 1: Group-Level Inclusive Value}

The inclusive value of group $g$, denoted $IV_g$, quantifies the expected maximum utility that agent $i$ obtains from selecting the best alternative within group $g$. Formally, this expectation is taken over the realizations of idiosyncratic shocks $\epsilon_{ij}$ for all alternatives $j \in B_g$, conditional on the group-specific shock $\zeta_{ig}$, where $B_g$ denotes the set of alternatives in group $g$.

Conditional on choosing group $g$, agent $i$ selects the alternative yielding the highest utility:
\begin{equation}
    \max_{j \in B_g} U_{ijg} = \max_{j \in B_g} \left[\delta_j + \zeta_{ig} + (1-\eta)\epsilon_{ij}\right].
\end{equation}
Since $\zeta_{ig}$ is common to all alternatives within the group, it factors out of the maximization:
\begin{equation}
    \max_{j \in B_g} U_{ijg} = \zeta_{ig} + \max_{j \in B_g} \left[\delta_j + (1-\eta)\epsilon_{ij}\right].
\end{equation}
The inclusive value is therefore the expectation of this maximum over the distribution of $\epsilon_{ij}$:
\begin{equation}
    IV_g = E_{\epsilon}\left[\max_{j \in B_g} \left(\delta_j + (1-\eta)\epsilon_{ij}\right)\right].
\end{equation}

\paragraph{Derivation}
I establish the closed form of $IV_g$ by exploiting properties of the Type-I extreme value distribution. The derivation proceeds in two stages: I first characterize the distribution of maximum utility for a general choice problem, then extend this result to accommodate arbitrary scale parameters before applying it to the nested logit framework.

\paragraph{General Choice Problem}
Consider a decision-maker choosing among $n$ alternatives with systematic utilities $V_1, ..., V_n$ and i.i.d. Type-I extreme value errors $\epsilon_1, ..., \epsilon_n$ with zero location and unit scale. The Type-I extreme value (Gumbel) distribution has cumulative distribution function $F(\epsilon) = \exp(-\exp(-\epsilon))$ and probability density function $f(\epsilon) = \exp(-\epsilon - \exp(-\epsilon))$. Define the maximum utility as $U^* = \max_k (V_k + \epsilon_k)$. The CDF of $U^*$ satisfies:
\begin{equation}
    P(U^* \leq u) = P(V_k + \epsilon_k \leq u ~\forall k) = \prod_{k=1}^n P(\epsilon_k \leq u - V_k) = \prod_{k=1}^n \exp(-\exp(-(u - V_k))).
\end{equation}
Exploiting properties of the exponential function, this simplifies to:
\begin{equation}
    P(U^* \leq u) = \exp\left(-\sum_{k=1}^n \exp(V_k - u)\right) = \exp\left(-\exp\left(-u + \ln\left(\sum_{k=1}^n \exp(V_k)\right)\right)\right).
\end{equation}
This reveals that $U^*$ follows a Type-I extreme value distribution with location parameter $\mu = \ln(\sum_{k=1}^n \exp(V_k))$ and unit scale. Since the expected value of a Type-I extreme value random variable with location $\mu$ and unit scale equals $\mu + \gamma$, where $\gamma \approx 0.5772$ is the Euler-Mascheroni constant, I obtain:
\begin{equation}
    E[\max_k (V_k + \epsilon_k)] = \ln\left(\sum_{k=1}^n \exp(V_k)\right) + \gamma.
\end{equation}

\paragraph{Arbitrary Scale}
I now extend this result to accommodate arbitrary scale parameters. Consider utilities of the form $V_k + \eta\epsilon_k$ with scale parameter $\eta > 0$. Factoring out $\eta$ from the maximization yields:
\begin{equation}
    \max_k (V_k + \eta\epsilon_k) = \eta \cdot \max_k \left(\frac{V_k}{\eta} + \epsilon_k\right).
\end{equation}
Taking expectations and applying the unit-scale result with rescaled systematic utilities $\tilde{V}_k = V_k/\eta$:
\begin{equation}
    E[\max_k (V_k + \eta\epsilon_k)] = \eta \cdot E\left[\max_k \left(\frac{V_k}{\eta} + \epsilon_k\right)\right] = \eta \ln\left(\sum_{k=1}^n \exp(V_k/\eta)\right) + \eta\gamma.
\end{equation}

\paragraph{Application to the Nested Logit Framework}
Applying this general formula to the within-group problem, where systematic utilities are $\delta_j$, the scale parameter is $(1-\eta)$, and errors $\epsilon_{ij}$ are i.i.d. Type-I extreme value, yields:
\begin{equation}
    IV_g = E_{\epsilon}\left[\max_{j \in B_g} \left(\delta_j + (1-\eta)\epsilon_{ij}\right)\right] = (1-\eta) \ln\left(\sum_{j \in B_g} \exp\left(\frac{\delta_j}{1-\eta}\right)\right) + (1-\eta)\gamma.
\end{equation}
The constant term $(1-\eta)\gamma$ is common across all groups and does not affect choice probabilities or welfare comparisons, since only differences in inclusive values influence these quantities. I therefore adopt the normalization that omits this constant:
\begin{equation}
    \tilde{IV}_g = (1-\eta) \ln\left(\sum_{j \in B_g} \exp\left(\frac{\delta_j}{1-\eta}\right)\right).
    \label{eq:group_iv}
\end{equation}
This expression recovers the familiar log-sum formula from multinomial logit models, scaled by the nesting parameter $(1-\eta)$. Using the normalized utilities $V_j = \delta_j/(1-\eta)$, I can equivalently express the inclusive value as $\tilde{IV}_g = (1-\eta) \ln(\sum_{j \in B_g} \exp(V_j))$.

\paragraph{Properties of the Group-Level Inclusive Value}
 The inclusive value $IV_g$ serves as a comprehensive measure of group attractiveness, aggregating the value of all alternatives in group $g$ in a manner that respects both the number of alternatives and their individual utilities. This aggregation exhibits strict monotonicity: $IV_g$ is strictly increasing in each $\delta_j$, ensuring that any improvement in an alternative's utility raises the group's overall attractiveness. 

The log-sum structure of $IV_g$ naturally captures the value of variety within groups. Holding individual utilities constant, adding alternatives to group $g$ increases $IV_g$, reflecting the agent benefit from expanded choice sets even when new alternatives offer similar utilities to existing ones. 

The nesting parameter $\eta$ governs the scale of the group-level inclusive value through the multiplicative factor $(1-\eta)$. As $\eta \to 1$, indicating stronger within-group correlation and greater substitutability among alternatives, the scale $(1-\eta) \to 0$ diminishes the importance of utility differences $\delta_j$ for group-level decisions. 

For welfare analysis at the group level, differences in the inclusive value provide a natural metric: $\Delta IV_g = IV_g' - IV_g$ represents the welfare gain from improving group $g$'s choice set, measured in utility units. This interpretation establishes the group-level inclusive value as the appropriate tool for evaluating policy interventions or market changes that affect the composition or quality of alternatives within a particular group.

\subsubsection{Step 2: Overall Inclusive Value}

Having established the group-level inclusive value, $IV_g$, I now derive the overall inclusive value, $IV$, which represents the expected maximum utility that agent $i$ obtains from the entire choice set across all groups.

Agent $i$ selects the alternative that yields the highest utility across all groups:
\begin{equation}
    \max_g\max_{j \in B_g} U_{ijg} = \max_g\max_{j \in B_g} \left[\delta_j + \zeta_{ig} + (1-\eta)\epsilon_{ij}\right].
\end{equation}
The overall inclusive value is the expected value of this maximum over the joint distribution of $\zeta_{ig}$ and $\epsilon_{ij}$:
\begin{equation}
    IV = E_{\zeta,\epsilon}\left[\max_g\max_{j \in B_g} \left(\delta_j + \zeta_{ig}+ (1-\eta)\epsilon_{ij}\right)\right].
\end{equation}

\paragraph{Derivation}
The derivation proceeds by leveraging the nested structure of the model. Since $\zeta_{ig}$ does not depend on the alternative $j$ within group $g$, I can factor the within-group maximization:
\begin{equation}
    IV = E_{\zeta,\epsilon}\left[\max_g \left(\zeta_{ig} +\max_{j \in B_g} \left(\delta_j + (1-\eta)\epsilon_{ij}\right)\right)\right].
\end{equation}

The key insight is to recognize the distributional properties of the within-group maximum. From the derivation in Step 1, the term $\max_{j \in B_g} \left(\delta_j + (1-\eta)\epsilon_{ij}\right)$ follows a Type-I extreme value distribution with location parameter $IV_g = (1-\eta) \ln\left(\sum_{j \in B_g} \exp\left(\frac{\delta_j}{1-\eta}\right)\right) + (1-\eta)\gamma$ and scale parameter $(1-\eta)$. Therefore, I can represent this maximum in terms of its location and a new Type-I extreme value random variable $\tilde{\epsilon}_{ig}$ with zero location and unit scale:
\begin{equation}
    \max_{j \in B_g} \left(\delta_j + (1-\eta)\epsilon_{ij}\right) \stackrel{d}{=} IV_g + (1-\eta)\tilde{\epsilon}_{ig},
\end{equation}
where $\stackrel{d}{=}$ denotes equality in distribution, and $\tilde{\epsilon}_{ig}$ represents the stochastic component of the within-group maximum utility.

Substituting this distributional representation into the overall inclusive value expression:
\begin{equation}
\begin{aligned}
    IV &=  E_{\zeta,\tilde{\epsilon}}\left[\max_g \left(IV_g + \zeta_{ig} + (1-\eta)\tilde{\epsilon}_{ig}\right)\right].
\end{aligned}
\end{equation}

Decomposing $IV_g$ into its normalized component plus the constant term:
\begin{equation}
    IV = E_{\zeta,\tilde{\epsilon}}\left[\max_g \left(\tilde{IV}_g + (1-\eta)\gamma + \zeta_{ig} + (1-\eta)\tilde{\epsilon}_{ig}\right)\right],
\end{equation}
where $\tilde{IV}_g = (1-\eta) \ln\left(\sum_{j \in B_g} \exp\left(\frac{\delta_j}{1-\eta}\right)\right)$. Since $(1-\eta)\gamma$ is constant across groups, it factors out of the maximization:
\begin{equation}
    IV = E_{\zeta,\tilde{\epsilon}}\left[\max_g \left(\tilde{IV}_g + \zeta_{ig} + (1-\eta)\tilde{\epsilon}_{ig}\right)\right] + (1-\eta)\gamma.
\end{equation}

The random variable $\zeta_{ig}$ follows a specific distribution such that the composite random variable $\zeta_{ig} + (1-\eta)\tilde{\epsilon}_{ig}$ follows a Type-I extreme value distribution with zero location and unit scale \citep{cardell1997variance}. This distributional property is the foundation of the model's tractability. Applying the standard result for the expected maximum of Type-I extreme value random variables (as derived in Step 1), where the systematic utilities are $\tilde{IV}_g$ and the errors have unit scale:
\begin{equation}
    E_{\zeta,\tilde{\epsilon}}\left[\max_g \left(\tilde{IV}_g + \zeta_{ig} + (1-\eta)\tilde{\epsilon}_{ig}\right)\right] = \ln\left(\sum_{g} \exp(\tilde{IV}_g)\right) + \gamma.
\end{equation}

Combining terms yields the overall inclusive value:
\begin{equation}
    IV = \ln\left(\sum_{g} \exp(\tilde{IV}_g)\right) + \gamma + (1-\eta)\gamma.
\end{equation}

Substituting the expression for $\tilde{IV}_g$ and simplifying:
\begin{equation}
\begin{aligned}
    IV &= \ln\left(\sum_{g} \exp\left((1-\eta) \ln\left(\sum_{j \in B_g} \exp\left(\frac{\delta_j}{1-\eta}\right)\right)\right)\right) + \gamma + (1-\eta)\gamma\\
    &= \ln\left(\sum_{g}\left(\sum_{j \in B_g} \exp\left(\frac{\delta_j}{1-\eta}\right)\right)^{(1-\eta)}\right) + \gamma + (1-\eta)\gamma.
\end{aligned}
\end{equation}

As with the group-level inclusive value, the constants involving the Euler-Mascheroni constant do not affect choice probabilities or welfare comparisons. Adopting the normalization that omits these constants:
\begin{equation}
    \tilde{IV} = \ln\left(\sum_{g}\left(\sum_{j \in B_g} \exp\left(\frac{\delta_j}{1-\eta}\right)\right)^{(1-\eta)}\right).
    \label{eq:overall_iv}
\end{equation}

This expression reveals the hierarchical structure of the nested logit model: the overall inclusive value aggregates group-level inclusive values through a log-sum formula, with the nesting parameter $(1-\eta)$ determining the weight placed on within-group variety relative to across-group differences.

\paragraph{Properties of the Overall Inclusive Value}
The overall inclusive value provides a complete welfare metric for the entire choice environment, capturing the value agents derive from both within-group variety and across-group differences. This aggregation inherits the monotonicity properties of the group-level measure while respecting the hierarchical structure: $IV$ is strictly increasing in each alternative's utility $\delta_j$, in each group-level inclusive value $IV_g$, and in the number of alternatives or groups available to the agent.

The nesting parameter $\eta$ governs how within-group and across-group diversity contribute to overall welfare, with important limiting cases clarifying the model's structure. As $\eta \to 0$, within-group correlation vanishes and the nested logit collapses to a standard logit, yielding $IV = \ln(\sum_g \sum_{j \in B_g} \exp(\delta_j)) + \gamma$. In this limit, the nested structure becomes irrelevant and all alternatives compete equally regardless of group membership. Conversely, as $\eta \to 1$, alternatives within groups become nearly perfect substitutes, and choices depend primarily on group-level attributes rather than individual alternative characteristics. The exponent $(1-\eta)$ in equation \eqref{eq:overall_iv} reflects this trade-off: smaller values amplify the importance of within-group variety, while larger values emphasize across-group differentiation.

For policy analysis and welfare evaluation, differences in the overall inclusive value provide the theoretically appropriate metric: $\Delta IV = IV' - IV$ measures the utility gain from any modification to the choice environment, whether through improving existing alternatives, adding new alternatives, or introducing new groups. 
\clearpage
\section{Derivation of $\frac{\partial(U^U_{jt}-U^U_{Ojt})}{\partial D_{jt}}$}\label{a:proof_deriv}

I derive the marginal effect of permanent worker employment on the union's surplus through sequential application of the chain rule to the nested logit structure.

\paragraph{Step 1: Express the union surplus}

I define the union's utility and disagreement utility as:
\begin{equation}
U^U_{jt} = \frac{1}{\gamma_{Dt}}D_{t}IV_{Dt}
\end{equation}
\begin{equation}
U^U_{Ojt} = \frac{1}{\gamma_{Dt}}D_{t}IV_{Dt \backslash j}
\end{equation}
where $IV_{Dt}$ denotes the inclusive value when plant $j$ operates and $IV_{Dt \backslash j}$ denotes the inclusive value when plant $j$ shuts down. The union's surplus becomes:
\begin{equation}
U^U_{jt}-U^U_{Ojt} = \frac{1}{\gamma_{Dt}}D_{t}(IV_{Dt} - IV_{Dt \backslash j})
\end{equation}

\paragraph{Step 2: Apply the product rule}
Differentiating with respect to $D_{jt}$:
\begin{equation}
\frac{\partial(U^U_{jt}-U^U_{Ojt})}{\partial D_{jt}} = \frac{1}{\gamma_{Dt}}D_{t}\frac{\partial(IV_{Dt} - IV_{Dt \backslash j})}{\partial D_{jt}}
\end{equation}
I assume that changes in plant-specific permanent employment $D_{jt}$ do not affect aggregate market employment, which remains fixed at ${D}_t$. This reflects workers' mobility across plants within the industry. 

Since $IV_{Dt \backslash j}$ excludes plant $j$ from the choice set, changes in plant $j$'s employment $D_{jt}$ do not affect this counterfactual inclusive value. Therefore $\frac{\partial IV_{Dt \backslash j}}{\partial D_{jt}} = 0$, and:
\begin{equation}
\frac{\partial(U^U_{jt}-U^U_{Ojt})}{\partial D_{jt}} = \frac{1}{\gamma_{Dt}}D_{t}\frac{\partial IV_{Dt}}{\partial D_{jt}}
\end{equation}

\paragraph{Step 3: Compute the inclusive value derivative}

The inclusive value is defined as:
\begin{equation}
IV_{Dt} = \log(G_t)
\end{equation}
where
\begin{equation}
G_t = \sum_{g}\left(\sum_{j \in B_g} \exp\left(\frac{\delta_{Djt}}{1-\eta_D}\right)\right)^{(1-\eta_D)}
\end{equation}

Applying the chain rule:
\begin{equation}
\frac{\partial IV_{Dt}}{\partial D_{jt}} = \frac{1}{G_t} \cdot \frac{\partial G_t}{\partial D_{jt}}
\end{equation}

\paragraph{Step 4: Derive $\frac{\partial G_t}{\partial D_{jt}}$}

Let $r$ denote the nest containing establishment $j$. I define:
\begin{equation}
I_r = \sum_{j \in B_r} \exp\left(\frac{\delta_{Djt}}{1-\eta_D}\right)
\end{equation}

Then $G_t = \sum_{g} I_g^{(1-\eta_D)}$. Since only nest $r$ contains establishment $j$, only this nest contributes to the derivative:
\begin{equation}
\frac{\partial G_t}{\partial D_{jt}} = \frac{\partial I_r^{(1-\eta_D)}}{\partial D_{jt}}
\end{equation}

Using the power rule:
\begin{equation}
\frac{\partial I_r^{(1-\eta_D)}}{\partial D_{jt}} = (1-\eta_D) I_r^{-\eta_D} \cdot \frac{\partial I_r}{\partial D_{jt}}
\end{equation}

Only establishment $j$ within nest $r$ depends on $D_{jt}$:
\begin{equation}
\frac{\partial I_r}{\partial D_{jt}} = \exp\left(\frac{\delta_{Djt}}{1-\eta_D}\right) \cdot \frac{1}{1-\eta_D} \cdot \frac{\partial \delta_{Djt}}{\partial D_{jt}}
\end{equation}

Since $\delta_{Djt} = \alpha_{Dt}t_{\text{trend}} +\gamma_{Dt} W_{jt,D}(D_{jt})+\xi_{Djt}$:
\begin{equation}
\frac{\partial \delta_{Djt}}{\partial D_{jt}} = \gamma_{Dt} \cdot \frac{\partial W_{jt,D}}{\partial D_{jt}}
\end{equation}

Combining these results:
\begin{equation}
\frac{\partial G_t}{\partial D_{jt}} = (1-\eta_D) I_r^{-\eta_D} \cdot \exp\left(\frac{\delta_{Djt}}{1-\eta_D}\right) \cdot \frac{1}{1-\eta_D} \cdot \gamma_{Dt} \cdot \frac{\partial W_{jt,D}}{\partial D_{jt}}
\end{equation}

Simplifying:
\begin{equation}
\frac{\partial G_t}{\partial D_{jt}} = I_r^{-\eta_D} \cdot \exp\left(\frac{\delta_{Djt}}{1-\eta_D}\right) \cdot \gamma_{Dt} \cdot \frac{\partial W_{jt,D}}{\partial D_{jt}}
\end{equation}

\paragraph{Step 5: Express in terms of market shares}

In the nested logit framework with $IV_{Dt} = \log(G_t)$, the market share of permanent workers at establishment $j$ decomposes as:
\begin{equation}
s_{Djt} = s_{Dj|rt} \cdot s_{Dr|t} = \frac{\exp\left(\frac{\delta_{Djt}}{1-\eta_D}\right)}{I_r} \cdot \frac{I_r^{(1-\eta_D)}}{G_t} = \frac{\exp\left(\frac{\delta_{Djt}}{1-\eta_D}\right) \cdot I_r^{-\eta_D}}{G_t}
\end{equation}

Substituting into the expression from Step 3:
\begin{equation}
\frac{\partial IV_{Dt}}{\partial D_{jt}} = \frac{1}{G_t} \cdot I_r^{-\eta_D} \cdot \exp\left(\frac{\delta_{Djt}}{1-\eta_D}\right) \cdot \gamma_{Dt} \cdot \frac{\partial W_{jt,D}}{\partial D_{jt}}
\end{equation}

Therefore:
\begin{equation}
\frac{\partial IV_{Dt}}{\partial D_{jt}} = s_{Djt} \cdot \gamma_{Dt} \cdot \frac{\partial W_{jt,D}}{\partial D_{jt}}
\end{equation}

\paragraph{Step 6: Combine results}
Substituting into the expression from Step 2:
\begin{equation}
\frac{\partial(U^U_{jt}-U^U_{Ojt})}{\partial D_{jt}} = D_{t} \cdot s_{Djt} \cdot \frac{\partial W_{jt,D}}{\partial D_{jt}}
\end{equation}

\clearpage
\section{Normalization via Geometric Means}\label{a:geo_means}

I employ the CES normalization approach developed in \cite{grieco2016production} and \cite{hong2024search}. This normalization permits interpretation of the factor share parameters as marginal returns to inputs for a plant whose inputs, productivity, and input prices are evaluated at the baseline point $\bar Z$, where each variable equals its geometric mean: \( \bar{X} = \left( \prod_{n=1}^N X_n \right)^{\frac{1}{N}} \). The normalization expresses the factor share parameters as functions of observables and common structural parameters while eliminating scale dependencies and unit-of-measurement effects from the estimation procedure. The zero-mean property of the measurement error \( \varepsilon_{jt} \) guarantees equality between the geometric means of observed and planned output: \( \bar{Q} = \bar{\tilde{Q}} \).

The choice of baseline point defines a family of CES functions characterized by the following system of equations from the profit maximization problem:
\begin{align}
& \bar{Q} = \left( \tilde{\alpha}_K \bar{K}^{\sigma^O} + \tilde{\alpha}_M \bar{M}^{\sigma^O} + \tilde{\alpha}_L \left(\exp(\bar{\omega}^{L})\bar{L}\right)^{\sigma^O} \right)^\frac{1}{\sigma^O}\exp(\bar{\omega}^{H}) \\
& \bar{L} = \left(\tilde{\alpha}_S \bar{S}^{\sigma^M} + \tilde{\alpha}_E\bar{E}^{\sigma^M}\right)^\frac{1}{\sigma^M}\\
& \bar{E} = \left(\tilde{\alpha}_C \bar{C}^{\sigma^I} + \tilde{\alpha}_D \bar{D}^{\sigma^I}\right)^\frac{1}{\sigma^I}\\
& \tilde{\alpha}_K + \tilde{\alpha}_M + \tilde{\alpha}_L = 1 \label{eq:basealpha1}\\
& \tilde{\alpha}_S + \tilde{\alpha}_E = 1 \label{eq:basealpha2}\\
& \tilde{\alpha}_C + \tilde{\alpha}_D = 1 \label{eq:basealpha3}\\
& \left(\frac{MP_C}{MP_D}\right)_{\bar Z} = \frac{\tilde{\alpha}_C\bar{C}^{\sigma^I}\bar{D}}{\tilde{\alpha}_D\bar{D}^{\sigma^I}\bar{C}}=\bar \mu_{CD} \label{eq:baseCD}\\
& \left(\frac{MP_S}{MP_E}\right)_{\bar Z} = \frac{\tilde{\alpha}_S\bar{S}^{\sigma^M}\bar{E}}{\tilde{\alpha}_E\bar{E}^{\sigma^M}\bar{S}}=\bar \mu_{SE} \label{eq:baseHB}\\
& \left(\frac{MP_M}{MP_L}\right)_{\bar Z} = \frac{\tilde{\alpha}_M\bar M^{\sigma^O}\bar L}{\tilde{\alpha}_L (\exp{(\bar\omega^L)}\bar L)^{\sigma^O}\bar M} = \bar \mu_{ML} \label{eq:baseML}\\
& \left(\frac{MP_M}{MP_K}\right)_{\bar Z} = \frac{\tilde{\alpha}_M\bar M^{\sigma^O}\bar K}{\tilde{\alpha}_K\bar K^{\sigma^O}\bar M} =\left(\frac{\bar{E}_K}{\tau\bar{E}_M}\right) \bar \mu_{MK}=\frac{\bar{K}}{\tau\bar{M}}\label{eq:baseMK}
\end{align}
Here, $\bar\omega^H$ and $\bar\omega^L$ denote the productivity levels for a representative plant producing $\bar Q$ with mean inputs $\bar K$, $\bar M$, and $\bar L$. Evaluated at the baseline point, $MP_C$, $MP_D$, $MP_S$, $MP_E$, $MP_M$, $MP_L$, and $MP_K$ denote the marginal products of contract workers, permanent workers, supervisors, the workers' aggregate nest, materials, the aggregate labor input, and capital, respectively, while $\bar \mu_{CD}$, $\bar \mu_{SE}$, $\bar \mu_{ML}$, and $\bar \mu_{MK}$ denote the optimal marginal rates of technical substitution between contract and permanent workers, supervisors and the workers' aggregate nest, materials and aggregate labor, and materials and capital. Here $\bar{E}_K$ and $\bar{E}_M$ denote the baseline expenditures on capital and materials, respectively.

Imperfect competition in the output market does not affect this baseline system, as price markups leave the marginal rates of technical substitution between inputs unchanged. In contrast, imperfect competition in input markets and non-wage marginal costs both influence these marginal rates. I therefore specify these rates to reflect market imperfections and non-wage employment costs.

For contract versus permanent workers, I define the marginal rate of technical substitution as $\bar{\mu}_{CD}=\frac{\phi_C\bar{W}_C}{\phi_D\bar{W}_D}$, where the parameters $\phi_C$ and $\phi_D$ capture mean wedges between marginal revenue products and wages arising from non-wage costs and imperfect competition in the respective labor markets. For supervisors versus the workers' aggregate, I specify $\bar{\mu}_{SE}=\frac{\bar{W}_S}{\bar{W}_E}$, where $\bar{W}_S$ denotes the baseline supervisor wage and $\bar{W}_E$ denotes the marginal cost index for the workers' aggregate, given by the exact CES input price index evaluated at the baseline:
\begin{equation}
\bar{W}_E=\left(\tilde{\alpha}_C^{\frac{1}{1-\sigma^I}} (\phi_C \bar{W}_C)^{\frac{\sigma^I}{\sigma^I-1}} + \tilde{\alpha}_D^{\frac{1}{1-\sigma^I}} (\phi_D \bar{W}_D)^{\frac{\sigma^I}{\sigma^I-1}}\right)^{\frac{\sigma^I-1}{\sigma^I}}
\end{equation}

For materials versus aggregate labor, I set $\bar{\mu}_{ML}=\frac{\bar{P}_M}{\bar{W}_L}$, where $\bar{P}_M$ denotes the geometric mean materials price and $\bar{W}_L$ denotes the baseline wage index for aggregate labor, given by the exact CES price index evaluated at the baseline:
\begin{equation}
\bar{W}_L=\left(\tilde{\alpha}_S^{\frac{1}{1-\sigma^M}} \bar{W}_S^{\frac{\sigma^M}{\sigma^M-1}} + \tilde{\alpha}_E^{\frac{1}{1-\sigma^M}}\bar{W}_E^{\frac{\sigma^M}{\sigma^M-1}}\right)^{\frac{\sigma^M-1}{\sigma^M}}
\end{equation}
For capital versus materials, the parameter $\tau$ operates through the expenditure ratio to capture capital's deviation from static optimality, reflecting capital's quasi-fixed nature within the period (\cite{grieco2016production}, Online Appendix 4).

Equations (\ref{eq:basealpha1})-(\ref{eq:baseMK}) constitute a system of seven equations in seven unknown factor share parameters, which solve as
\begin{align}
& \tilde{\alpha}_C = \frac{\frac{\phi_C\bar W_C\bar C}{\bar C^{\sigma^I}}}{\frac{\phi_C\bar W_C\bar C}{\bar C^{\sigma^I}} + \frac{\phi_D\bar W_D\bar D}{\bar D^{\sigma^I}}}\\
& \tilde{\alpha}_D = 1 - \tilde{\alpha}_C \\
& \tilde{\alpha}_S = \frac{\frac{\bar W_S\bar S}{\bar S^{\sigma^M}}}{\frac{\bar W_S\bar S}{\bar S^{\sigma^M}} + \frac{\bar W_E\bar E}{\bar E^{\sigma^M}}} = \frac{\frac{\bar W_S\bar S}{\bar S^{\sigma^M}}}{\frac{\bar W_S\bar S}{\bar S^{\sigma^M}} + \frac{\phi_C\bar W_C\bar C+\phi_D\bar W_D\bar D}{\bar E^{\sigma^M}}}\label{eq:alpha_S}\\
& \tilde{\alpha}_E = 1 - \tilde{\alpha}_S\\
& \tilde{\alpha}_M = \frac{\frac{\bar P_M\bar M}{\bar M^{\sigma^O}}}{\frac{\bar P_M\bar M}{\bar M^{\sigma^O}} + \frac{\bar W_L\bar L}{(\exp(\bar \omega_L)\bar L)^{\sigma^O}} + \frac{\tau\bar P_M\bar M}{\bar K^{\sigma^O}}}=\frac{\frac{\bar P_M\bar M}{\bar M^{\sigma^O}}}{\frac{\bar P_M\bar M}{\bar M^{\sigma^O}} + \frac{\bar W_S\bar S+\phi_C\bar W_C\bar C+\phi_D\bar W_D\bar D}{(\exp(\bar \omega_L)\bar L)^{\sigma^O}} + \frac{\tau\bar P_M\bar M}{\bar K^{\sigma^O}}} \label{eq:alpha_M}\\
& \tilde{\alpha}_K = \frac{\frac{\tau\bar P_M\bar M}{\bar K^{\sigma^O}}}{\frac{\bar P_M\bar M}{\bar M^{\sigma^O}} + \frac{\bar W_L\bar L}{(\exp(\bar \omega_L)\bar L)^{\sigma^O}} + \frac{\tau\bar P_M\bar M}{\bar K^{\sigma^O}}}=\frac{\frac{\tau\bar P_M\bar M}{\bar K^{\sigma^O}}}{\frac{\bar P_M\bar M}{\bar M^{\sigma^O}} + \frac{\bar W_S\bar S+\phi_C\bar W_C\bar C+\phi_D\bar W_D\bar D}{(\exp(\bar \omega_L)\bar L)^{\sigma^O}} + \frac{\tau\bar P_M\bar M}{\bar K^{\sigma^O}}} \label{eq:alpha_K}\\
& \tilde{\alpha}_L = 1 - \tilde{\alpha}_M - \tilde{\alpha}_K
\end{align}
where the second equality in \eqref{eq:alpha_S} derives from applying Shephard's lemma to the corresponding input nest cost minimization problem, and the second equalities in \eqref{eq:alpha_M} and \eqref{eq:alpha_K} derive from applying Shephard's lemma twice—once at the workers' aggregate level and once at the aggregate labor level. This reparameterization yields the normalized production function:
\begin{equation}
Q_{jt} = \bar Q\left( \alpha_K\left(\frac{K_{jt}}{\bar K}\right)^{\sigma^O} + \alpha_M\left( \frac{M_{jt}}{\bar M}\right)^{\sigma^O} +  \alpha_L\left(\exp(\ddot\omega_{jt}^{L}) \frac{L_{jt}}{\bar L}\right)^{\sigma^O} \right)^\frac{1}{\sigma^O} \label{eq:qnorm1}
\end{equation}
where
\begin{equation}
    L_{jt} = \bar L\left(\alpha_S\left( \frac{S_{jt}}{\bar S}\right)^{\sigma^M} + \alpha_E \left( \frac{E_{jt}}{\bar E}\right)^{\sigma^M}\right)^\frac{1}{\sigma^M} \label{eq:lnorm1}
\end{equation}
\begin{equation}
    E_{jt} = \bar E\left(\alpha_C\left( \frac{C_{jt}}{\bar C}\right)^{\sigma^I} + \alpha_D \left( \frac{D_{jt}}{\bar D}\right)^{\sigma^I}\right)^\frac{1}{\sigma^I} \label{eq:lnorm2}
\end{equation}
\begin{equation}
   \ddot\omega^i_{jt} = \omega^i_{jt}-\bar\omega^i\quad\forall i \in\{H,L\}
\end{equation}
and 
\begin{align}
& \alpha_C = \frac{\phi_C\bar W_C \bar C}{\phi_C\bar W_C \bar C + \phi_D\bar W_D \bar D}\\
& \alpha_D = 1 - \alpha_C \\
& \alpha_S = \frac{\bar W_S \bar S}{\bar W_S \bar S + \phi_C\bar W_C \bar C + \phi_D\bar W_D \bar D}\\
& \alpha_E = 1 - \alpha_S \\
& \alpha_M = \frac{\bar P_M \bar M}{\bar P_M \bar M + \bar W_S \bar S + \phi_C\bar W_C \bar C + \phi_D\bar W_D \bar D + \tau\bar P_M \bar M} \label{alphaM}\\
& \alpha_K = \frac{\tau\bar P_M \bar M}{\bar P_M \bar M +\bar W_S \bar S + \phi_C\bar W_C \bar C + \phi_D\bar W_D \bar D + \tau\bar P_M \bar M} \label{alphaK}\\
& \alpha_L = 1-\alpha_K-\alpha_M \label{alphaL}\\
\end{align}
This normalization ensures each $\alpha_X$ measures the marginal returns of input $X$ evaluated at geometric mean levels of inputs, productivity, and prices. The normalized productivity terms $\ddot{\omega}^H_{jt}$ and $\ddot{\omega}^L_{jt}$ preserve their original demeaned AR(1) dynamics, maintaining the temporal structure of productivity evolution.

\subsection{Normalized Input Allocation Problem for Contract Workers, Supervisors, and Materials}
{\small
\begin{equation}
\begin{aligned}
\label{eq:Input_problem_norm}
\max_{\{C_{jt},S_{jt},M_{jt}\}}&\quad\Pi_{jt} \\
\text{s.t.}\quad\Pi_{jt}&=P_{jt}(Q_{jt})Q_{jt}-W_{jt,C}(C_{jt})C_{jt} - W^*_{jt,D}D^*_{jt}-W_{jt,S}S_{jt}-P_{jt,M}M_{jt}-\Phi_{jt}\\
Q_{jt} &= \bar{Q}\left(\alpha_K(\tau, \phi_C, \phi_D) \ddot{K}_{jt}^{\sigma^O} + \alpha_M(\tau, \phi_C, \phi_D) \ddot{M}_{jt}^{\sigma^O} + \alpha_L(\tau, \phi_C, \phi_D) (\exp(\ddot{\omega}_{jt}^{L}) \ddot{L}_{jt})^{\sigma^O} \right)^\frac{1}{\sigma^O}\exp(\ddot{\omega}_{jt}^{H})\\
\ddot{L}_{jt} &= \left( \alpha_S(\phi_C, \phi_D) \ddot{S}_{jt}^{\sigma^M}+ \alpha_E(\phi_C, \phi_D) \ddot{E}_{jt}^{\sigma^M} \right)^\frac{1}{\sigma^M}\\
\ddot{E}_{jt} &= \left( \alpha_C(\phi_C, \phi_D) \ddot{C}_{jt}^{\sigma^I} + \alpha_D(\phi_C, \phi_D) \ddot{D}_{jt}^{\sigma^I} \right)^\frac{1}{\sigma^I}\\
\ddot{\omega}_{jt}^{i} &= \ddot{\iota}_{i,t} + \rho_i \ddot{\omega}_{jt-1}^{i} \xi_{jt}^{i}\quad\forall~i\in\{H,L\}\\
U_{iCjt}&=\alpha_{C}t_{\text{trend}} +\gamma_{Ct} W_{jt,C}+\xi_{Cjt}+\zeta_{iCrt}+(1-\eta_C)\epsilon_{iCjt}
\end{aligned}
\end{equation}}

\clearpage
\section{First Order Conditions}\label{a:FOCs}
\begin{align}
&C:\quad\left(\frac{\partial W_{jt,C}}{\partial C_{jt}}\frac{C_{jt}}{W_{jt,C}}+1+\frac{\partial \Phi_{jt}}{\partial C_{jt}}\frac{1}{W_{jt,C}}\right)\frac{W_{jt,C}}{P_{jt}}\nonumber\\
&=\left(\frac{\partial P_{jt}}{\partial Q_{jt}}\frac{Q_{jt}}{P_{jt}}+1\right)\frac{Q_{jt}}{C_{jt}}\frac{\alpha_L(\exp(\ddot{\omega}_{jt}^{L}) \ddot{L}_{jt})^{\sigma^O}}{\left(\alpha_K \ddot{K}_{jt}^{\sigma^O} + \alpha_M \ddot{M}_{jt}^{\sigma^O} + \alpha_L (\exp(\ddot{\omega}_{jt}^{L}) \ddot{L}_{jt})^{\sigma^O} \right)}\frac{\alpha_E\ddot{E}_{jt}^{\sigma^M}}{\ddot{L}_{jt}^{\sigma^M}}\frac{\alpha_C\ddot{C}_{jt}^{\sigma^I}}{\ddot{E}_{jt}^{\sigma^I}}
\\[1.5em]
&S:\quad  \frac{W_{jt,S}}{P_{jt}}=\left(\frac{\partial P_{jt}}{\partial Q_{jt}}\frac{Q_{jt}}{P_{jt}}+1\right)\frac{Q_{jt}}{S_{jt}}\frac{\alpha_L(\exp(\ddot{\omega}_{jt}^{L}) \ddot{L}_{jt})^{\sigma^O}}{\left(\alpha_K \ddot{K}_{jt}^{\sigma^O} + \alpha_M \ddot{M}_{jt}^{\sigma^O} + \alpha_L(\exp(\ddot{\omega}_{jt}^{L}) \ddot{L}_{jt})^{\sigma^O} \right)}\frac{\alpha_S\ddot{S}_{jt}^{\sigma^M}}{\ddot{L}_{jt}^{\sigma^M}}
\\[1.5em]
&M:\quad \frac{P_{jt,M}}{P_{jt}}=\left(\frac{\partial P_{jt}}{\partial Q_{jt}}\frac{Q_{jt}}{P_{jt}}+1\right)\frac{Q_{jt}}{M_{jt}}\frac{\alpha_M \ddot{M}_{jt}^{\sigma^O}}{\left(\alpha_K \ddot{K}_{jt}^{\sigma^O} + \alpha_M \ddot{M}_{jt}^{\sigma^O} + \alpha_L (\exp(\ddot{\omega}_{jt}^{L}) \ddot{L}_{jt})^{\sigma^O} \right)}
\end{align}

\clearpage
\section{Step 2 - Identification Proof}\label{a:step1_proof}
I define the instrument vector as
\begin{equation}
   \mathbf{Z}_L=[\mathbf{I}\{year=t\}, \mathbf{Z}_1, \mathbf{Z}_2, \mathbf{Z}_3, \mathbf{Z}_4, \mathbf{Z}_5, \ddot{\omega}^L_{jt-1}(\sigma^O,\sigma^M,\sigma^I, \phi_C, \phi_D)]',
\end{equation}
where $\mathbf{Z}_1$, $\mathbf{Z}_2$, $\mathbf{Z}_3$, $\mathbf{Z}_4$, and $\mathbf{Z}_5$ represent distinct, exogenous, and relevant instrument sets. I define the parameter vector as
\begin{equation}
   \theta_L=\{\ddot{\iota}_{L,t},\sigma^O,\sigma^M,\sigma^I,\rho_L,\phi_C, \phi_D\}.
\end{equation}
Since instruments are parameter-invariant, the Jacobian of the population moment mapping satisfies
\begin{equation}
   J_L=\mathbb{E}\left[\frac{\partial \mathbf{Z_L}\otimes \xi^L_{jt}(\theta_L)}{\partial\theta_L}\right]=\mathbb{E}\left[\mathbf{Z_L}\otimes\frac{\partial  \xi^L_{jt}(\theta_L)}{\partial\theta_L}\right].
\end{equation}

\noindent\textbf{Proposition.} The Jacobian matrix $J_L$ achieves full column rank:
\begin{equation}
   \text{rank}(J_L)=\text{dim}(\theta_L)=\text{dim}(\ddot{\iota}_{L,t})+6.
\end{equation}

\noindent\textbf{Proof.} I establish identification through systematic evaluation of the Jacobian components. I define the transformations $\delta = 1/\sigma^O$, $\gamma = 1/\sigma^M$, and $\kappa = 1/\sigma^I$ to simplify notation. I denote $\Omega_{jt} = \alpha_S \ddot{S}_{jt}^{1/\gamma} + \alpha_E \Phi_{jt}^{\kappa/\gamma}$, where $\Phi_{jt} = \alpha_C\ddot{C}_{jt}^{1/\kappa} + \alpha_D\ddot{D}_{jt}^{1/\kappa}$.

\textbf{Time fixed effects ($\ddot{\iota}_{L,t}$).}
For each $t'$,
\begin{equation}
\mathbb{E}\!\left[
\frac{\partial}{\partial \ddot{\iota}_{L,t'}}\bigl(\mathbf{I}\{year=t\}\otimes \xi^L_{jt}\bigr)
\right]
=
\mathbb{E}\!\left[
-\mathbf{I}\{year=t\}\,\mathbf{I}\{t=t'\}
\right],
\end{equation}
which yields an identity submatrix of rank $\operatorname{dim}(\ddot{\iota}_{L,t})$.

\textbf{Outer nest elasticity ($\sigma^O$):} Labor-augmenting productivity depends on $\delta = 1/\sigma^O$ through multiple channels. The partial derivative yields
\begin{equation}
\frac{\partial \ddot{\omega}^L_{jt}}{\partial \delta} = -\log\alpha_S + \log\left(\frac{W_{jt,S}S_{jt}}{P_{jt,M}M_{jt}}\right) + \log\left(\frac{\bar{P}_M\bar{M}}{{\bar W_S \bar S + \phi_C\bar W_C \bar C + \phi_D\bar W_D \bar D}}\right) + \log\Omega_{jt} - \frac{1}{\gamma}\log\ddot{S}_{jt}.
\end{equation}
The corresponding Jacobian component is
\begin{equation}
\mathbb{E}\left[\frac{\partial}{\partial\delta}\left[\mathbf{Z}_1\otimes\xi^L_{jt}\right]\right] = \mathbb{E}\left[\mathbf{Z}_1 \otimes \left[\frac{\partial\ddot{\omega}^L_{jt}}{\partial\delta} - \rho_L\frac{\partial\ddot{\omega}^L_{jt-1}}{\partial\delta}\right]\right].
\end{equation}

\textbf{Middle nest elasticity ($\sigma^M$):} The parameter $\gamma = 1/\sigma^M$ enters through the nested CES aggregator. Applying the chain rule yields
\begin{equation}
\frac{\partial \ddot{\omega}^L_{jt}}{\partial \gamma} = \frac{\partial}{\partial \gamma}\left[(\delta - \gamma)\log\Omega_{jt}\right] - \frac{\partial}{\partial \gamma}\left[\frac{\delta}{\gamma}\log\ddot{S}_{jt}\right].
\end{equation}
For the first term, I apply the product rule:
\begin{equation}
\frac{\partial}{\partial \gamma}\left[(\delta - \gamma)\log\Omega_{jt}\right] = -\log\Omega_{jt} + (\delta - \gamma)\frac{\partial\log\Omega_{jt}}{\partial\gamma}.
\end{equation}
I compute $\frac{\partial\log\Omega_{jt}}{\partial\gamma}$ as
\begin{equation}
\frac{\partial\log\Omega_{jt}}{\partial\gamma} = \frac{1}{\Omega_{jt}}\frac{\partial\Omega_{jt}}{\partial\gamma} = -\frac{1}{\gamma^2\Omega_{jt}}\left[\alpha_S \ddot{S}_{jt}^{1/\gamma}\log\ddot{S}_{jt} + \kappa\alpha_E \Phi_{jt}^{\kappa/\gamma}\log\Phi_{jt}\right].
\end{equation}
For the second term,
\begin{equation}
\frac{\partial}{\partial \gamma}\left[\frac{\delta}{\gamma}\log\ddot{S}_{jt}\right] = -\frac{\delta}{\gamma^2}\log\ddot{S}_{jt}.
\end{equation}
Therefore,
\begin{equation}
\frac{\partial \ddot{\omega}^L_{jt}}{\partial \gamma} = -\log\Omega_{jt} - \frac{(\delta - \gamma)}{\gamma^2\Omega_{jt}}\left[\alpha_S \ddot{S}_{jt}^{1/\gamma}\log\ddot{S}_{jt} + \kappa\alpha_E \Phi_{jt}^{\kappa/\gamma}\log\Phi_{jt}\right] + \frac{\delta}{\gamma^2}\log\ddot{S}_{jt}.
\end{equation}
The corresponding Jacobian component is
\begin{equation}
\mathbb{E}\left[\frac{\partial}{\partial\gamma}\left[\mathbf{Z}_2\otimes\xi^L_{jt}\right]\right] = \mathbb{E}\left[\mathbf{Z}_2 \otimes \left[\frac{\partial\ddot{\omega}^L_{jt}}{\partial\gamma} - \rho_L\frac{\partial\ddot{\omega}^L_{jt-1}}{\partial\gamma}\right]\right].
\end{equation}

\textbf{Inner nest elasticity ($\sigma^I$):} The parameter $\kappa = 1/\sigma^I$ affects productivity through the bottom-nest aggregator:
\begin{equation}
\frac{\partial \ddot{\omega}^L_{jt}}{\partial \kappa} = (\delta - \gamma)\frac{\partial\log\Omega_{jt}}{\partial\kappa}.
\end{equation}
I compute the derivative of $\log\Omega_{jt}$ as
\begin{equation}
\frac{\partial\log\Omega_{jt}}{\partial\kappa} = \frac{1}{\Omega_{jt}}\frac{\partial\Omega_{jt}}{\partial\kappa} = \frac{1}{\Omega_{jt}}\alpha_E\frac{\partial}{\partial\kappa}\left[\Phi_{jt}^{\kappa/\gamma}\right].
\end{equation}
Applying the chain rule to $\Phi_{jt}^{\kappa/\gamma}$, where $\Phi_{jt}$ also depends on $\kappa$, I obtain
\begin{equation}
\frac{\partial}{\partial\kappa}\left[\Phi_{jt}^{\kappa/\gamma}\right] = \Phi_{jt}^{\kappa/\gamma}\left[\frac{1}{\gamma}\log\Phi_{jt} + \frac{\kappa}{\gamma}\frac{1}{\Phi_{jt}}\frac{\partial\Phi_{jt}}{\partial\kappa}\right],
\end{equation}
where
\begin{equation}
\frac{\partial\Phi_{jt}}{\partial\kappa} = -\frac{1}{\kappa^2}\left[\alpha_C\ddot{C}_{jt}^{1/\kappa}\log\ddot{C}_{jt} + \alpha_D\ddot{D}_{jt}^{1/\kappa}\log\ddot{D}_{jt}\right].
\end{equation}
Therefore,
\begin{equation}
\frac{\partial\log\Omega_{jt}}{\partial\kappa} = \frac{\alpha_E\Phi_{jt}^{\kappa/\gamma}}{\Omega_{jt}}\left[\frac{1}{\gamma}\log\Phi_{jt} - \frac{1}{\gamma\kappa\Phi_{jt}}\left(\alpha_C\ddot{C}_{jt}^{1/\kappa}\log\ddot{C}_{jt} + \alpha_D\ddot{D}_{jt}^{1/\kappa}\log\ddot{D}_{jt}\right)\right].
\end{equation}
The corresponding Jacobian component is
\begin{equation}
\mathbb{E}\left[\frac{\partial}{\partial\kappa}\left[\mathbf{Z}_3\otimes\xi^L_{jt}\right]\right] = \mathbb{E}\left[\mathbf{Z}_3 \otimes \left[\frac{\partial\ddot{\omega}^L_{jt}}{\partial\kappa} - \rho_L\frac{\partial\ddot{\omega}^L_{jt-1}}{\partial\kappa}\right]\right].
\end{equation}

\textbf{Share parameters ($\phi_C$ and $\phi_D$):} The parameters $\phi_C$ and $\phi_D$ enter labor-augmenting productivity through two distinct channels: the share functions $\alpha_C(\phi_C,\phi_D)$, $\alpha_D(\phi_C,\phi_D)$, $\alpha_E(\phi_C,\phi_D)$, and $\alpha_S(\phi_C,\phi_D)$, and the aggregate labor costs $\bar{W}_L\bar{L}=\bar W_S \bar S + \phi_C\bar W_C \bar C + \phi_D\bar W_D \bar D$.

For the contract-permanent worker shares within the innermost nest, I obtain
\begin{equation}
\frac{\partial \alpha_C}{\partial \phi_C} = \frac{\alpha_C\alpha_D}{\phi_C}, \quad 
\frac{\partial \alpha_C}{\partial \phi_D} = -\frac{\alpha_C\alpha_D}{\phi_D}, \quad
\frac{\partial \alpha_D}{\partial \phi_C} = -\frac{\alpha_C\alpha_D}{\phi_C}, \quad
\frac{\partial \alpha_D}{\partial \phi_D} = \frac{\alpha_C\alpha_D}{\phi_D}.
\end{equation}
For the supervisor-worker aggregate shares in the middle nest, I compute
\begin{equation}
\frac{\partial \alpha_S}{\partial \phi_C} = -\alpha_S \frac{\bar{W}_C\bar{C}}{\bar{W}_L\bar{L}}, \quad
\frac{\partial \alpha_S}{\partial \phi_D} = -\alpha_S \frac{\bar{W}_D\bar{D}}{\bar{W}_L\bar{L}},
\end{equation}
\begin{equation}
\frac{\partial \alpha_E}{\partial \phi_C} = \alpha_S \frac{\bar{W}_C\bar{C}}{\bar{W}_L\bar{L}}, \quad
\frac{\partial \alpha_E}{\partial \phi_D} = \alpha_S \frac{\bar{W}_D\bar{D}}{\bar{W}_L\bar{L}}.
\end{equation}
For the aggregate baseline payroll, the derivatives are
\begin{equation}
\frac{\partial (\bar{W}_L\bar{L})}{\partial \phi_C} = \bar W_C \bar C, \quad
\frac{\partial (\bar{W}_L\bar{L})}{\partial \phi_D} = \bar W_D \bar D.
\end{equation}

Applying the chain rule to equation \eqref{eq:charac_L}, the partial derivative of labor-augmenting productivity with respect to $\phi_C$ decomposes into three terms:
\begin{equation}
\frac{\partial \ddot{\omega}^L_{jt}}{\partial \phi_C} = -\delta\frac{1}{\alpha_S}\frac{\partial \alpha_S}{\partial \phi_C} - \delta\frac{1}{\bar{W}_L\bar{L}}\frac{\partial (\bar{W}_L\bar{L})}{\partial \phi_C} + (\delta - \gamma)\frac{\partial \log\Omega_{jt}}{\partial \phi_C}.
\end{equation}
Substituting the derivatives from Step 1 reveals an exact cancellation between the first two terms:
\begin{equation}
\begin{aligned}
\frac{\partial \ddot{\omega}^L_{jt}}{\partial \phi_C} &= -\delta\frac{1}{\alpha_S}\left(-\alpha_S \frac{\bar{W}_C\bar{C}}{\bar{W}_L\bar{L}}\right) - \delta\frac{1}{\bar{W}_L\bar{L}}\left(\bar W_C \bar C\right) + (\delta - \gamma)\frac{\partial \log\Omega_{jt}}{\partial \phi_C}\\
&= \delta\frac{\bar{W}_C\bar{C}}{\bar{W}_L\bar{L}} - \delta\frac{\bar W_C \bar C}{\bar{W}_L\bar{L}} + (\delta - \gamma)\frac{\partial \log\Omega_{jt}}{\partial \phi_C}\\
&= (\delta - \gamma)\frac{\partial \log\Omega_{jt}}{\partial \phi_C}.
\end{aligned}
\end{equation}
This cancellation demonstrates that $\phi_C$ affects productivity exclusively through the nested CES aggregator $\Omega_{jt}$, with magnitude proportional to the elasticity wedge $(\delta - \gamma)$. By symmetry, the partial derivative with respect to $\phi_D$ satisfies
\begin{equation}
\frac{\partial \ddot{\omega}^L_{jt}}{\partial \phi_D} = (\delta - \gamma)\frac{\partial \log\Omega_{jt}}{\partial \phi_D}.
\end{equation}

I now characterize $\frac{\partial \log\Omega_{jt}}{\partial \phi_C}$ by decomposing it into two economically distinct components. The derivative of the innermost aggregator with respect to $\phi_C$ is
\begin{equation}
\frac{\partial \Phi_{jt}}{\partial \phi_C} = \frac{\partial \alpha_C}{\partial \phi_C}\ddot{C}_{jt}^{1/\kappa} + \frac{\partial \alpha_D}{\partial \phi_C}\ddot{D}_{jt}^{1/\kappa} = \frac{\alpha_C\alpha_D}{\phi_C}\left[\ddot{C}_{jt}^{1/\kappa} - \ddot{D}_{jt}^{1/\kappa}\right].
\end{equation}
Applying the chain rule to $\log\Omega_{jt} = \log\left(\alpha_S \ddot{S}_{jt}^{1/\gamma} + \alpha_E \Phi_{jt}^{\kappa/\gamma}\right)$ yields
\begin{equation}
\frac{\partial \log\Omega_{jt}}{\partial \phi_C} = \frac{1}{\Omega_{jt}}\left[\frac{\partial \alpha_S}{\partial \phi_C}\ddot{S}_{jt}^{1/\gamma} + \frac{\partial \alpha_E}{\partial \phi_C}\Phi_{jt}^{\kappa/\gamma} + \alpha_E\frac{\kappa}{\gamma}\Phi_{jt}^{\kappa/\gamma - 1}\frac{\partial \Phi_{jt}}{\partial \phi_C}\right].
\end{equation}
Substituting the derivatives and collecting terms, I obtain
\begin{equation}
\begin{aligned}
\frac{\partial \log\Omega_{jt}}{\partial \phi_C} &= \frac{1}{\Omega_{jt}}\left[-\alpha_S \frac{\bar{W}_C\bar{C}}{\bar{W}_L\bar{L}}\ddot{S}_{jt}^{1/\gamma} + \alpha_S \frac{\bar{W}_C\bar{C}}{\bar{W}_L\bar{L}}\Phi_{jt}^{\kappa/\gamma} + \alpha_E\frac{\kappa}{\gamma}\Phi_{jt}^{\kappa/\gamma - 1}\frac{\alpha_C\alpha_D}{\phi_C}\left[\ddot{C}_{jt}^{1/\kappa} - \ddot{D}_{jt}^{1/\kappa}\right]\right]\\
&= \frac{\alpha_S}{\Omega_{jt}}\frac{\bar{W}_C\bar{C}}{\bar{W}_L\bar{L}}\left[\Phi_{jt}^{\kappa/\gamma} - \ddot{S}_{jt}^{1/\gamma}\right] + \frac{\alpha_E\alpha_C\alpha_D}{\Omega_{jt}}\frac{\kappa}{\gamma\phi_C}\Phi_{jt}^{\kappa/\gamma - 1}\left[\ddot{C}_{jt}^{1/\kappa} - \ddot{D}_{jt}^{1/\kappa}\right].
\end{aligned}
\end{equation}
By parallel reasoning, the derivative with respect to $\phi_D$ satisfies
\begin{equation}
\frac{\partial \log\Omega_{jt}}{\partial \phi_D} = \frac{\alpha_S}{\Omega_{jt}}\frac{\bar{W}_D\bar{D}}{\bar{W}_L\bar{L}}\left[\Phi_{jt}^{\kappa/\gamma} - \ddot{S}_{jt}^{1/\gamma}\right] + \frac{\alpha_E\alpha_C\alpha_D}{\Omega_{jt}}\frac{\kappa}{\gamma\phi_D}\Phi_{jt}^{\kappa/\gamma - 1}\left[\ddot{D}_{jt}^{1/\kappa} - \ddot{C}_{jt}^{1/\kappa}\right].
\end{equation}

The corresponding Jacobian components are
\begin{equation}
\mathbb{E}\left[\frac{\partial}{\partial\phi_C}\left[\mathbf{Z}_4\otimes\xi^L_{jt}\right]\right] = \mathbb{E}\left[\mathbf{Z}_4 \otimes \left[\frac{\partial\ddot{\omega}^L_{jt}}{\partial\phi_C} - \rho_L\frac{\partial\ddot{\omega}^L_{jt-1}}{\partial\phi_C}\right]\right],
\end{equation}
\begin{equation}
\mathbb{E}\left[\frac{\partial}{\partial\phi_D}\left[\mathbf{Z}_5\otimes\xi^L_{jt}\right]\right] = \mathbb{E}\left[\mathbf{Z}_5 \otimes \left[\frac{\partial\ddot{\omega}^L_{jt}}{\partial\phi_D} - \rho_L\frac{\partial\ddot{\omega}^L_{jt-1}}{\partial\phi_D}\right]\right].
\end{equation}

Each derivative $\frac{\partial \log\Omega_{jt}}{\partial \phi_C}$ and $\frac{\partial \log\Omega_{jt}}{\partial \phi_D}$ comprises two economically interpretable terms. The first term varies with the difference $\Phi_{jt}^{\kappa/\gamma} - \ddot{S}_{jt}^{1/\gamma}$, measuring the divergence between the workers' aggregate and supervisors, weighted by the respective baseline payroll shares $\frac{\bar{W}_C\bar{C}}{\bar{W}_L\bar{L}}$ and $\frac{\bar{W}_D\bar{D}}{\bar{W}_L\bar{L}}$. The second term captures within-nest divergence between contract and permanent workers. Critically, these second terms exhibit opposite signs depending on whether $\ddot{C}_{jt}^{1/\kappa} \gtrless \ddot{D}_{jt}^{1/\kappa}$, ensuring linear independence between $\frac{\partial \log\Omega_{jt}}{\partial \phi_C}$ and $\frac{\partial \log\Omega_{jt}}{\partial \phi_D}$. Furthermore, the distinct baseline payroll weights $\frac{\bar{W}_C\bar{C}}{\bar{W}_L\bar{L}} \neq \frac{\bar{W}_D\bar{D}}{\bar{W}_L\bar{L}}$ in the first terms provide additional independent variation. Under nondegenerate support of the input ratios $\ddot{C}_{jt}/\ddot{D}_{jt}$ and $\Phi_{jt}/\ddot{S}_{jt}$, strictly positive share parameters, appropriate instrument relevance, and $\delta \neq \gamma$, these derivatives contribute rank two to the Jacobian.\\

\textbf{Autoregressive parameter ($\rho_L$):}
The persistence parameter contributes the following component of the Jacobian:
\begin{equation}
\mathbb{E}\!\left[
\frac{\partial}{\partial \rho_L}
\bigl(\ddot{\omega}^L_{jt-1}\otimes \xi^L_{jt}\bigr)
\right]
= -\,\mathbb{E}\!\left[
\ddot{\omega}^L_{jt-1}\otimes \ddot{\omega}^L_{jt-1}
\right].
\end{equation}

\textbf{Rank condition.} The Jacobian matrix $J_L$ achieves full column rank through contributions from distinct sources of identifying variation. Time fixed effects deliver an identity submatrix of rank $\operatorname{dim}(\ddot{\iota}_{L,t})$ through the year dummy instruments. The three elasticity parameters ($\sigma^O$, $\sigma^M$, $\sigma^I$) are identified through their respective instruments $\mathbf{Z}_1$, $\mathbf{Z}_2$, and $\mathbf{Z}_3$, contributing rank three under nondegenerate input support, strictly positive nest weights, and $\delta \neq \gamma$. The wedge parameters $\phi_C$ and $\phi_D$ are separately identified through instruments $\mathbf{Z}_4$ and $\mathbf{Z}_5$ via two distinct channels: within-nest variation driven by heterogeneity in $\ddot{C}_{jt}/\ddot{D}_{jt}$, and cross-nest variation weighted by distinct baseline payroll shares, jointly contributing rank two. The autoregressive parameter $\rho_L$, instrumented by lagged productivity, generates a Jacobian component equal to the negative second moment matrix, contributing rank one under strictly positive variance. While these parameter blocks share the lagged productivity instrument—creating off-diagonal Jacobian elements—the distinct functional forms through which each parameter enters the productivity characterization, combined with appropriate instrument relevance from $\mathbf{Z}_1$ through $\mathbf{Z}_5$, ensures linear independence of the Jacobian columns:
\[
\operatorname{rank}(J_L)=\operatorname{dim}(\ddot{\iota}_{L,t})+3+2+1=\operatorname{dim}(\theta_L).\quad\qed
\]

\clearpage
\section{Step 3 – Identification Proof}\label{a:step3_proof}
Let the instrument vector be
\begin{equation}
    \mathbf{Z}_H=\big[\mathbf{I}\{year=t\},\, \mathbf{Z}_a,\, \mathbf{Z}_b\big]',
\end{equation}
where \(\mathbf{Z}_a\) and \(\mathbf{Z}_b\) denote distinct, exogenous, and relevant instrument sets. 
Define the parameter vector
\begin{equation}
    \theta_H=\{\ddot{\iota}_{H,t},\,\tau,\,\rho_H\}.
\end{equation}
Since instruments are parameter–invariant, the Jacobian of the population moment mapping is
\begin{equation}
    J_H
    \equiv
    \mathbb{E}\!\left[
        \frac{\partial\big(\mathbf{Z}_H\otimes \chi_{jt}(\theta_H)\big)}{\partial\theta_H}
    \right]
    =
    \mathbb{E}\!\left[
        \mathbf{Z}_H\otimes \frac{\partial \chi_{jt}(\theta_H)}{\partial\theta_H}
    \right].
\end{equation}

\noindent\textbf{Proposition.} The Jacobian matrix \(J_H\) has full column rank:
\begin{equation}
    \operatorname{rank}(J_H)=\operatorname{dim}(\theta_H)
    =\operatorname{dim}(\ddot{\iota}_{H,t})+2.
\end{equation}

\noindent\textbf{Proof.} I establish full column rank by evaluating the Jacobian columns. 
Hold the Step~2 estimates \((\hat{\sigma}^O,\hat{\sigma}^M,\hat{\sigma}^I,\hat{\phi}_C, \hat{\phi}_D,\hat{\ddot{\omega}}^{L}_{jt})\) fixed and define the CES component
\[
f_{jt}(\tau)
=\Big(\alpha_K(\tau)\,\ddot{K}_{jt}^{\hat{\sigma}^O}
+\alpha_M(\tau)\,\ddot{M}_{jt}^{\hat{\sigma}^O}
+\alpha_L(\tau)\,(\exp(\hat{\ddot{\omega}}^{L}_{jt})\,\hat{\ddot{L}}_{jt})^{\hat{\sigma}^O}\Big)^{1/\hat{\sigma}^O},
\]
and the non–capital composite
\[
\Phi_{jt}
\equiv \ddot{M}_{jt}^{\hat{\sigma}^O}
+\frac{\bar W_S \bar S + \hat{\phi}_C\bar W_C \bar C + \hat{\phi}_D\bar W_D \bar D}{\bar{P}_M\bar{M}}\,
\big(\exp(\hat{\ddot{\omega}}^{L}_{jt})\,\hat{\ddot{L}}_{jt}\big)^{\hat{\sigma}^O}.
\]
Replacing \(\alpha_K,\alpha_M,\alpha_L\) via \eqref{eq:alpha_tau} yields
\[
\log f_{jt}(\tau)
=\frac{1}{\hat{\sigma}^O}\log\!\left(
\frac{\bar{P}_M\bar{M}}{\bar{P}_M\bar{M}+\bar W_S \bar S + \hat{\phi}_C\bar W_C \bar C + \hat{\phi}_D\bar W_D \bar D+\tau\,\bar{P}_M\bar{M}}
\right)
+\frac{1}{\hat{\sigma}^O}\log\!\left(\tau\,\ddot{K}_{jt}^{\hat{\sigma}^O}+\Phi_{jt}\right).
\]
Let \(A\equiv \bar{P}_M\bar{M}\) and \(B\equiv \bar W_S \bar S + \hat{\phi}_C\bar W_C \bar C + \hat{\phi}_D\bar W_D \bar D\). Then
\begin{equation}
\label{eq:dlogf_dtau}
\frac{\partial \log f_{jt}(\tau)}{\partial \tau}
=\frac{1}{\hat{\sigma}^O}\left[
-\frac{A}{A+B+\tau A}
+\frac{\ddot{K}_{jt}^{\hat{\sigma}^O}}{\tau\,\ddot{K}_{jt}^{\hat{\sigma}^O}+\Phi_{jt}}
\right].
\end{equation}
\newline
\textbf{Time fixed effects \((\ddot{\iota}_{H,t})\).}
For each \(t'\),
\begin{equation}
\mathbb{E}\!\left[
\frac{\partial}{\partial \ddot{\iota}_{H,t'}}\bigl(\mathbf{I}\{year=t\}\otimes \chi_{jt}\bigr)
\right]
=
\mathbb{E}\!\left[
-\mathbf{I}\{year=t\}\,\mathbf{I}\{t=t'\}
\right],
\end{equation}
which delivers an identity submatrix of rank \(\operatorname{dim}(\ddot{\iota}_{H,t})\).\\
\newline
\textbf{Capital–allocation parameter \((\tau)\).}
Because instruments are parameter–invariant,
\begin{equation}
\mathbb{E}\!\left[
\frac{\partial}{\partial \tau}\bigl(\mathbf{Z}_a\otimes \chi_{jt}\bigr)
\right]
=\mathbb{E}\!\left[
\mathbf{Z}_a\otimes
\left(-\,\frac{\partial \log f_{jt}(\tau)}{\partial \tau}
+\rho_H\,\frac{\partial \log f_{jt-1}(\tau)}{\partial \tau}\right)
\right],
\end{equation}
with \(\partial \log f_{jt}(\tau)/\partial \tau\) given by \eqref{eq:dlogf_dtau}. Variation in \(\ddot{K}_{jt}\) relative to \(\Phi_{jt}\) across \((j,t)\) ensures that this column is nondegenerate when \(\tau>0\) and \(\hat{\sigma}^O\neq 1\).\\
\newline
\textbf{Autoregressive parameter \((\rho_H)\).}
From
\[
\chi_{jt}
= \log \ddot{\tilde Q}_{jt} 
- \ddot{\iota}_{H,t}
- \log f_{jt}(\tau)
- \rho_H\big(\log \ddot{\tilde Q}_{jt-1}-\log f_{jt-1}(\tau)\big),
\]
it follows that
\[
\frac{\partial \chi_{jt}}{\partial \rho_H}
=-\big(\log \ddot{\tilde Q}_{jt-1}-\log f_{jt-1}(\tau)\big)
=-(\ddot{\omega}^H_{jt-1}+\varepsilon_{jt-1}),
\]
and therefore
\begin{equation}
\mathbb{E}\!\left[
\frac{\partial}{\partial \rho_H}\bigl(\mathbf{Z}_b\otimes \chi_{jt}\bigr)
\right]
= -\,\mathbb{E}\!\left[
\mathbf{Z}_b\otimes \big(\ddot{\omega}^H_{jt-1}+\varepsilon_{jt-1}\big)
\right].
\end{equation}\\
\newline
\textbf{Rank condition.} Time indicators deliver an identity block of rank \(\operatorname{dim}(\ddot{\iota}_{H,t})\).
The \(\tau\) column is nondegenerate by \eqref{eq:dlogf_dtau} under \(\tau>0\), \(\hat{\sigma}^O\neq 1\), and nondegenerate support of \((\ddot{K}_{jt},\Phi_{jt})\), contributing rank one.
For \(\rho_H\), letting \(u_{jt-1}\equiv \log \ddot{\tilde Q}_{jt-1}-\log f_{jt-1}(\tau)\), the corresponding column is \(-\,\mathbb{E}[\mathbf{Z}_b\otimes u_{jt-1}]\), which contributes rank one under the standard relevance condition \(\mathbb{E}[\mathbf{Z}_b\,u_{jt-1}]\neq \mathbf{0}\).
Since these blocks load on disjoint instrument coordinates in \(\mathbf{Z}_H\), they are mutually linearly independent and independent of the time–effects block. Hence the Jacobian is block–triangular and ranks add:
\[
\operatorname{rank}(J_H)
=\operatorname{dim}(\ddot{\iota}_{H,t})+2
=\operatorname{dim}(\theta_H).
\quad\qed
\]


\clearpage
\section{Kalman Procedures}\label{a:kalman}

This appendix details the Kalman filtering procedure used to compute the likelihood function for the state-space model in equations \eqref{eq:state_space_measurement} and \eqref{eq:state_space_transition}, and the Kalman smoother algorithm to obtain point estimates of the unobservable state vector. 

The state-space representation of the estimating equation from Step 3 takes the form:
\begin{align}
    & \widetilde{\omega}_{jt} = H \zeta_{jt} + A Y_{jt} + \hat{\ddot{\iota}}_{H,t}, \label{app_eq:state_space_measurement}  \\
    & \zeta_{jt} = F \zeta_{jt-1} + G u_{jt}, \label{app_eq:state_space_transition}
\end{align}
where the measurement equation \eqref{app_eq:state_space_measurement} links observed noisy productivity to the latent state vector, and the transition equation \eqref{app_eq:state_space_transition} governs the evolution of the state vector over time.

The system matrices are defined as follows. The state and disturbance vectors of the transition equation are:
\begin{align*}
    \zeta_{jt} = \left[\begin{array}{ccc} \xi_{jt}^{H} & \varepsilon_{jt} & \varepsilon_{jt-1} \end{array}\right]', \quad 
    u_{jt} = \left[\begin{array}{cc} \xi_{jt}^{H} & \varepsilon_{jt} \end{array}\right]',
\end{align*}
while the observation components are:
\begin{align*}
    Y_{jt} = \left[\begin{array}{cccc} \widetilde{\omega}_{jt-1}  \end{array}\right]'.
\end{align*}
The system matrices governing the state-space dynamics are:
\begin{align*}
    & H = \left[\begin{array}{ccc} 1 & 1 & - \hat{\rho}_{H} \end{array}\right], \quad 
    A = \left[\begin{array}{cccc} \hat{\rho}_{H}  \end{array}\right], \\
    & F = \left[\begin{array}{ccc} 0 & 0 & 0 \\ 0 & 0 & 0 \\ 0 & 1 & 0 \end{array}\right], \quad 
    G = \left[\begin{array}{cc} 1 & 0 \\ 0 & 1 \\ 0 & 0 \end{array}\right].
\end{align*}
The unknown variance-covariance matrix of the disturbance vector $u_{jt}$ is given by:
\begin{equation}
    \Sigma=\begin{bmatrix}
        (\sigma^H)^2 & 0 \\
        0 & (\sigma^\varepsilon)^2
    \end{bmatrix}.
\end{equation}
For each plant \( j \), assuming that $\sigma^H$ and $\sigma^\varepsilon$ are known, the Kalman filtering steps proceed as follows (\cite{hamilton1994filter}):
\begin{align*}
    & \zeta_{j,0|0} = \mathbb{E}(\zeta_{j0}) = \vec{0}_{3 \times 1}, \\
    & \mathbf{P}_{j,0|0} = \mathbb{E}(\zeta_{j,0|0} \zeta_{j,0|0}'), \\
    & \zeta_{j,t+1|t} = F \zeta_{j, t|t}, \\
    & \mathbf{P}_{j,t+1|t} = F \mathbf{P}_{j, t|t} F'  + G \Sigma G', \\
    & \nu_{jt+1} = \widetilde{\omega}_{jt+1} - H \zeta_{j,t+1|t} - A Y_{jt+1} - \hat{\ddot{\iota}}_{H,t+1} , \\
    & N_{j,t+1} = H \mathbf{P}_{j,t+1|t} H', \\
    & O_{j,t+1} = \mathbf{P}_{j,t+1|t} H' N_{j,t+1}^{-1}, \\
    & \zeta_{j,t+1|t+1} =  \zeta_{j,t+1|t}  -  O_{j,t+1}  \nu_{jt+1}, \\
    & \mathbf{P}_{j,t+1|t+1} = (I - O_{j,t+1} H) \mathbf{P}_{j,t+1|t}.
\end{align*}
At each iteration, I compute the log-likelihood of observing \(\tilde{\omega}_{jt+1}\) using \(\nu_{jt+1}\) and \(N_{j,t+1}\):
\begin{align*}
    \log{f(\tilde{\omega}_{jt+1} | \Omega_{t})} = - \log{(\text{det}(N_{j,t+1}))} - \nu_{jt+1}' N_{j,t+1}^{-1} \nu_{jt+1}.
\end{align*}
Let $T_j$ be the maximum number of periods $t$ for which plant $j$ is observed. The contribution of plant $j$ to the overall likelihood is then:
\begin{align*}
    L_{j}(\sigma^{H},\sigma^{\varepsilon}) = - \sum_{t=0}^{T_{j}} \log{(\text{det}(N_{j,t}))} - \sum_{t=0}^{T_{j}} \nu_{jt}' N_{j,t}^{-1} \nu_{jt},
\end{align*}
and the total likelihood function is
\begin{align*}
    \mathcal{L}(\sigma^{H},\sigma^{\varepsilon}) = \sum_{j=1}^{J} L_{j}(\sigma^{H},\sigma^{\varepsilon}).
\end{align*}
Upon estimating $\sigma^{H}$ and $\sigma^{\varepsilon}$ via quasi-maximum likelihood estimation, I recover the estimates of \(\xi_{jt}^{H}\) and \(\varepsilon_{jt}\) using the Kalman smoother \citep{hamilton1994filter}, based on all available information over the sample period. Specifically, given the Kalman-filtered variables \((\hat{\zeta}_{j,t|t}, \hat{\mathbf{P}}_{j,t|t}, \hat{\zeta}_{j,t|t+1}, \hat{\mathbf{P}}_{j,t+1|t})\), I recursively apply the Kalman smoother from \(T_{j}\) back to \(t=1\):
\begin{align*}
    & C_{jt} = \hat{\mathbf{P}}_{j,t|t} F' \hat{\mathbf{P}}_{j,t+1|t}^{-1}, \\
    & \hat{\zeta}_{j,t|T_{j}} = \hat{\zeta}_{j,t|t} + C_{jt}(\hat{\zeta}_{j,t+1|T_{j}} - \hat{\zeta}_{j,t+1|t}), \\
    & \hat{\mathbf{P}}_{j,t|T_{j}} = \hat{\mathbf{P}}_{j,t|t} + C_{jt}(\hat{\mathbf{P}}_{j,t+1|T_{j}} - \hat{\mathbf{P}}_{j,t+1|t}) C_{jt}'.
\end{align*}
The smoothed estimates \(\hat{\xi}_{jt}^{H}\) and \(\hat{\varepsilon}_{jt}\) correspond to the first and second elements of \(\hat{\zeta}_{j,t|T_{j}}\), respectively.

%
\clearpage
\section{Bargaining Parameter – Identification Proof}\label{a:bargaining_proof}
Let the instrument vector be
\begin{equation}
    \mathbf{Z}_B=\big[\mathbf{1},\, \mathbf{Z}_\alpha,\, \mathbf{Z}_\beta\big]',
\end{equation}
where \(\mathbf{Z}_\alpha\) and \(\mathbf{Z}_\beta\) denote distinct, exogenous, and relevant instrument sets, and $\mathbf{1}$ denotes the unit vector. 
Define the parameter vector
\begin{equation}
    \theta_B=\{\beta,\,\theta_0,\,\theta_1\}.
\end{equation}
Since instruments are parameter–invariant, the Jacobian of the population moment mapping is
\begin{equation}
    J_B
    \equiv
    \mathbb{E}\!\left[
        \frac{\partial\big(\mathbf{Z}_B\otimes \varphi_{jt}(\theta_B)\big)}{\partial\theta_B}
    \right]
    =
    \mathbb{E}\!\left[
        \mathbf{Z}_B\otimes \frac{\partial \varphi_{jt}(\theta_B)}{\partial\theta_B}
    \right].
\end{equation}

\noindent\textbf{Proposition.} The Jacobian matrix \(J_B\) has full column rank:
\begin{equation}
    \operatorname{rank}(J_B)=\operatorname{dim}(\theta_B)
    =3.
\end{equation}

\noindent\textbf{Proof.} Define the residual function
\begin{equation}
\varphi_{jt}(\theta_B) = -\theta_0 - \theta_1 \log D_{jt} + \log G_{jt}(\beta),
\end{equation}
where
\begin{equation}
G_{jt}(\beta) = \left(\frac{\partial P_{jt}}{\partial Q_{jt}}\frac{Q_{jt}}{P_{jt}}+1\right)P_{jt}\frac{\partial Q_{jt}}{\partial D_{jt}} + \frac{\beta}{1-\beta}\,H_{jt} - \left(\frac{\partial W_{jt,D}}{\partial D_{jt}}\frac{D_{jt}}{W_{jt,D}}+1\right)W_{jt,D},
\end{equation}
and
\begin{equation}
H_{jt} \equiv \left[\frac{\Pi_{jt}}{D_{jt}}\frac{\partial (U^U_{jt}-U^U_{Ojt})}{\partial D_{jt}}\frac{D_{jt}}{U^U_{jt}-U^U_{Ojt}}\right].
\end{equation}
I establish full column rank by evaluating each Jacobian column.\\
\newline
\textbf{Intercept parameter \((\theta_0)\):}
Because $\varphi_{jt}$ enters $\theta_0$ linearly with coefficient $-1$,
\begin{equation}
\frac{\partial \varphi_{jt}}{\partial \theta_0} = -1.
\end{equation}
Therefore,
\begin{equation}
\mathbb{E}\left[\frac{\partial}{\partial \theta_0}\big(\mathbf{1} \otimes \varphi_{jt}\big)\right] = \mathbb{E}\left[-\mathbf{1}\right] = -\mathbf{1},
\end{equation}
which delivers a column of constant $-1$ values (rank one).\\
\newline
\textbf{Log-employment coefficient \((\theta_1)\):}
From the linear term $-\theta_1 \log D_{jt}$ in $\varphi_{jt}$,
\begin{equation}
\frac{\partial \varphi_{jt}}{\partial \theta_1} = -\log D_{jt}.
\end{equation}
Hence,
\begin{equation}
\mathbb{E}\left[\frac{\partial}{\partial \theta_1}\big(\mathbf{Z}_\beta \otimes \varphi_{jt}\big)\right] = \mathbb{E}\left[\mathbf{Z}_\beta \otimes (-\log D_{jt})\right] = -\,\mathbb{E}\left[\mathbf{Z}_\beta \cdot \log D_{jt}\right].
\end{equation}
Under the relevance condition $\mathbb{E}[\mathbf{Z}_\beta \cdot \log D_{jt}] \neq \mathbf{0}$, this column is nondegenerate and contributes rank one.\\
\newline
\textbf{Bargaining parameter \((\beta)\):}
By the chain rule,
\begin{equation}
\frac{\partial \varphi_{jt}}{\partial \beta} = \frac{\partial}{\partial \beta}\log G_{jt}(\beta) = \frac{1}{G_{jt}(\beta)}\,\frac{\partial G_{jt}(\beta)}{\partial \beta}.
\end{equation}
The only $\beta$-dependent term in $G_{jt}(\beta)$ is $\frac{\beta}{1-\beta}\,H_{jt}$. Compute:
\begin{equation}
\frac{\partial}{\partial \beta}\left(\frac{\beta}{1-\beta}\right) = \frac{\partial}{\partial \beta}\big[\beta(1-\beta)^{-1}\big] = (1-\beta)^{-1} + \beta(1-\beta)^{-2} = \frac{1}{(1-\beta)^2}.
\end{equation}
Thus,
\begin{equation}
\frac{\partial G_{jt}}{\partial \beta} = \frac{H_{jt}}{(1-\beta)^2},
\end{equation}
and
\begin{equation}
\frac{\partial \varphi_{jt}}{\partial \beta} = \frac{1}{G_{jt}(\beta)}\cdot\frac{H_{jt}}{(1-\beta)^2}.
\end{equation}
Consequently,
\begin{equation}
\mathbb{E}\left[\frac{\partial}{\partial \beta}\big(\mathbf{Z}_\alpha \otimes \varphi_{jt}\big)\right] = \mathbb{E}\left[\mathbf{Z}_\alpha \otimes \frac{H_{jt}}{(1-\beta)^2\,G_{jt}(\beta)}\right].
\end{equation}
This column is nondegenerate under $\beta \in (0,1)$, nondegenerate support of $H_{jt}$, and the relevance condition $\mathbb{E}[\mathbf{Z}_\alpha \cdot H_{jt}/G_{jt}] \neq \mathbf{0}$. Hence this column contributes rank one.\\
\newline
\textbf{Rank condition.}
The Jacobian has the block structure:
\begin{equation}
J_B = \begin{bmatrix}
\mathbb{E}[\mathbf{1}] \cdot (-1) & \mathbb{E}[\mathbf{1} \cdot (-\log D_{jt})] & \mathbb{E}\left[\mathbf{1} \cdot \frac{H_{jt}}{(1-\beta)^2 G_{jt}}\right] \\[6pt]
0 & \mathbb{E}[\mathbf{Z}_\alpha \cdot (-\log D_{jt})] & \mathbb{E}\left[\mathbf{Z}_\alpha \cdot \frac{H_{jt}}{(1-\beta)^2 G_{jt}}\right] \\[6pt]
0 & \mathbb{E}[\mathbf{Z}_\beta \cdot (-\log D_{jt})] & 0
\end{bmatrix}.
\end{equation}
Column 1 ($\theta_0$) loads only on the constant (rank 1), column 2 ($\theta_1$) loads on $\mathbf{Z}_\beta$ via $-\log D_{jt}$ (rank 1 under relevance), and column 3 ($\beta$) loads on $\mathbf{Z}_\alpha$ via $H_{jt}/[(1-\beta)^2 G_{jt}]$ (rank 1 under relevance and variation). Since $\mathbf{Z}_\alpha$ and $\mathbf{Z}_\beta$ are distinct instrument sets, the three columns are mutually linearly independent. The Jacobian is block-triangular, and ranks add:
\begin{equation}
\operatorname{rank}(J_B) = 1 + 1 + 1 = 3 = \dim(\theta_B). \quad \qed
\end{equation}

\clearpage
\section{Production Function Estimates: Bootstrap Inference}\label{a:bootstrap_pf}

Table \ref{tab:bootstrap_stats} presents comprehensive summary statistics for non-parametric bootstrap replications of the production function parameters. For each parameter, I report the mean, median, standard deviation, 10th and 90th percentiles, alongside the effective simulation count. I initialize each bootstrap algorithm with 3,600 replications and then exclude simulations that converged to boundary solutions. This filtering procedure explains the variation in $N_{sim}$ across parameters, with final sample sizes below the initial 3,600 replications reflecting the exclusion of boundary-converged estimates. The density distributions of bootstrapped parameter estimates between the 1st and 99th percentiles, displayed below the summary table, reveal the complete distributional characteristics underlying these summary measures.

\begin{table}[ht]
\centering
\caption{Bootstrap Statistics Summary} 
\label{tab:bootstrap_stats}
\begin{tabular}{lrrrrrr}
  \toprule
Parameter & Mean & Median & SD & P10 & P90 & $N_{sim}$ \\ 
  \midrule
    $\log(\tau)$ & 0.254 & 0.345 & 1.900 & -2.390 & 2.745 & 2624 \\ 
    $\phi_C$ & 2.146 & 1.769 & 0.952 & 1.222 & 3.797 & 649 \\ 
  $\phi_D$ & 2.216 & 1.935 & 0.864 & 1.323 & 3.751 & 544 \\ 
$\sigma^O$ & 0.589 & 0.679 & 0.690 & 0.343 & 0.913 & 1510 \\ 
  $\sigma^M$ & -0.258 & -0.060 & 1.056 & -1.291 & 0.671 & 2871 \\ 
  $\sigma^I$ & -1.436 & -1.308 & 1.529 & -2.912 & 0.315 & 2719 \\ 
    $\alpha_L$ & 0.051 & 0.051 & 0.035 & 0.003 & 0.097 & 3600 \\ 
  $\alpha_M$ & 0.487 & 0.493 & 0.332 & 0.032 & 0.905 & 3600 \\ 
  $\alpha_K$ & 0.558 & 0.595 & 0.330 & 0.054 & 0.971 & 2978 \\ 
  $\alpha_E$ & 0.640 & 0.640 & 0.021 & 0.613 & 0.667 & 3600 \\ 
  $\alpha_S$ & 0.360 & 0.360 & 0.021 & 0.333 & 0.387 & 3600 \\ 
  $\alpha_D$ & 0.755 & 0.756 & 0.030 & 0.716 & 0.794 & 3600 \\ 
  $\alpha_C$ & 0.245 & 0.244 & 0.030 & 0.206 & 0.284 & 3600 \\ 
  $c_L$ & -0.120 & -0.136 & 0.319 & -0.386 & 0.116 & 3600 \\ 
  $\rho_L$ & 0.900 & 0.903 & 0.040 & 0.848 & 0.949 & 3600 \\ 
    $c_H$ & 0.393 & 0.325 & 0.354 & 0.044 & 0.837 & 3600 \\ 
  $\rho_H$ & 0.877 & 0.879 & 0.088 & 0.765 & 0.988 & 3600 \\ 
  $\sigma^H$ & 0.936 & 0.945 & 0.129 & 0.762 & 1.090 & 3600 \\ 
  $\sigma^\varepsilon$ & 0.305 & 0.317 & 0.114 & 0.142 & 0.447 & 1890 \\ 
   \bottomrule
\end{tabular}
\end{table}

\begin{figure}[h!]
    \centering
    \begin{subfigure}[b]{0.4\textwidth}
        \centering
        \includegraphics[width=\linewidth]{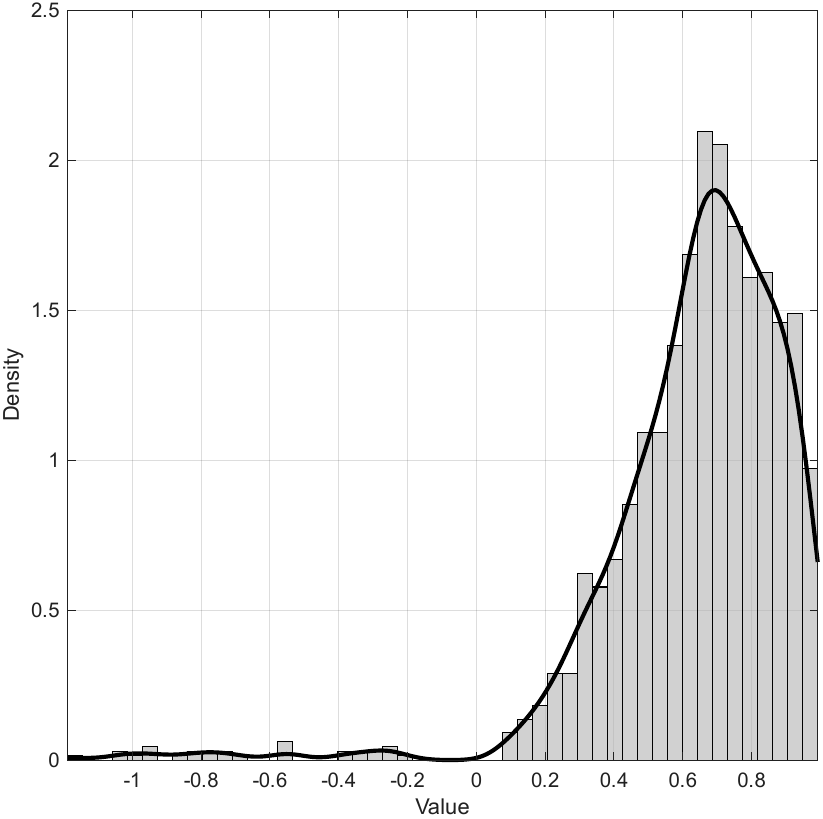}
        \caption*{$\sigma^O$}
    \end{subfigure}
    \hfill
    \begin{subfigure}[b]{0.4\textwidth}
        \centering
        \includegraphics[width=\linewidth]{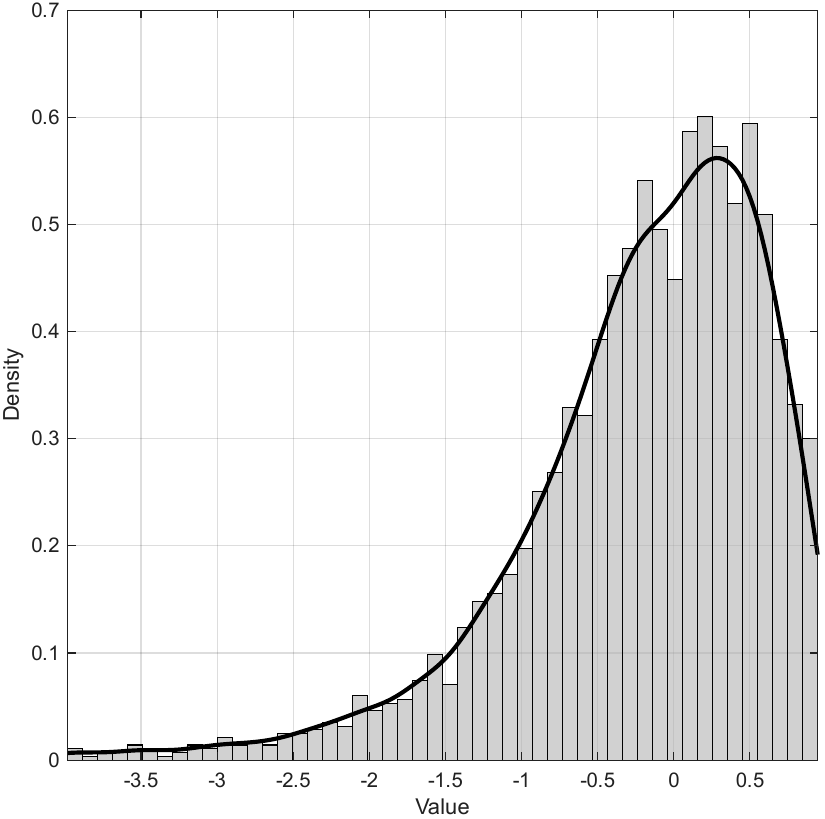}
        \caption*{$\sigma^M$}
    \end{subfigure}

    \vspace{0.1cm}
    \begin{subfigure}[b]{0.4\textwidth}
        \centering
        \includegraphics[width=\linewidth]{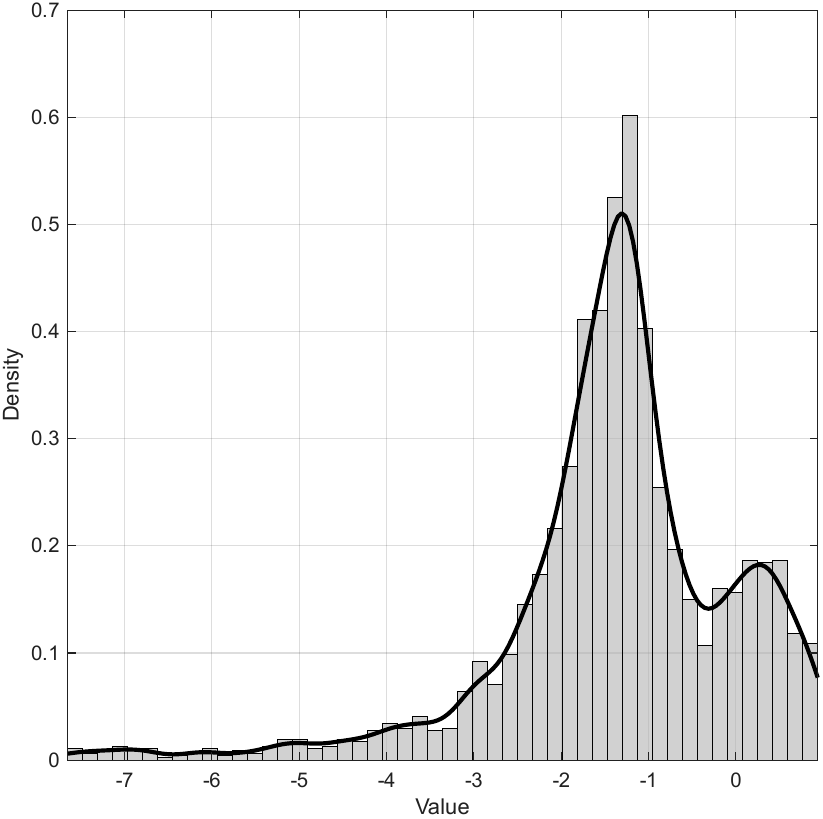}
        \caption*{$\sigma^I$}
    \end{subfigure}
    \hfill
    \begin{subfigure}[b]{0.4\textwidth}
        \centering
        \includegraphics[width=\linewidth]{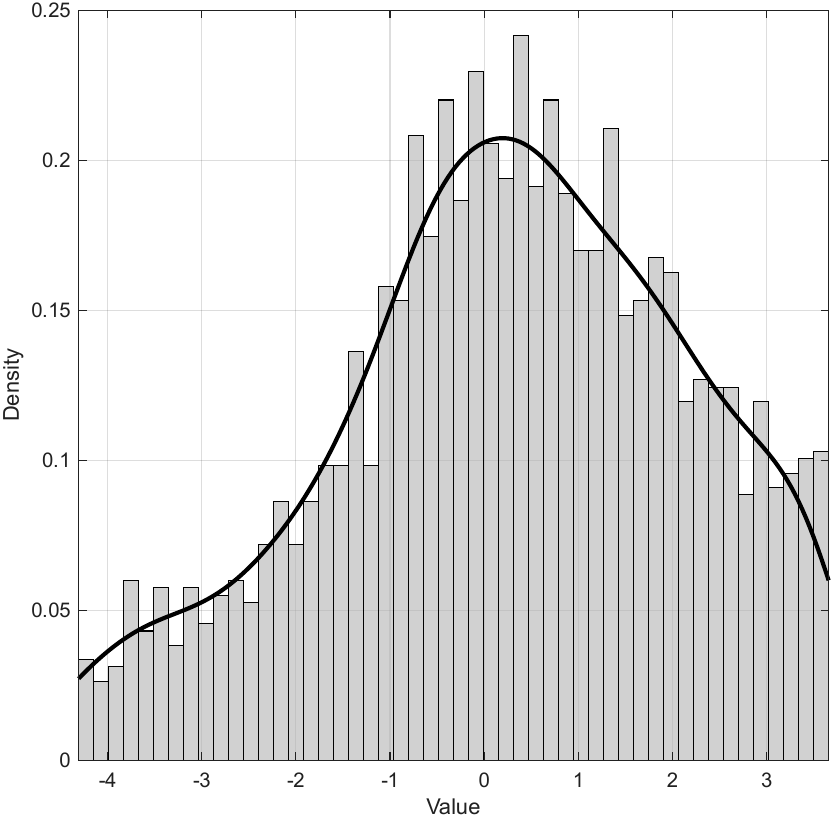}
        \caption*{$\log(\tau)$}
    \end{subfigure}
    
    \vspace{0.1cm}
        \begin{subfigure}[b]{0.4\textwidth}
        \centering
        \includegraphics[width=\linewidth]{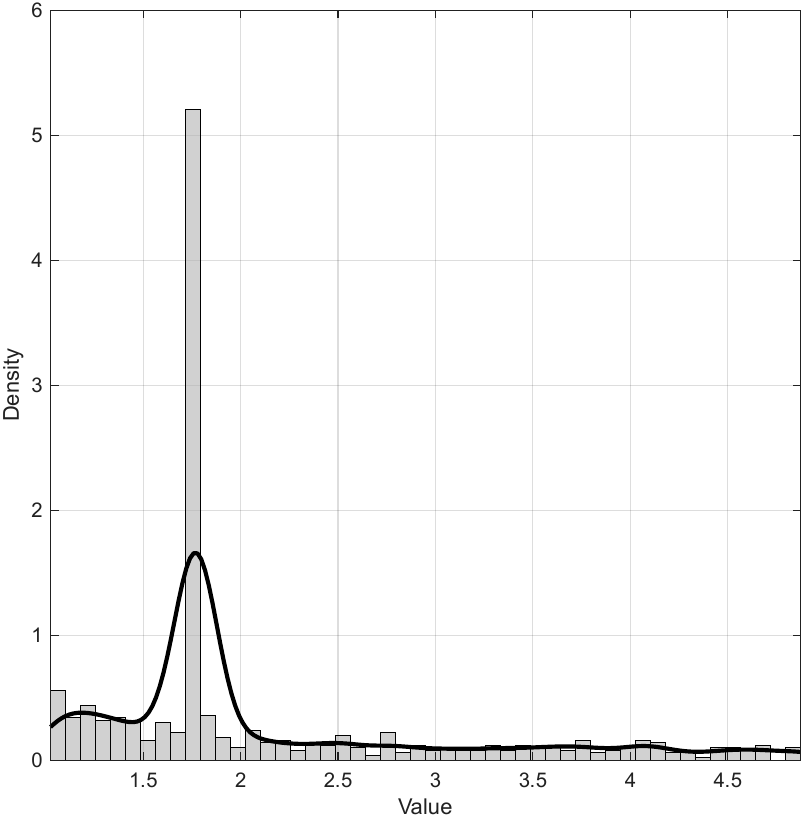}
        \caption*{$\phi_C$}
    \end{subfigure}
    \hfill
            \begin{subfigure}[b]{0.4\textwidth}
        \centering
        \includegraphics[width=\linewidth]{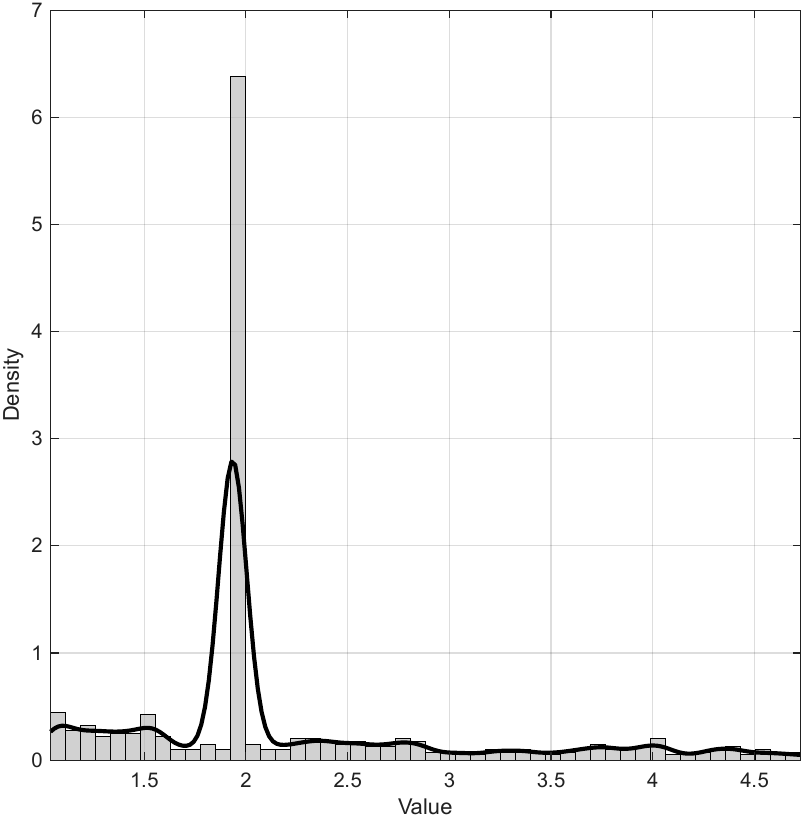}
        \caption*{$\phi_D$}
    \end{subfigure}
\end{figure}

\begin{figure}[h]
    \centering
        \begin{subfigure}[b]{0.4\textwidth}
        \centering
        \includegraphics[width=\linewidth]{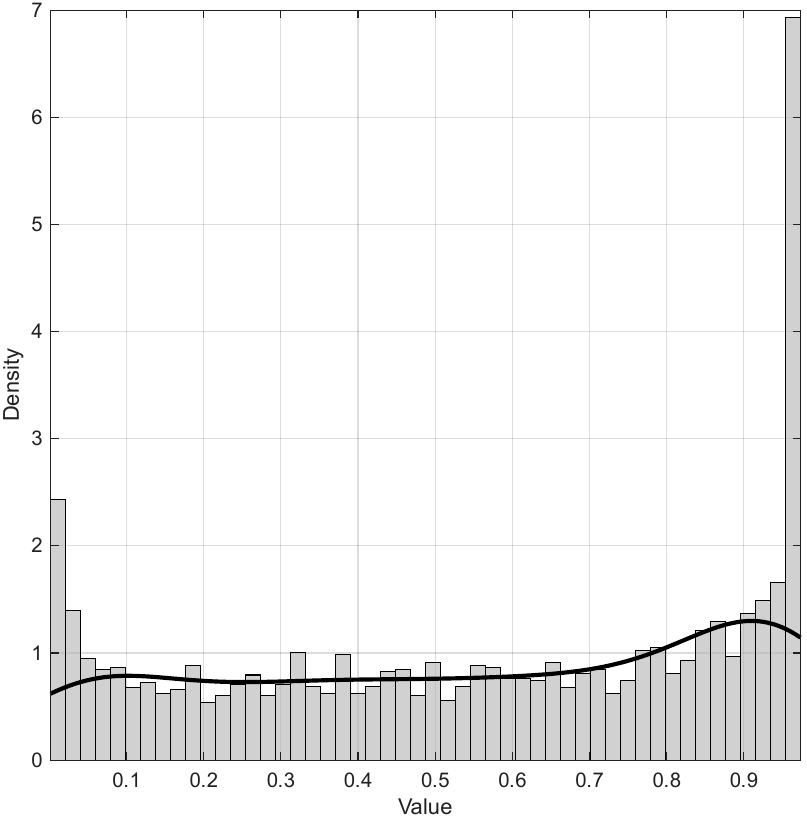}
        \caption*{$\alpha_K$}
    \end{subfigure}
    \hfill
    \begin{subfigure}[b]{0.4\textwidth}
        \centering
        \includegraphics[width=\linewidth]{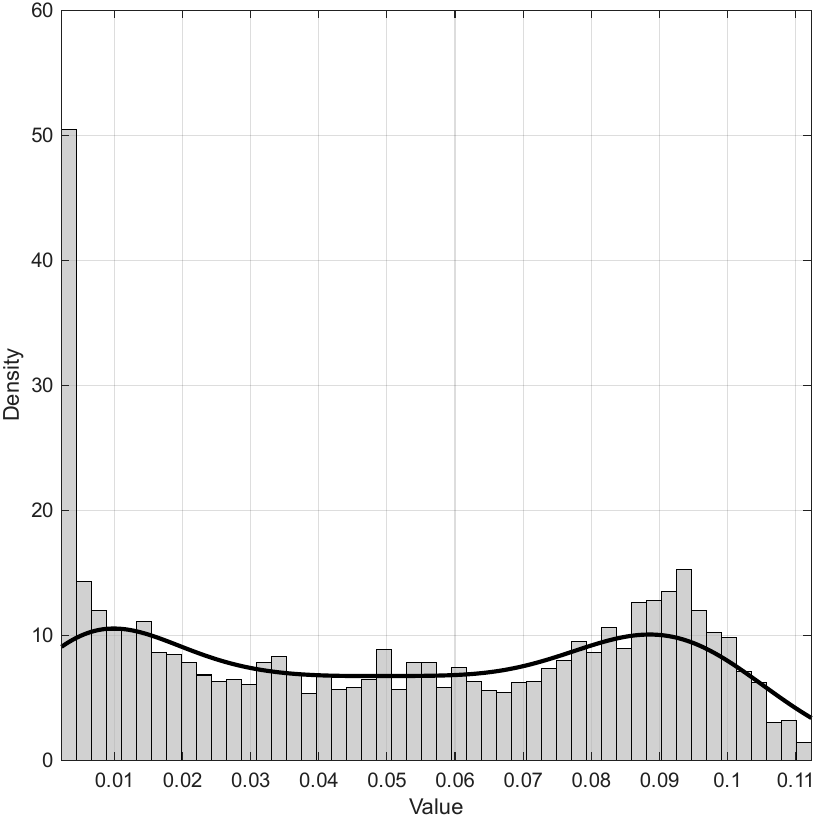}
        \caption*{$\alpha_L$}
    \end{subfigure}
    \vspace{0.1cm}
    \begin{subfigure}[b]{0.4\textwidth}
        \centering
        \includegraphics[width=\linewidth]{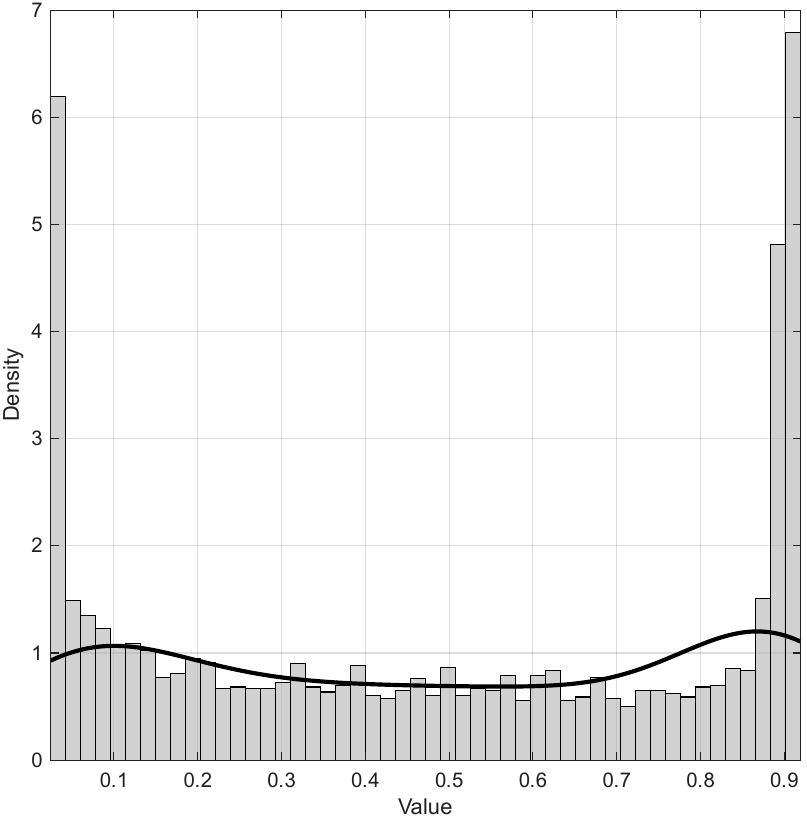}
        \caption*{$\alpha_M$}
    \end{subfigure}
    \hfill
        \begin{subfigure}[b]{0.4\textwidth}
        \centering
        \includegraphics[width=\linewidth]{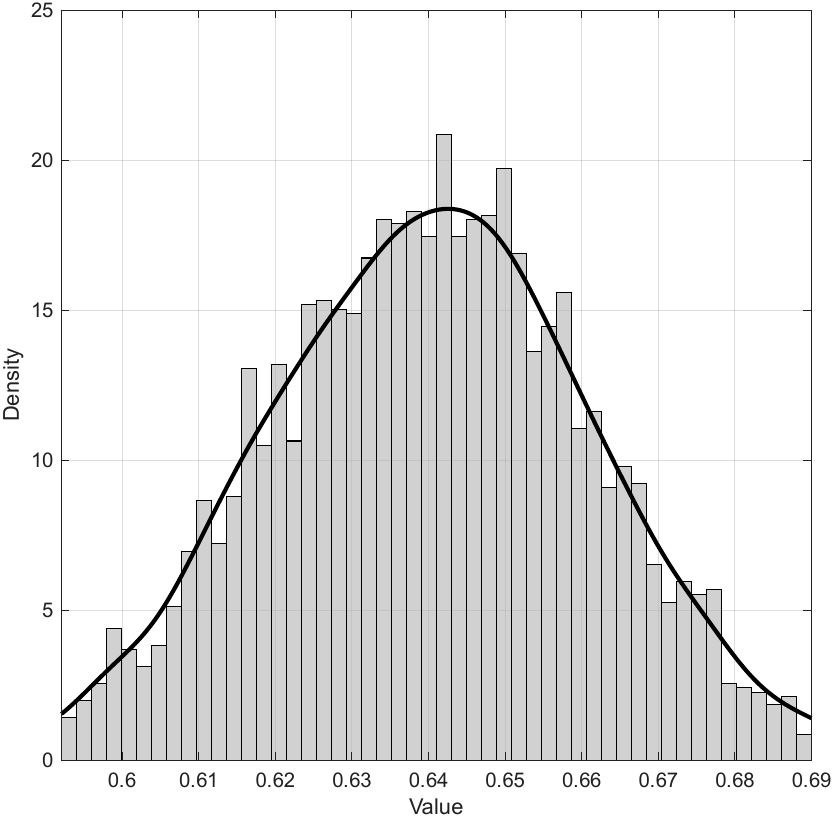}
        \caption*{$\alpha_E$}
    \end{subfigure}
    \vspace{0.1cm}
    \begin{subfigure}[b]{0.4\textwidth}
        \centering
        \includegraphics[width=\linewidth]{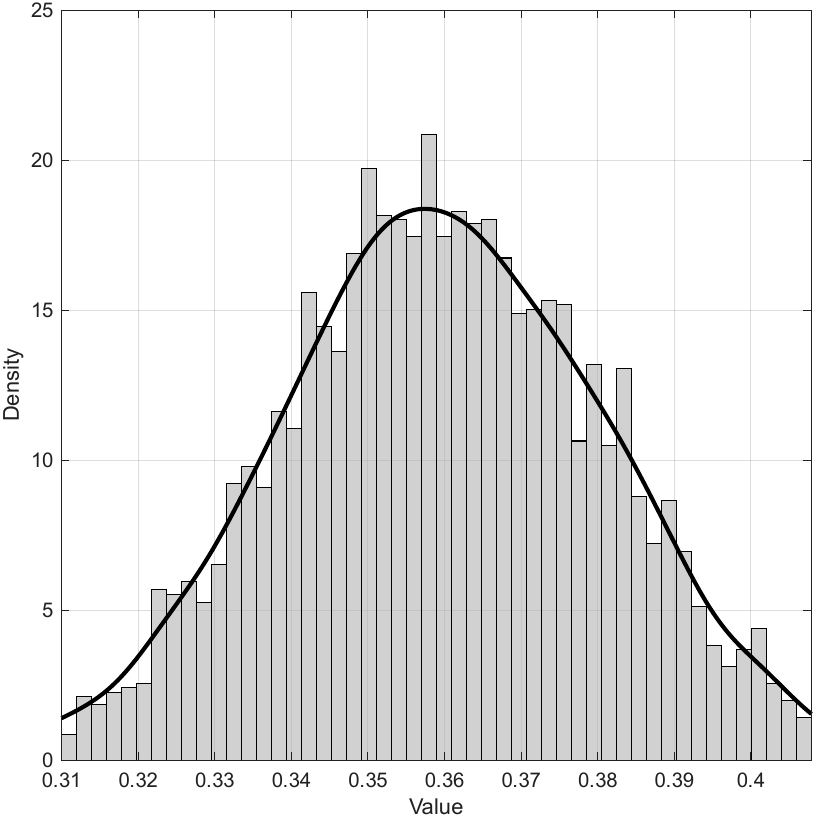}
        \caption*{$\alpha_S$}
    \end{subfigure}
    \hfill
        \begin{subfigure}[b]{0.4\textwidth}
        \centering
        \includegraphics[width=\linewidth]{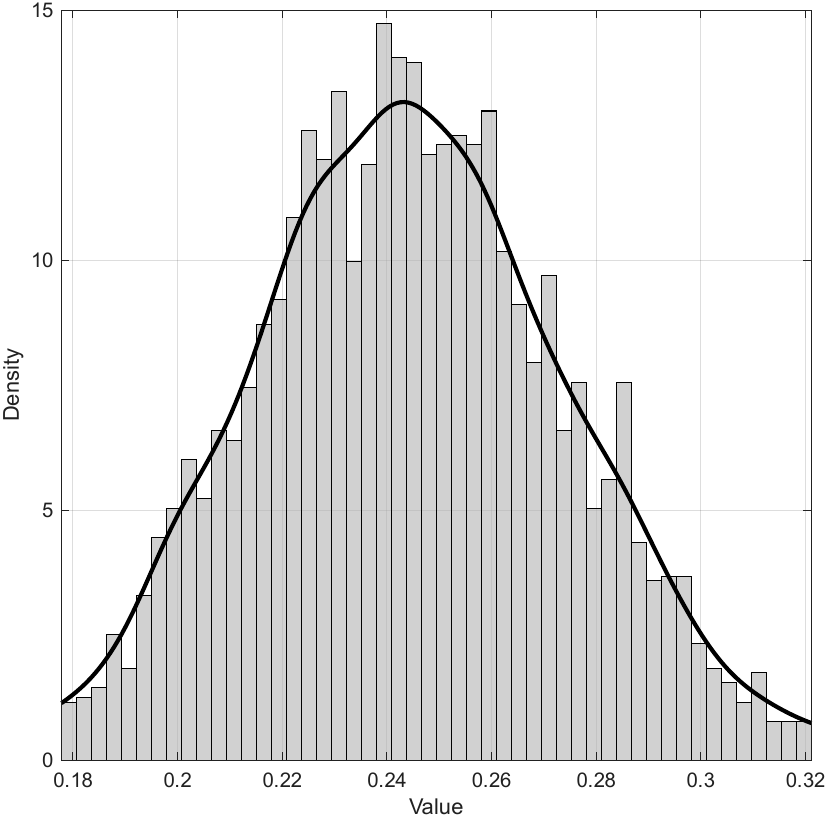}
        \caption*{$\alpha_C$}
    \end{subfigure}
  \end{figure}

\begin{figure}[h]
        \begin{subfigure}[b]{0.4\textwidth}
        \centering
        \includegraphics[width=\linewidth]{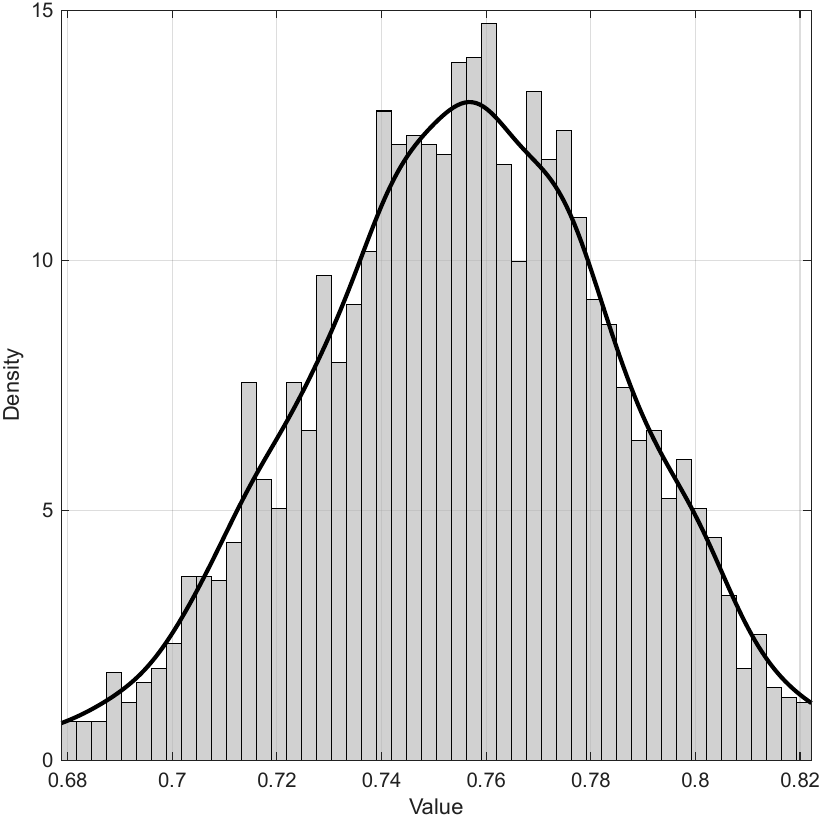}
        \caption*{$\alpha_D$}
    \end{subfigure}\\
        \vspace{0.1cm}
    \begin{subfigure}[b]{0.4\textwidth}
        \centering
        \includegraphics[width=\linewidth]{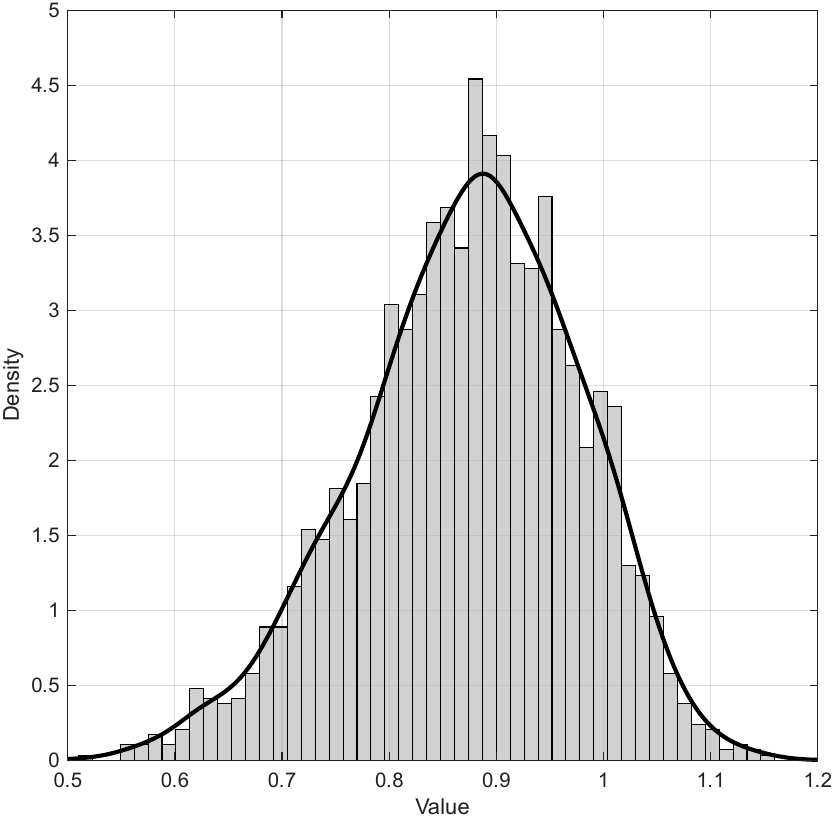}
        \caption*{$\sigma^H$}
    \end{subfigure}
    \hfill
    \begin{subfigure}[b]{0.4\textwidth}
        \centering
        \includegraphics[width=\linewidth]{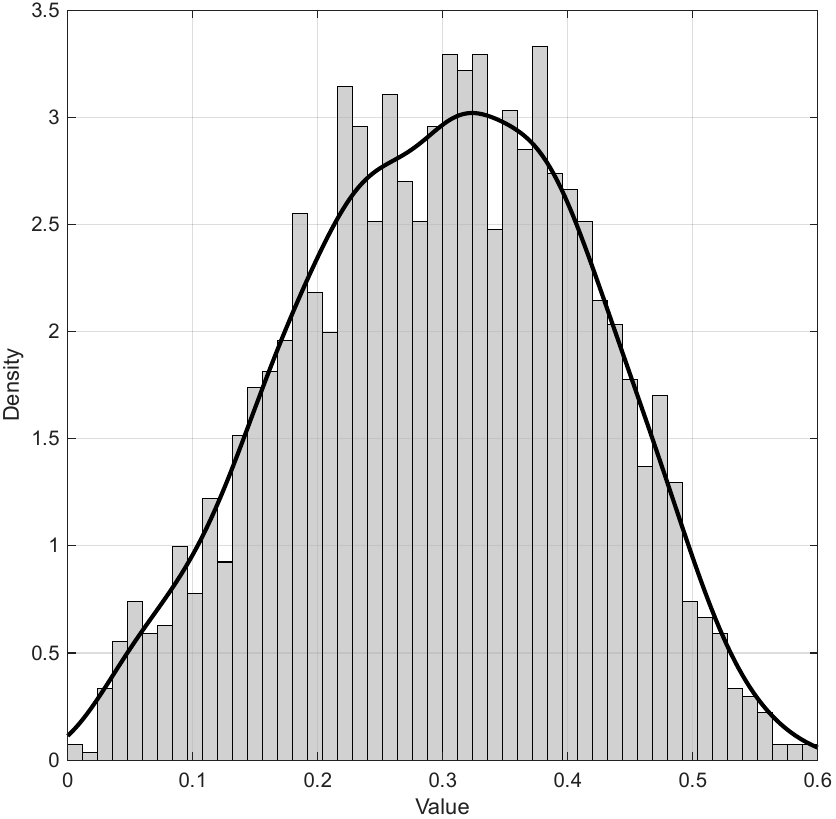}
        \caption*{$\sigma^\varepsilon$}
    \end{subfigure}
\end{figure}
\clearpage


\section{Bargaining Parameter Estimate: Bootstrap Inference}\label{a:bootstrap_barg}

Table \ref{tab:bootstrap_beta_summary} presents comprehensive summary statistics for non-parametric bootstrap replications of the bargaining and non-wage marginal cost parameters. For each parameter, I report the mean, median, standard deviation, 10th and 90th percentiles, alongside the effective simulation count. I initialize each bootstrap algorithm with 3,600 replications and then exclude simulations that converged to boundary solutions. This filtering procedure explains the variation in $N_{sim}$ across parameters, with final sample sizes below the initial 3,600 replications reflecting the exclusion of boundary-converged estimates. The density distributions of bootstrapped parameter estimates between the 1st and 99th percentiles, displayed below the summary table, reveal the complete distributional characteristics underlying these summary measures.

\begin{table}[ht]
\centering
\caption{Bootstrap Statistics Summary} 
\label{tab:bootstrap_beta_summary}
\begin{tabular}{lrrrrrr}
  \toprule
Parameter & Mean & Median & SD & P10 & P90 & $N_{sim}$\\ 
  \midrule
$\beta$ & 0.003 & 0.003 & 0.001 & 0.003 & 0.004 & 3600 \\ 
  $\theta_1$ & 0.028 & 0.015 & 0.161 & -0.164 & 0.238 & 3600 \\ 
  $\theta_0$ & 6.683 & 6.859 & 1.880 & 4.215 & 8.918 & 3600 \\ 
   \bottomrule
\end{tabular}
\end{table}

\begin{figure}[h!]
    \centering
     \begin{subfigure}[b]{0.48\textwidth}
        \centering
        \includegraphics[width=\linewidth]{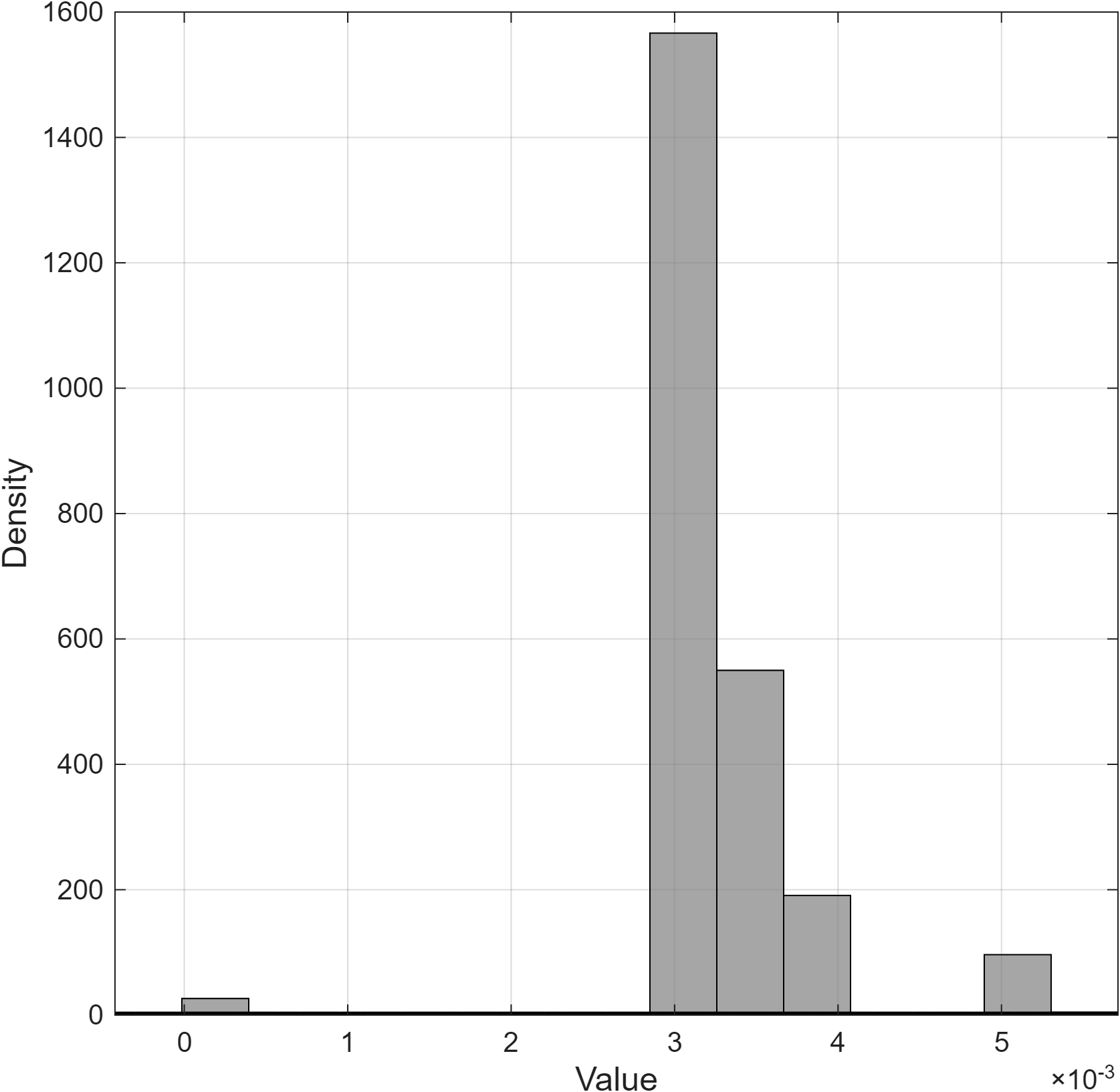}
        \caption*{$\beta$}
    \end{subfigure}\\
        \vspace{0.1cm}
    \begin{subfigure}[b]{0.48\textwidth}
        \centering
        \includegraphics[width=\linewidth]{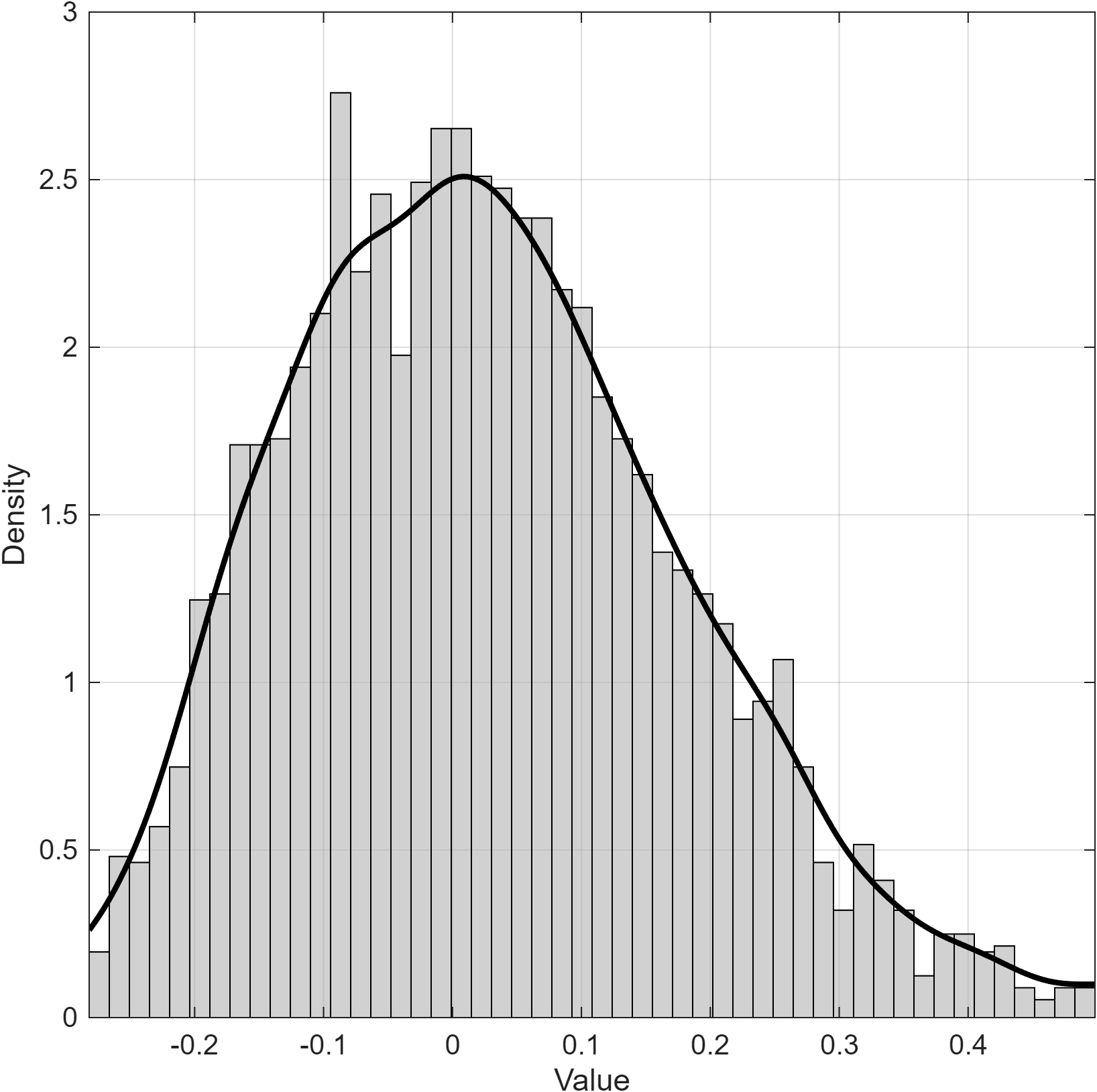}
        \caption*{$\theta_1$}
    \end{subfigure}
    \hfill
    \begin{subfigure}[b]{0.48\textwidth}
        \centering
        \includegraphics[width=\linewidth]{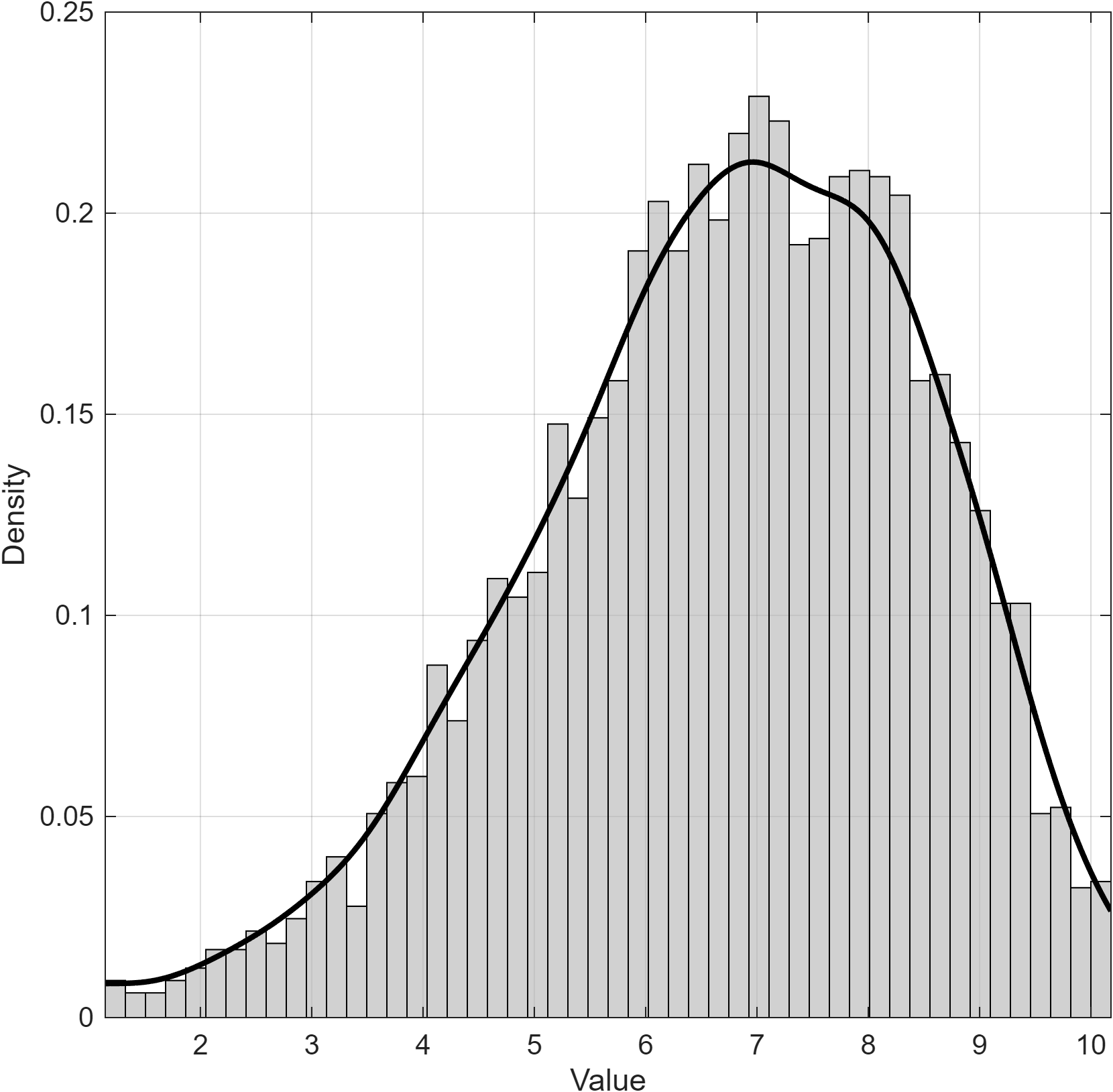}
        \caption*{$\theta_0$}
    \end{subfigure}

\end{figure}
\clearpage

\section{Section 5.3: Additional Figure}\label{sec:sec53_addfig}
\FloatBarrier
\begin{figure}[ht]
    \centering
    \includegraphics[width=0.8\linewidth]{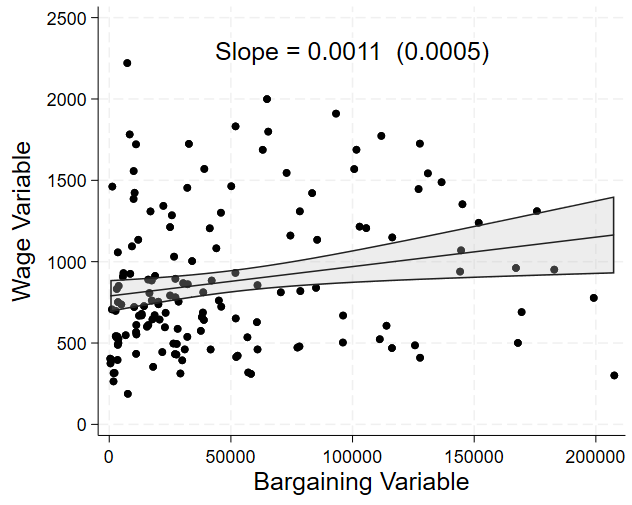}
    \caption{OLS - Second Stage, Positive Profits}
    \label{fig:OLS_0}
\caption*{\footnotesize \textit{Note:} The figure plots the left-hand side variable (the "Wage Variable") of equation \eqref{eq:Nash_bargaining_FOC}, $\left(\frac{\partial W_{jt,D}}{\partial D_{jt}}\frac{D_{jt}}{W_{jt,D}}+1\right)W_{jt,D}$, against the bargaining variable. Observations with zero reported profits are dropped. Fitted regression lines with 95\% confidence intervals are displayed.}
\end{figure}

\clearpage
\section{Empirical Distributions of Wage and Bargaining Variables' Components}\label{sec:wage_barg_comp}

\begin{figure}[ht]
\centering
\caption{Wage and Bargaining Variables Components Analysis}\label{fig:comp_analysis}
\begin{subfigure}[h]{0.42\textwidth}
    \centering
    \includegraphics[width=\linewidth]{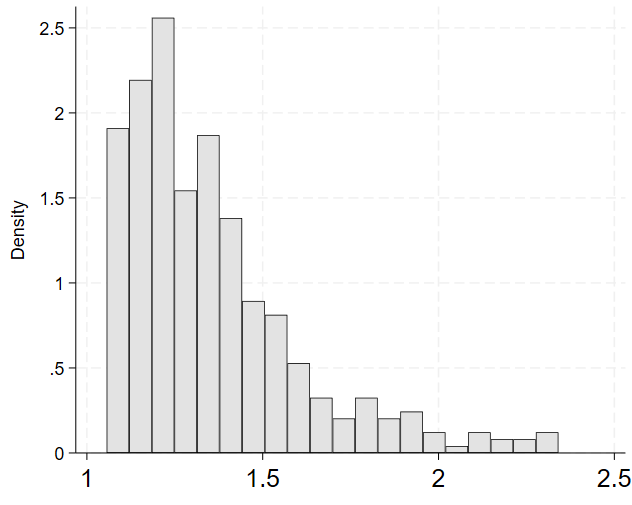}
    \caption{$\left(\frac{\partial W_{jt,D}}{\partial D_{jt}}\frac{D_{jt}}{W_{jt,D}}+1\right)$}\label{fig:elasD}
\end{subfigure}
\hfill
\begin{subfigure}[h]{0.42\textwidth}
    \centering
    \includegraphics[width=\linewidth]{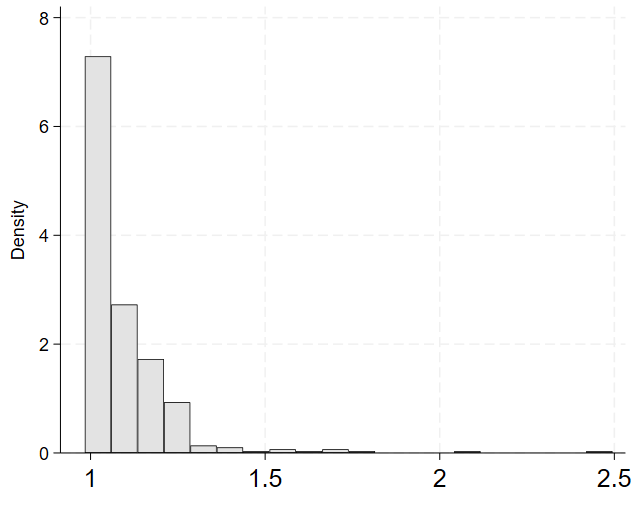}
    \caption{$\frac{\partial (U^U_{jt}-U^U_{Ojt})}{\partial D_{jt}}\frac{D_{jt}}{(U^U_{jt}-U^U_{Ojt})}$}\label{fig:elasU}
\end{subfigure}
\vspace{0.5cm}
\begin{subfigure}[h]{0.42\textwidth}
    \centering
    \includegraphics[width=\linewidth]{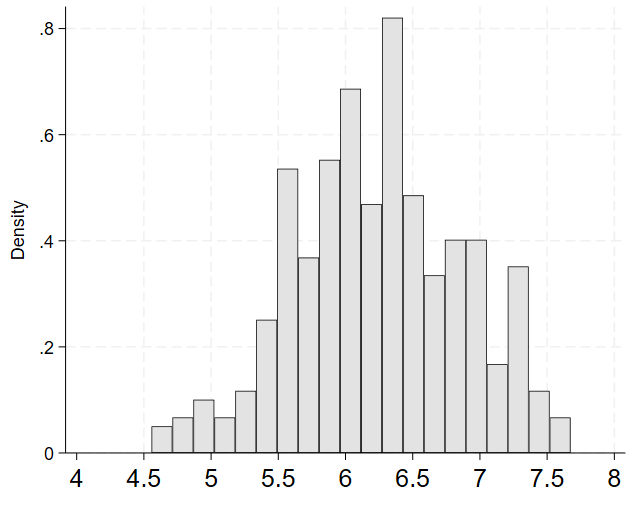}
    \caption{$\log\left(W_{jt,D}\right)$}\label{fig:WD}
\end{subfigure}
\hfill
\begin{subfigure}[h]{0.42\textwidth}
    \centering
    \includegraphics[width=\linewidth]{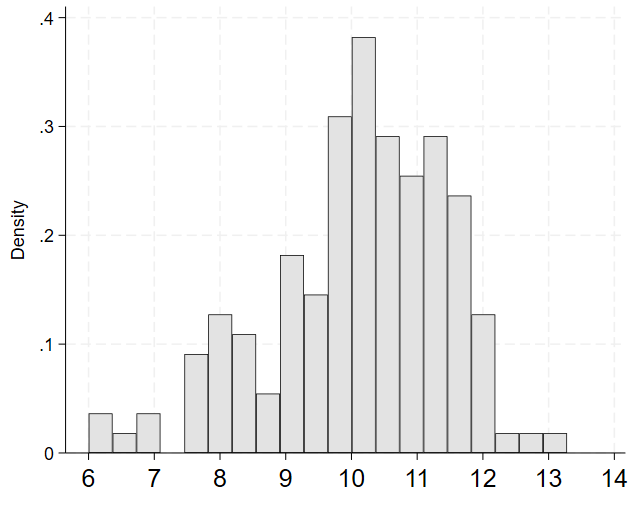}
    \caption{$\log\left(\frac{\Pi_{jt}}{D_{jt}}\right)$}\label{fig:PI}
\end{subfigure}
\caption*{\scriptsize \textit{Note:} Panel (A) plots the empirical distribution of one plus the inverse elasticity of labor supply for permanent workers, $\left(\frac{\partial W_{jt,D}}{\partial D_{jt}}\frac{D_{jt}}{W_{jt,D}}+1\right)$; panel (B) plots the elasticity of union surplus to permanent employment, $\frac{\partial (U^U_{jt}-U^U_{Ojt})}{\partial D_{jt}}\frac{D_{jt}}{(U^U_{jt}-U^U_{Ojt})}$; panel (C) plots log wages for permanent workers, $\log\left(W_{jt,D}\right)$; panel (D) plots log profits per permanent worker workday, $\log\left(\frac{\Pi_{jt}}{D_{jt}}\right)$.}
\end{figure}

Panel \ref{fig:elasD} displays the distribution of one plus the inverse elasticity of labor supply for permanent workers, $\left(\frac{\partial W_{jt,D}}{\partial D_{jt}}\frac{D_{jt}}{W_{jt,D}}+1\right)$, which enters the wage variable, while panel \ref{fig:elasU} shows the elasticity of union surplus to permanent employment, $\frac{\partial (U^U_{jt}-U^U_{Ojt})}{\partial D_{jt}}\frac{D_{jt}}{(U^U_{jt}-U^U_{Ojt})}$, which enters the bargaining variable. Both distributions exhibit similar ranges: 1 to 2.5 for the former and 1 to 1.5 for the latter.

The scale difference emerges predominantly from the gap between wages and profitability for permanent workers. Panels \ref{fig:WD} and \ref{fig:PI} present the empirical distributions of these variables in logs. Log wages for permanent workers, $\log\left(W_{jt,D}\right)$,  have an average value of 6.24, while log profits per permanent worker workday, $\log\left(\frac{\Pi_{jt}}{D_{jt}}\right)$, average 10.11. This four-log-point difference implies that average permanent worker profitability exceeds wages by a factor of 55.

